\documentstyle[aps,prb,epsf,floats]{revtex}
\begin{document}
\twocolumn[\hsize\textwidth\columnwidth\hsize\csname@twocolumnfalse%
\endcsname
\title{Bond operator theory of doped antiferromagnets:\\
from Mott insulators with bond-centered charge order,\\ to
superconductors with nodal fermions}
\author{Kwon Park$^1$ and Subir Sachdev$^{1,2}$}
\address{$^1$Department of Physics, Yale University, P.O. Box 208120, New
Haven, CT 06520-8120\\
$^2$Department of Physics, Harvard University, Cambridge MA 02138}
\date{August 1, 2001}

\maketitle

\begin{abstract}
The ground states and excitations of two-dimensional insulating
and doped Mott insulators are described by a bond operator
formalism. While the method represents the degrees of freedom of
an arbitrary antiferromagnet exactly, it is especially suited to
systems in which there is a natural pairing of sites into bonds,
as in states with spontaneous or explicit spin-Peierls order (or
bond-centered charge order). In the undoped insulator, as
discussed previously, we obtain both paramagnetic and
magnetically-ordered states. We describe the evolution of
superconducting order in the ground state with increasing
doping---at low doping, the superconductivity is weak, can
co-exist with magnetic order, and there are no gapless spin 1/2
fermionic excitations; at high doping, the magnetic order is
absent and we obtain a BCS $d$-wave superconductor with gapless
spin 1/2, nodal fermions. We present the critical theory
describing the onset of these nodal fermionic excitations. We
discuss the evolution of the spin spectrum, and obtain regimes
where a spin 1 exciton contributes a sharp resonance in the
dynamic spin susceptiblity. We also discuss the experimental
consequences of low-energy, dynamically fluctuating, spin-Peierls
order in an isotropic CuO$_2$ plane---we compute consequences for
the damping and dispersion of an optical phonon involving
primarily the O ions, and compare the results with recent neutron
scattering measurements of phonon spectra.
\end{abstract}
\pacs{PACS numbers:} ]


\section{Introduction}
\label{intro}
By now, it is reasonably well established that the
doped antiferromagnets found in the cuprate compounds have a
superconducting ground state with a $d$-wave pairing symmetry.
Moreover, many low temperature ($T$) properties appear to be well
described in the framework of the conventional BCS theory of
$d$-wave superconductors.  There are a number of fascinating
properties at temperatures above $T_c$ (the critical temperature
for the onset of superconductivity) which are not well understood,
but there are numerous plausible candidate theories for these,
involving crossovers between different competing orders in doped
antiferromagnets \cite{science}.

However, there are some low $T$ properties of the superconducting
state that do not appear naturally in the traditional BCS
framework. Among these are ({\em a\/}) the appearance of a $S=1/2$
moment near non-magnetic Zn or Li impurties in the underdoped
region \cite{bobroff,alloulold}, ({\em b\/}) the presence of low
energy collective spin excitations (a $S=1$ spin exciton) at
$(\pi,\pi)$ and nearby incommensurate wavevectors, and ({\em c\/})
instabilities to various co-existing spin and charge density wave
states. While it is possible to ``cook-up'' a microscopic
Hamiltonian, and a corresponding Hartree-Fock treatment, to
generate any of these physical properties in the superconducting
state, a proper understanding of the physics should require that
they emerge naturally from some deeper principle.

In a series of
papers~\cite{rs1,sr,rodolfo,vojtaprl,sbv,lt22,icmp}, it has been
proposed that all of these unusual properties of the
superconductor emerge very naturally in a theory of the doped Mott
insulator. The theory is best understood in two distinct steps. In
the first step, we disrupt the magnetic N\'{e}el long-range order
in the insulator by adding some frustrating exchange interactions.
Then, in step two, we dope the paramagnetic Mott insulator
obtained in step one with mobile charge carriers.

Considerable theoretical progress has been made in understanding
the first step: the quantum transition involving the destruction
of N\'{e}el long-range order in the half-filled insulator. It has
been argued that the first paramagnetic state on the other side of
such a quantum critical point generically has the following
properties:
 \\
(i) a stable $S=1$ particle excitation (an exciton) and permanent
confinement of $S=1/2$ spinons,\\
(ii) broken translational symmetry due to the appearance of a
bond-centered charge ({\em e.g.} spin-Peierls) order, and\\
(iii) confinement of a $S=1/2$ moment in
the vicinity of non-magnetic Zn/Li impurities.

Note the similarity of the features (i)-(iii) of the insulator to
the properties ({\em a})-({\em c}) of the superconductor. The
essence of step two is then very simple: properties (i)-(iii)
survive for a finite range of doping the paramagnetic Mott
insulator, and this provides a ``natural'' understanding of
properties ({\em a})-({\em c}) of the superconductor.

We now present a somewhat more detailed discussion of step two:
doping the confined Mott insulator with mobile charge carriers.
Apart from a possible insulating Wigner crystal ground states at
very low doping (such a Wigner crystal state must appear in the
presence of long-range Coulomb interactions), the ground state is
expected to be superconducting \cite{sr,fradkiv,vojtaprl}, and
this will also be the case in the calculations in the present
paper. The pairing amplitude is found to be $d$-wave
like\cite{sr,vojtaprl} {\em i.e.\/} the pairing amplitude has
opposite signs in the horizontal and vertical directions. By
adjusting the strength of the Coulomb repulsion between the holes,
we can modify the characteristic size of the hole pairs. Only for
the larger hole pairs does the superconducting ground state posses
gapless nodal fermionic excitations. We discuss the fate of
(i)-(iii) in the doped antiferromagnet, along with connections to
the
experiments in turn:\\
(i) \underline{$S=1$ collective spin exciton}: In the confined
insulator, the lowest energy excitation above the spin gap is a
stable, $S=1$, bosonic exciton with a minimum in its dispersion at
$(\pi,\pi)$. Upon doping to the superconductor, additional gapped
$S=1/2$ fermionic excitations are expected to appear
\cite{troyer,poil} (these are the Bogoliubov quasiparticles). For
sufficiently large doping, the fermionic spectral becomes gapless
at four nodal points in the Brillouin zone, and so the global spin
gap vanishes. In general, the $S=1$ boson will be unstable to
decay in into two of these fermionic excitations. However,
constraints from momentum conservation can (and do) protect the
integrity of the $S=1$ exciton in certain regions of the Brillouin
zone. In particular, the $S=1$ exciton is likely to be stable near
momenta $(\pi, \pi)$; this can happen even if the fermionic
excitations are gapless, provided the spacing between the nodal
points does not equal $(\pi,\pi)$. Experimentally a $S=1$ neutron
scattering resonance is indeed observed in the $d$-wave
superconductors \cite{rossat1,mook1,tony3,bourges,he}, and it is
our contention that this excitation is continuously connected to
that in the spin-Peierls insulator. We will obtain explicit
results for the evolution of the bosonic and fermionic spin
excitations in this paper, from the spin-Peierls insulator, to the
superconductor with gapless
nodal excitations.\\
(ii) \underline{Broken translational symmetry}: In the simplest
scenario \cite{sr,vojtaprl}, which can be realized for a range of
couplings in models without long-range Coulomb interactions, the
bond-centered charge (or spin Peierls) order of the paramagnetic
insulator survives in the doped superconductor all the way upto a
critical doping at which full square lattice symmetry is restored
in a transition to the $d$-wave superconductor; with Coulomb
interactions, Wigner crystal states also appear for a range of
very small hole doping, but otherwise the situation is similar.
The critical theory of the vanishing of the bond-centered (or
site-centered) CDW order in the $d$-wave superconductor has been
discussed elsewhere\cite{vojtaprl}. For other parameters, more
complex striped states are also possible, with a period larger
than 2 sites and with modulation of the hole density on the sites.
Experimentally, charge stripe states of period 4 have been clearly
observed \cite{period4,ek}, but it is not yet established whether
the modulation of the spin density is site or bond centered. More
recently, the observation of McQueeney {\em et al.} \cite{egami}
in optimally doped and superconducting LSCO have been interpreted
using a picture of bond-centered charge stripes of period 2, like
those found in the spin-Peierls state . We will discuss
these observations further in Section~\ref{phonon}.\\
(iii) \underline{$S=1/2$ moment near Zn/Li impurities}: In
principle, it is possible that the $S=1/2$ moment confined near a
Zn/Li impurity in the paramagnetic Mott insulator disappears
immediately at an infinitesimal hole doping concentration: one
hole can be trapped near the Zn/Li impurity, and this
configuration is compatible with the global preservation of the
spin-Peierls (or other bond-centered charge-) order. However the
kinetic energy cost, makes this unlikely. Barring this
uninteresting possibility, the $S=1/2$ moment will survive in the
superconductor, and this offers a natural explanation for the NMR
experiments\cite{bobroff}. Eventually, the fermionic $S=1/2$
excitations of the superconductor will Kondo screen the
moment\cite{kondo,tolya,zhu}, but because of the linearly
vanishing fermionic density of states at the Fermi level, this
happens only above finite values of the impurity exchange coupling
and particle-hole asymmetry ({\em i.e.} above a critical doping).
Moreover, there is no fundamental reason for this Kondo screening
transition to co-incide with the point at which translational
symmetry is restored (this transition was discussed above in (i));
the two transitions could occur in either order as a function of
increasing doping

The purpose of this paper is to present a theory of doped
antiferromagents which displays the crossover from an insulator at
zero doping (with or without long-range magnetic order) to a
superconductor with gapless, nodal, fermionic $S=1/2$ excitations
at some moderate doping. Further, we require the theory to obey
properties (i)-(iii) directly at the mean-field level. While a
large number of previous theories of doped antiferromagnets have
been presented previously, none of them satisfy all of these
requirements. The studies of Ref~\onlinecite{vojtaprl} were able
to examine the intricate competition between different charge
ordered states and superconductivity---however, the ground states
were well away from a region of magnetic order, and there was no
sharp, collective, $S=1$ excitation in the Gaussian fluctuations
about the mean-field theory. Conversely, approaches which do yield
confinement of spinons and collective $S=1$ excitations at zero or
low doping \cite{rs1,csy,nagaosa} are not easily extended to reach
a superconducting state with $S=1/2$ gapless nodal excitations at
moderate doping.

We now outline the remainder of the paper. In Section~\ref{sec:bo}
we introduce the central formalism of bond operators, and its
application to doped antiferromagnets. The main results of bond
operator theory of the square lattice antiferromagnet are
presented in Section~\ref{sec:res}, with most details of the
mean-field calculations being relegated to Appendices~\ref{2leg},~
ref{square}, and~\ref{coulomb}: the important phase diagram is in
Fig~\ref{phasediagram}. The critical theory for the onset of nodal
fermion excitations is in Section~\ref{nodal}.
Section~\ref{phonon} differs from the remainder of the paper in
that it considers systems in which the bond-centered charge order
is {\em not} present. Instead it considers the case when the
spin-Peierls order is dynamically fluctuating and describes its
influence on the optical phonon spectrum of the CuO$_2$ plane---we
find that the calculated phonon spectrm is indeed strikingly
similar to recent neutron scattering observations. The main
conclusions are stated in Section~\ref{conc}. Finally, in
Appendix~\ref{nematic} we describe the interplay between the
spin-Peierls states considered in this paper, and states with
electronic ``nematic'' order.

\section{Bond operators} \label{sec:bo}

Our approach is a generalization of the bond operator theory of
Ref~\onlinecite{sb} (and the related work of
Ref~\onlinecite{andrey}) to doped antiferromagnets. The earlier
work \cite{sb} was designed for insulating systems, and has since
been applied succesfully in a number of studies of spin ladder and
related compounds \cite{rgs,grs,ps,valeri,mgi,tmu,cb,so5}, and
also in bilayer quantum Hall systems\cite{eugene,sommer} . Here we
shall extend the formalism to doped antiferromagnets; a closely
related extension was discussed by Lee {\em et al.} \cite{lee} but
only applied to one-dimensional systems. Eder\cite{eder}, Sushkov
\cite{sushkov2}, and Vojta and Becker \cite{vb} have also
considered doped systems by associated methods. Recently, Jurecka
and Brenig\cite{brenig} have used a bond operator method to
analyze a Kondo lattice model using a formulation that bears some
similarity to ours.

While our bond operator formalism can, in principle, be applied to
an arbitrary doped antiferromagnet, it induces a bias by using a
basis of states which explicitly refer to a preferred, disjoint
pairing of all the sites. A full and exact solution of the bond
operator Hamiltonian should restore the full symmetry of the
underlying Hamiltonian in which this preferred pairing may be
absent. However, in practice, this restoration of symmetry is
difficult to achieve, and this is the principal shortcoming of the
bond operator method. So the main utility of the approach lies in
treating systems in which there is a natural pairing of sites in
the ground state, either imposed by the Hamiltonian, or by a
spontaneous symmetry breaking. In this paper, we will restrict
consideration to the case where the ground state possesses
bond-centered charge density wave (spin-Peierls) order: there is
evidently a natural pairing of sites in such a structure. The
spin-Peierls order can either be spontaneous or explicit; the
latter is the case in the doped two-leg ladder compounds like
SrCu$_2$O$_3$ and Sr$_{14-x}$Ca$_{x}$Cu$_{24}$O$_{41}$. In the
presence of this background spin Peierls order, the method can
then address the competition between magnetic N\'{e}el order and
superconductivity, and follow the evolution of the fermionic
$S=1/2$ and the bosonic $S=1$ excitations.

We will now introduce the formalism by showing the exact mappings
between bond and site operators of a pair of sites. In hole-doped
antiferromagnets, we can project out all states with 2 electrons
on one site, and this leaves a total of 9 states in a pair of
sites. Let $c^{\dagger}_{1a}$ and $c^{\dagger}_{2a}$ ($a =
\uparrow, \downarrow$) be the electron creation operators on the
two sites. Then, as in the insulator \cite{sb}, we introduce four
bond boson creation operators, $s^{\dagger}$ and
$t_{\alpha}^{\dagger}$ ($\alpha =x,y,z$) which are defined by
($\sigma^{\alpha}_{ab}$ are the Pauli matrices, and
$\varepsilon_{ab}$ is the second-rank antisymmetric tensor with
$\varepsilon_{\uparrow\downarrow} = 1$)
\begin{eqnarray}
s^{\dagger} |v \rangle &=& \frac{1}{\sqrt{2}} \varepsilon_{ab}
c^{\dagger}_{1a} c^{\dagger}_{2b}| 0
\rangle \nonumber \\
t_{\alpha}^{\dagger} |v \rangle &=& \frac{1}{\sqrt{2}}
\sigma^{\alpha}_{bc} \varepsilon_{ca} c^{\dagger}_{1a}
c^{\dagger}_{2b} | 0 \rangle, \label{e1}
\end{eqnarray}
where $|0 \rangle$ is the electron vacuum {\em i.e.} the state
with no electrons on the two sites, while $|v \rangle$ is an
unphysical state in which none of the bond bosons or fermions are
present. To describe the remaining 5 states of the doped
antiferromagnet, we introduce the bond fermionic operators
$h^{\dagger}_{1a}$ and $h^{\dagger}_{2a}$, and the additional bond
bosonic operator $d^{\dagger}$ which are defined by
\begin{eqnarray}
h^{\dagger}_{1a} |v \rangle &=& c_{1a}^{\dagger} |0 \rangle
\nonumber \\
h^{\dagger}_{2a} |v \rangle &=& c_{2a}^{\dagger} |0 \rangle
\nonumber \\
d^{\dagger} |v \rangle &=&  |0 \rangle. \label{e2}
\end{eqnarray}
The operators $s$, $d$, $t_{\alpha}$ all obey the canonical boson
commutation relations, while the $h_{1a}$, $h_{2a}$ obey canonical
fermion relations. Of course, the total space of states in Fock
space of these 5 bosons and 4 fermions is much larger than the 9
states allowed in the doped antiferromagnet. To restrict to the
physical subspace we must impose the single constraint
\begin{equation}
s^{\dagger}s + t^{\dagger}_{\alpha}t_{\alpha}+ h^{\dagger}_{1
a}h_{1 a} +h^{\dagger}_{2 a}h_{2 a} +d^{\dagger}d = 1. \label{e3}
\end{equation}

In the subspace constrained by (\ref{e3}), we can now write down
exact expressions for arbitrary electron operators in terms of the
bond operators. First, for the electron spin operators
\begin{eqnarray}
S_{1\alpha}&=& \frac{1}{2} c^{\dagger}_{1a} \sigma^{\alpha}_{ab} c_{1b} \nonumber \\
S_{2\alpha}&=& \frac{1}{2} c^{\dagger}_{2a} \sigma^{\alpha}_{ab}
c_{2b}, \label{e4}
\end{eqnarray}
we have the following expressions which generalize those in
Ref~\onlinecite{sb}
\begin{eqnarray}
S_{1\alpha} &=&
\frac{1}{2}(s^{\dagger}t_{\alpha}+t^{\dagger}_{\alpha}s -i
\epsilon_{\alpha\beta\gamma} t^{\dagger}_{\beta}t_{\gamma})
+\frac{1}{2}\sigma^{\alpha}_{ab}h^{\dagger}_{1a}h_{1b} \nonumber \\
&\equiv& \tilde{S}_{1\alpha} +\frac{1}{2}\sigma^{\alpha}_{ab}h^{\dagger}_{1a}h_{1b}
\nonumber \\
S_{2\alpha} &=&
-\frac{1}{2}(s^{\dagger}t_{\alpha}+t^{\dagger}_{\alpha}s +i
\epsilon_{\alpha\beta\gamma} t^{\dagger}_{\beta}t_{\gamma})
+\frac{1}{2}\sigma^{\alpha}_{ab}h^{\dagger}_{2a}h_{2b}\nonumber \\
&\equiv & \tilde{S}_{2\alpha}
+\frac{1}{2}\sigma^{\alpha}_{ab}h^{\dagger}_{2a}h_{2b}, \label{e5}
\end{eqnarray}
where $\epsilon_{\alpha\beta\gamma}$ is the third-rank
antisymmetric tensor with $\epsilon_{xyz} =1$. By considering
various matrix elements of the electron creation operators, we can
also obtain
\begin{eqnarray}
c^{\dagger}_{1a}&=& h^{\dagger}_{1a}d+\frac{1}{\sqrt{2}}
\varepsilon_{ab}s^{\dagger}h_{2b}
-\frac{1}{\sqrt{2}}\varepsilon_{ac}\sigma^{\alpha}_{cb}t^{\dagger}_{\alpha}
h_{2b} \nonumber \\
c^{\dagger}_{2a}&=& h^{\dagger}_{2a}d+\frac{1}{\sqrt{2}}
\varepsilon_{ab}s^{\dagger}h_{1b}
+\frac{1}{\sqrt{2}}\varepsilon_{ac}\sigma^{\alpha}_{cb}t^{\dagger}_{\alpha}
h_{1b}. \label{e6}
\end{eqnarray}
Indeed, the expression (\ref{e5}) can be obtained (\ref{e6}) after
repeated application of the constraint (\ref{e3}). In a similar
manner, we can also obtain from (\ref{e6}) (or by direct
consideration of matrix elements) the useful expressions
\begin{eqnarray}
{S}_{1\alpha} { S}_{2\alpha} &=& \tilde{{ S}}_{1\alpha} \tilde{{
S}}_{2\alpha} = -\frac{3}{4} s^{\dagger} s
+\frac{1}{4}t^{\dagger}_{\alpha}t_{\alpha} \nonumber \\
c^{\dagger}_{1a}c_{2a}+c^{\dagger}_{2a}c_{1a}&=&
h^{\dagger}_{1a}h_{2a}+h^{\dagger}_{2a}h_{1a} \nonumber \\
c^{\dagger}_{1a}c_{1a}&=& 1-h^{\dagger}_{2a}h_{2a}-d^{\dagger}d\nonumber \\
c^{\dagger}_{2a}c_{2a}&=& 1-h^{\dagger}_{1a}h_{1a}-d^{\dagger}d.
 \label{e7}
\end{eqnarray}
We have now completed the definition of the bond operator
formalism. By application of (\ref{e5},\ref{e6},\ref{e7}) an
arbitrary Hamiltonian can be written down in terms of the bond
operators. Notice that the right hand sides of all of these
equations commute with constraint (\ref{e3}) and so will the
resulting Hamiltonian. For most systems of interest, the
Hamiltonian will only contain terms which are quartic in the bond
operators--we will merely apply the simplest possible
Hartree-Fock-BCS theory to solve such a model.

At this point, it is worth pointing an important distinction
between the present approach and the familiar ``slave'' boson and
fermion theories. A crucial advantage of our approach is that
there are no long-range gauge forces associated with the
fluctuations about our simple Hartree-Fock-BCS theory: the $s$
boson is strongly condensed in all the phases, and so the $U(1)$
gauge symmetry associated with (\ref{e3}) is always badly broken.
Consequently, the quantum numbers of the bare excitations are also
those of the renormalized quasiparticles; this is rarely the case
in the slave particle approaches. In particular, in the insulating
phases, the main elementary excitation will be the bosonic
$t_{\alpha}$ particle; in the paramagnet this is the $S=1$
exciton, while in the N\'{e}el state it reduces to the two spin
waves. In the superconducting states, we will find that the
$t_{\alpha}$ excitations persist as the $S=1$ spin exciton, and
the $h_{1a}$, $h_{2a}$ quanta turn into the fermionic, $S=1/2$
Bogoliubov quasiparticles. So there is a direct and transparent
relationship between the quantum numbers of the bond operators and
those of the elementary excitations of all the ground states. This
is main benefit of our approach, and we are not aware of any
previous theory which has satisfied these criteria. Of course,
some of the quantum numbers of our confining theory can also be
obtained in a direct Hartree-Fock-BCS theory of an extended
Hubbard model of the electrons. However, this has to be followed
by an RPA analysis of collective modes to obtain the $S=1$
exciton; moreover, in such theories, the largest energy scale in
the problem, the charge gap in the Mott insulator, is unphysically
related to the strength of the magnetic or charge order, and is
hence grossly underestimated.

We conclude this section by noting how the bond operator theory
satisfies criteria (i)-(iii) of Section~\ref{intro}. (i) As just
noted, the $t_{\alpha}$ quanta are gapped $S=1$ excitons in the
insulating paramagnet, and persist as sharp excitations in the
superconductor for a finite range of doping. (ii) Any mean-field
theory using the bond operator formalism will prefer the bonds on
which the operators reside over the others, and this naturally
leads to a broken lattice symmetry for symmetric Hamiltonians.
Indeed, the primary weakness of the bond operator formalism is
that there is no simple way to restore this symmetry. (iii)
Placing a Zn/Li impurity means that the partner of one site has
been removed. The excitations of this site (and only this site)
therefore cannot be described by the bond operators above: instead
we need a fermionic spinor, $h_a$, to create the $S=1/2$ state
with one electron, and a spinless boson, $b$, to represent the
hole. At zero doping, the $h_a$ particle constitutes the free
$S=1/2$ moment near the impurity. This particle will be bound
near the impurity site for a finite range of doping, and so the
moment will persist for a while in the superconductor.

\section{Results}
\label{sec:res}

\begin{figure}
\epsfxsize=3.5in \centerline{\epsffile{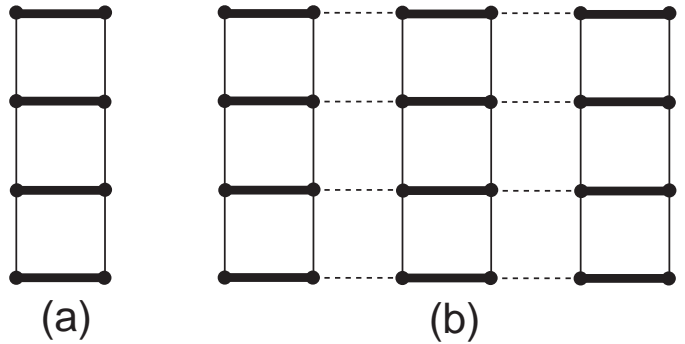}} \caption{
Definition of the Hamiltonian $H$ in (\protect\ref{e8}). The
two-leg ladder is in (a), while the square lattice is viewed as a
set of coupled two-leg ladders in (b). The thick links have
$t_{ij} = t$, $J_{ij} = J$, $V_{ij} = V$, the thin vertical links
have $t_{ij} =  t$, $J_{ij} = \lambda J$, and $V_{ij} = V$, while
the dashed links have $t_{ij} = t'$, $J_{ij} = \lambda' J$, and
$V_{ij} = V$.} \label{fig5}
\end{figure}

We applied the bond-operator method to the $t$-$J$-$V$ model
defined by the Hamiltonian
\begin{eqnarray}
H = \sum_{\langle ij \rangle}&& \left[ -t_{ij} \left(
c_{ia}^{\dagger} c_{ja} + c_{ja}^{\dagger} c_{ia} \right) + J_{ij}
S_{i \alpha} S_{j \alpha} \right. \nonumber \\
&&~~~~~~~~\left. + V_{ij} c_{ia}^{\dagger} c_{ia} c_{jb}^{\dagger}
c_{jb} \right] - \mu \sum_i c^{\dagger}_{ia} c_{ia} , \label{e8}
\end{eqnarray}
where the sum $\langle ij \rangle$ extends over nearest neighbor
pairs on a two-leg ladder or a square lattice, and it is implied
that all states with two electrons on any site have been projected
out. The $t_{ij}$ are the electron hopping matrix elements, the
$J_{ij} > 0$ are the antiferromagnetic exchange interactions, the
$V_{ij}>0$ are nearest neighbor repulsive Coulomb interactions,
and  $\mu$ is the chemical potential. The values of the $t_{ij}$,
$J_{ij}$ and $V_{ij}$ are indicated in Fig~\ref{fig5} for both the
two-leg ladder and the square lattice.
For the two-leg ladder we have $t_{ij} = t$, $J_{ij} = J$, $V_{ij}
= V$ on the horizontal links, and $t_{ij} =  t$, $J_{ij} = \lambda
J$, and $V_{ij} = V$ on the vertical links. We view the square
lattice as a collection of adjacent two-leg ladders: then the
couplings on each ladder are the same as before, while on the
links between the ladders we have $t_{ij} = t'$, $J_{ij} =
\lambda' J$, and $V_{ij} = V$. This parameterization is chosen so
that in both cases the exchange interaction decouples into
disconnected pairs of sites (``dimers'') at $\lambda=\lambda' =0$.
Further, at $\lambda=\lambda'=1$, $H$ has the full symmetry of the
square lattice. However, as discussed earlier, our mean-field
theory will continue to have a ground state with the symmetry of
Fig~\ref{fig5}b even at these values of $\lambda$, $\lambda'$:
this implies the presence of spontaneous bond-centered charge
order of period 2 in the ground state.

The calculation proceeds by the substitution of the operator
representations in Section~\ref{sec:bo} into (\ref{e8}), followed
by a Hartree-Fock-BCS treatment of all the quartic terms. The
procedure is quite lengthy, but the computations are quite similar
to those in earlier work. Details of the calculation are presented
in Appendices~\ref{2leg},~\ref{square} and~\ref{coulomb}.

Here we will discuss the results of such a calculation. All of the
phases can be characterized by a specification of the non-zero
expectation values of various combinations of the bond operators
in the ground state.

All states have the non-zero expectation values
\begin{equation}
\langle s \rangle \neq 0 ~~;~~ \langle t_{\alpha}^{\dagger}
t_{\alpha} \rangle \neq 0;~~ \langle t_{\alpha} t_{\alpha} \rangle
\neq 0. \label{e9}
\end{equation}
These non-zero values do not break any physical symmetries of the
Hamiltonian, and merely serve to break the U(1) gauge symmetry
associated with the constraint (\ref{e3}). As noted earlier, this
is fortunate as no long-range gauge forces then appear in the
fluctuations about our ground states. In addition, all phases will
also have non-zero expectation values of the operators $
t^{\dagger}_{\alpha} t_{\alpha}$ and $h_{1,2a}^{\dagger}
h_{1,2a}$, which do not break any symmetries of the Hamiltonian,
and also commute with the constrains (\ref{e3}).

At zero and non-zero doping we have find also find magnetically
ordered states. These are characterized by the non-zero
expectation value
\begin{equation}
\langle t_{\alpha} \rangle \neq 0 ; \label{e10}
\end{equation}
so condensation of single $t_{\alpha}$ bosons (which are $S=1$
particles) leads to the appearance of magnetic order. This also
implies that the magnetic ordering transition will be described by
the field theory a 3-component spin-vector order parameter.
Further, provided the ordering wavevector is not exactly equal to
the spacing between the points of gapless fermionic excitations,
the universality class of the transition is the same as that in
the O(3) non-linear sigma model \cite{vojtaprl,book}. This is to
be contrasted with magnetic ordering transitions in Schwinger
boson theories, which are associated with the condensation of
$S=1/2$ particles, possibly interacting with each other via gauge
forces.

The singlet superconducting states have the anomalous expectation
values
\begin{equation}
\langle d \rangle \neq 0~~~~;~~~~\varepsilon_{ab} \langle h_{1,2a}
h_{1,2b} \rangle \neq 0 \label{e11}.
\end{equation}
The spatial pattern of these anomalous condensates determines the
symmetry of the Cooper pair wavefunction, and whether there are
any nodal fermionic quasiparticles.

Finally, we also found states with co-existing magnetic order and
superconductivity, in which both (\ref{e10}) and (\ref{e11}) hold.

\begin{figure}
\epsfxsize=3.5in \centerline{\epsffile{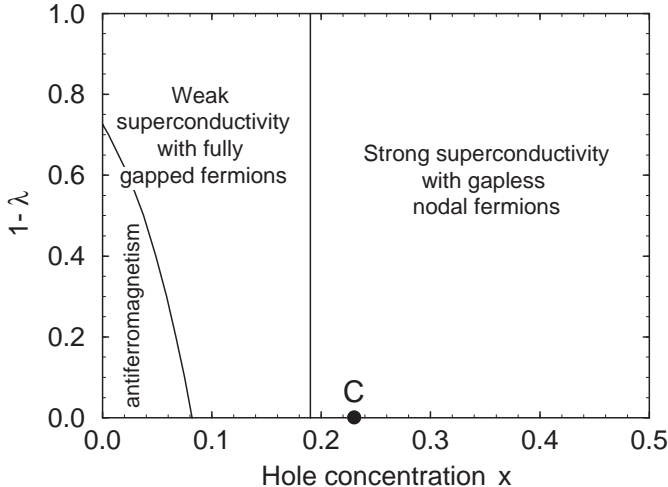}}
\caption{Phase diagram of the square lattice Hamiltonian $H$ in
(\protect\ref{e8}) for the configuration shown in
Fig~\protect\ref{fig5}b. Results are shown here for $t'=t$,
$\lambda=1$ and $t/J=3.0$ as a function of the hole concentration
$x$ and $1-\lambda'$. The model has full square lattice symmetry
at $1-\lambda'=0$, and the exchange interaction separates into
that on decoupled two-leg ladders at $1-\lambda'=1$. The
nearest-neighbor Coulomb repulsion, $V$ is taken to have a very
large positive value, and the results are indistinguishable from
those at $V=\infty$. The phase boundary towards antiferromagnetism
was determined by the point at the $t_{\alpha}$ excitation
spectrum had a vanishing spin gap, signaling the onset of a phase
in which (\protect\ref{e10}) was valid. The ``weak'' and
``strong'' superconductivity distinction is qualitative, and is
indicated by the rapid change in the physical superconducting
order parameter in Fig~\protect\ref{pairingamplitude} around
$x=0.2$. The antiferromagnetic phase co-exists with weak
superconductivity at non-zero $x$. The nearly vertical phase
boundary indicates the quantum phase transition at which nodal
fermions first appear: this transition is discussed in
Section~\protect\ref{nodal}. C denotes the point where the
spontaneous bond-centered charge order of the square lattice is
expected to disappear with increasing $x$; the bond-centered
charge order is explicitly present in the Hamiltonian for
$1-\lambda' > 0$, and is spontaneous only for non-zero $x$ before
the point C and $1-\lambda'=0$. The present bond operator approach
was not used to determine the position of C; nevertheless,
following Ref~\protect\onlinecite{vojtaprl} we are able to present
a theory of the critical properties of C in
Appendix~\protect\ref{nematic}} \label{phasediagram}
\end{figure}

Our main results are summarized in Fig~\ref{phasediagram}.
First, let us review the results at $x=0$, in the insulator. Here
the calculations are very similar to those already considered in
Ref~\onlinecite{sb}. For small $\lambda'$, the ground state is an
insulating paramagnet: the symmetry of the Hamiltonian and the
ground state is that of Fig~\ref{fig5}b, and the $t_{\alpha}$
excitations have an energy gap: these excitations constitute a
$S=1$ collective spin resonance (or a $S=1$ exciton). At a
critical value of $\lambda'$, the excitation gap vanishes, and
long-range N\'{e}el order sets in.

Now we turn to non-zero doping. In our Hartree-Fock-BCS
mean-field theory, we find that superconductivity appears at any
non-zero doping for small $\lambda'$. As one of us has discussed
in Ref.~\onlinecite{vojtaprl} (along with references to the
earlier literature), this is surely an artifact of our
approximation at very small $x$: in the presence of long-range
Coulomb interactions, Wigner crystal states will be present for
very small hole concentration. Even in our theory without
long-range Coulomb interactions, long-range charge
inhomogeneities can also appear in our mean-field framework
\cite{vojtaprl}; however, we will neglect these here for
simplicity. As indicated in Fig~\ref{phasediagram}, the
superconductivity persists in a non-magnetic state even at
$\lambda'=1$ for $x > 0.082$. In our present calculation, this
small $x$ superconducting state has bond-centered charge order of
period 2, but more complex bond-centered charge-ordered states
are also possible \cite{vojtaprl,zacher}. We also expect that the
superconductivity will co-exist with antiferromagnetism, but we
did not undertake a complete solution of the mean-field equations
within the antiferromagnetic phase because of the complexity of
the calculation. As in the insulator, the onset of
antiferromagnetim was determined by the point where the
$t_{\alpha}$ boson had a vanishing gap in its spectrum.

\begin{figure}
\epsfxsize=3.5in \centerline{\epsffile{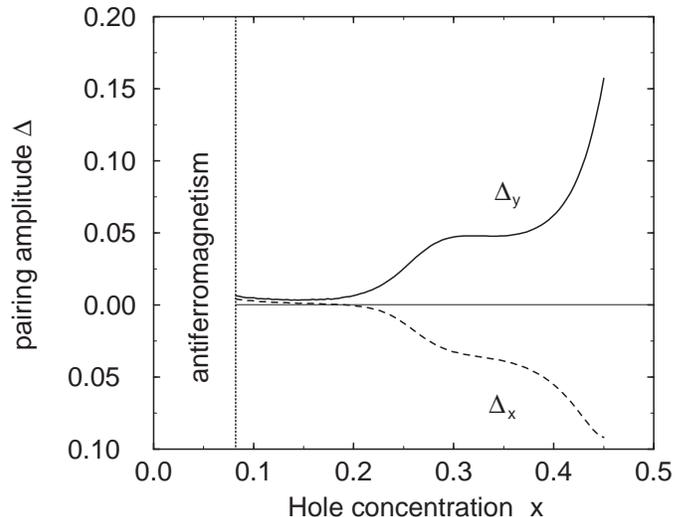}}
\caption{ Pairing amplitudes ($\Delta$) as a function of hole
concentration ($x$) when $t/J =3.0$, and
$\lambda=\lambda^{'}=1.0$, i.e. the isotropic case. The pairing
amplitude in the vertical direction of Fig~\ref{fig5}b is
$\Delta_y$, and that in the horizontal direction is $\Delta_x$.
For $x \leq 0.082$, there is instability to long-range
antiferromagnetism due to the collapse of the gap of $S=1$
particle excitation, $t_{\alpha}$. When $0.082 < x \leq 0.192$,
non-magnetic superconducting state is stable, and the pairing
amplitudes are of the same sign, but of very small value. For
doping concentration larger than roughly 0.2, the pairing
amplitudes have opposite signs, which, in turn, gives rise to
nodal fermions. } \label{pairingamplitude}
\end{figure}

We characterized the superconducting state by determining pairing
amplitudes, $\Delta_{x,y}$, which characterize the pairing of the
holes, $h_{1,2}$, in the $x$ and $y$ directions. These quantities
are described more precisely in Appendices~\ref{2leg}
and~\ref{square}. Their values are plotted as a function of the
hole concentration, $x$, in Fig~\ref{pairingamplitude}.
Note that $\Delta_x$ and $\Delta_y$ are roughly of the same
magnitude: this means that the superconductivity is genuinely
two-dimensional, and there is no regime in which a
quasi-one-dimensional Luther-Emery liquid-like behavior holds.
Indeed, it is even possible for $\Delta_x$ to be larger than
$\Delta_y$, which would make the superconductivity stronger in
the $x$ direction rather than in the vertical ``stripe''
direction. This latter phenomenon appears to be similar in spirit
to the ananlyses of
Refs~\onlinecite{campbell1,campbell2,campbell3}.

For small $x$, the pairing amplitudes are non-zero but quite
small; moreover, the pairing amplitudes in the horizontal and
vertical directions have the same sign in our mean field theory.
We will refer to this as the weak superconductivity regime. It is
possible that, upon including quantum fluctuations, the ground
state in this weak pairing regime is easily susceptible to a
quantum transition to an insulating state with some additional
translational symmetry breaking. Also note that the pairing
amplitude is nonzero at the boundary of the onset of
antiferromagnetic order: we therefore expect the
superconductivity to survive within the antiferromagnetic phase.

For larger $x$, the pairing amplitude in
Fig~\ref{pairingamplitude} increases rapidly and we reach a
``strong superconductivity'' regime. The pairing amplitudes now
have opposite signs in the horizontal and vertical directions, and
this permits gapless nodal fermionic quasiparticle excitations, as
shown explicitly in our results below. However, there is one
important feature associated with the appearance of the nodal
particles that is worth emphasizing. Originally, our calculations
were carried out with the nearest-neighbor Coulomb repulsion
$V=0$. In this case we found that the pairing of the holes
occurred primarily by the condensation of the $d$ bosons. Because
of the resulting very short-range pairing we found no nodal
points, even though the pairing amplitude had opposite signs in
the horizontal and vertical directions. It was only when we had
turned on a $V$ of the order of the bandwidth, which significantly
reduced the amplitude of the $d$ boson condensate and made the
hole pairing more long-ranged, did we find the appearance of the
nodal excitations.

\begin{figure}
\epsfxsize=3.5in \centerline{\epsffile{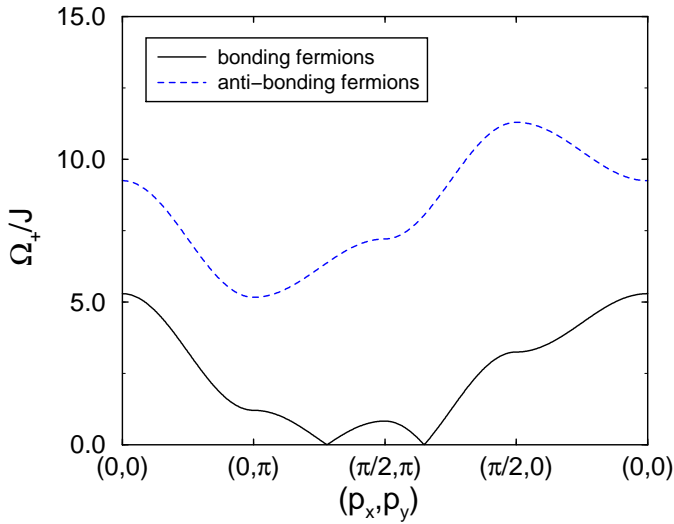}}
\caption{Dispersion curves of the bonding and anti-bonding $=1/2$
fermionic excitations in the weak pairing regime with $x=0.1$,
$t/J=3.0$, and $\lambda=\lambda^{'}=t^{'}/t=1$. There appear to
be fermionic excitations at energies very close to 0, but the
blow-up in Fig~\protect\ref{hweak2} shows that there is indeed a
gap in the fermionic spectrum. The $x$-axis is chosen to take a
representative straight path connecting points in the first
Brillouin zone: from $(p_x,p_y)= (0,0)$ to $(0,\pi)$ to
$(\pi/2,\pi)$ to $(\pi/2,0)$ to $(0,0)$. } \label{hweak1}
\end{figure}

\begin{figure}
\epsfxsize=3.5in \centerline{\epsffile{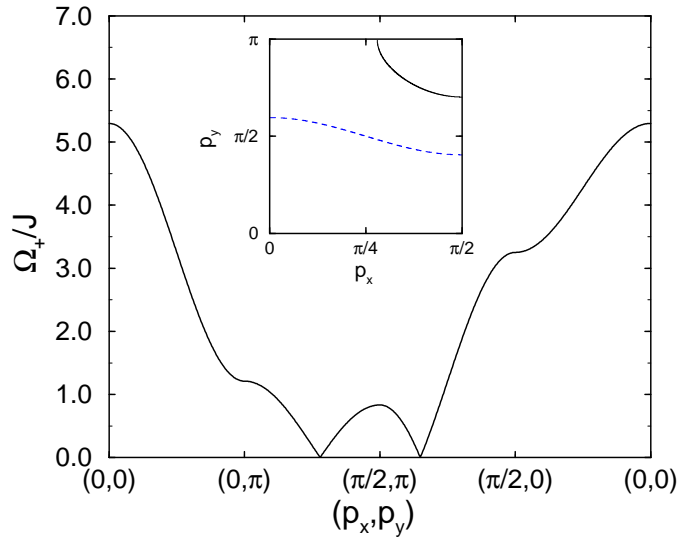}}
\caption{Blow-up of the fermionic dispersion in
Fig~\protect\ref{hweak1} at the lowest non-zero energies. Notice
that there is indeed a finite gap in the fermionic spectrum. This
is also made clear in the inset: the solid line shows the line
where the fermion kinetic energy vanishes (the incipient Fermi
surface) while the dashed line indicates where the fermion pairing
amplitude vanishes. For gapless excitations, both quantities have
to vanish, and the absence of a crossing point between the lines
indicates that a gap is always present. The solid line in the
inset is also the location of the incipient Fermi surface which
has been quenched by pairing---the position of this surface is
similar to that in Ref~\protect\onlinecite{sushkov2}.}
\label{hweak2}
\end{figure}

We now describe our results for the elementary excitations of the
superconducting states in our phase diagram. First, we discuss the
fermionic $S=1/2$ excitations; in principle, these are observable
in photoemission or tunneling experiments. Note that the unit cell
of Fig~\ref{fig5}b has two sites, and so the first Brillouin zone
extends between $-\pi/2$ and $\pi/2$ in the $x$ direction, and
between $-\pi$ and $\pi$ (as usual) in the $y$ direction.
Consequently, half the fermionic excitations in the first
Brillouin zone of the original square lattice will be folded back,
by a Bragg reflection in the vertical planes at $\pm \pi/2$, into
a second band in the first Brillouin zone of the lattice of
Fig~\ref{fig5}b. We will refer to these two bands as the `bonding'
and `anti-bonding' bands, based upon their wavefunctions within
the dimers, as discussed in Appendix~\ref{2leg}. We show the
$S=1/2$ fermionic bands in the weak-pairing regime in
Figs~\ref{hweak1} and~\ref{hweak2}.
There is a gap across the entire Brillouin zone. However, because
of the very small pairing amplitude, the fermionic excitation
energy becomes very small along an incipient ``Fermi surface'' in
the Brillouin zone. This Fermi surface line is indicated by the
solid line in the inset of Fig~\ref{hweak2}. The position of this
Fermi surface is similar to that in the computation by
Sushkov~\cite{sushkov2}, who also discussed its relationship to
photoemission experiments.

\begin{figure}
\epsfxsize=3.5in \centerline{\epsffile{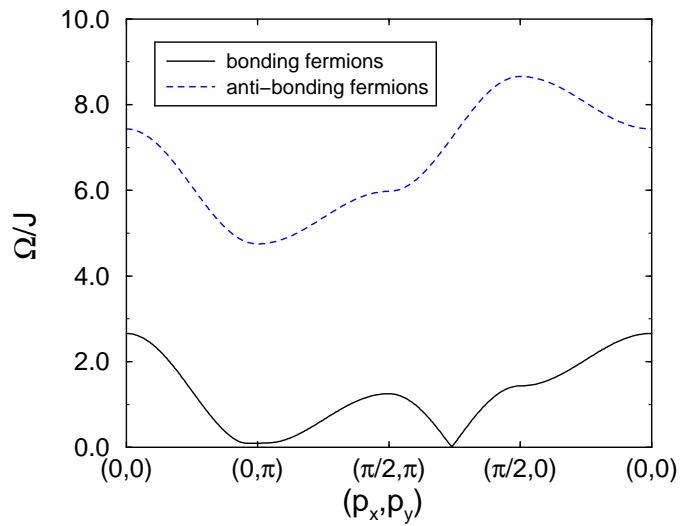}}
\caption{Dispersion curves of the bonding and anti-bonding
$S=1/2$ fermionic excitations with $x=0.3$, $t/J=3.0$, and
$\lambda=\lambda^{'}=t^{'}/t=1$ (as in Fig~\protect\ref{hweak1}
but at a larger doping). At this value of $x$, there is a nodal
quasiparticle excitation not too far from $(\pi/2,\pi/2)$.}
\label{hstrong1}
\end{figure}

\begin{figure}
\epsfxsize=3.5in \centerline{\epsffile{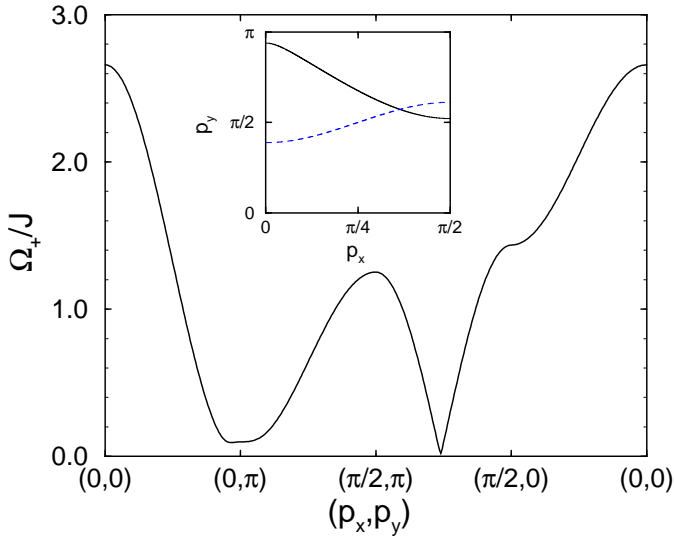}}
\caption{ Blow-up of the fermionic dispersion in
Fig~\protect\ref{hstrong1} at the lowest non-zero energies (as in
Fig~\protect\ref{hweak2} but at a larger doping). Notice that
there is clearly a gapless nodal point not too far from
$(\pi/2,\pi/2)$, as is also clear from the inset, where the two
curves now cross. The pairing gap near the ``antinodal point'',
$(\pi,0)$, remains finite.} \label{hstrong2}
\end{figure}

Next, we turn to a discussion of the fermionic excitations at
larger $x$, where the superconductivity is stronger. The $S=1/2$
excitation spectra are now shown in Figs~\ref{hstrong1}
and~\ref{hstrong2}.
Overall, the results at higher energies, are quite similar to
those in Figs~\ref{hweak1} and~\ref{hweak2}, but there is now a
dramatic difference at lower energies. There is a gapless
nodal-point near $(\pi/2, \pi/2)$, and this is made possible by
the $d$-wave-like pairing amplitudes. The spectra in
Fig~\ref{hweak2} and~\ref{hstrong2} cannot be continuously
connected, and there must be a singular phase transition point at
which the nodal fermion first appears. The position of this phase
boundary is indicated in Fig~\ref{phasediagram}, and the nature
of the phase transition will be discussed below in
Section~\ref{nodal}.

\begin{figure}
\epsfxsize=3.5in \centerline{\epsffile{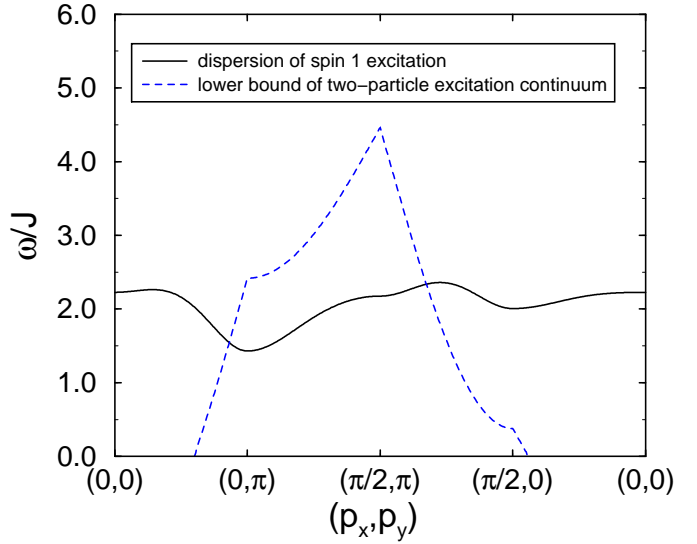}}
\caption{Dispersion curve (full line) of $S=1$, $t_{\alpha}$
particle excitation with $x=0.1$, $t/J=3.0$, and
$\lambda=\lambda^{'}=t^{'}/t=1$. The pairing is weak for this
value of doping. Also shown (dashed line) is the lower bound of
the two-particle continuum made up of a pair of $S=1/2$
excitations from Fig~\protect\ref{hweak1} and~\ref{hweak2}. Note
that the minimum of the $t_{\alpha}$ excitation is at $(0,\pi)$,
but because of the halving of the Brillouin zone this wavevector
is crystalographically equivalent to $(\pi,\pi)$.} \label{tweak}
\end{figure}

\begin{figure}
\epsfxsize=3.5in \centerline{\epsffile{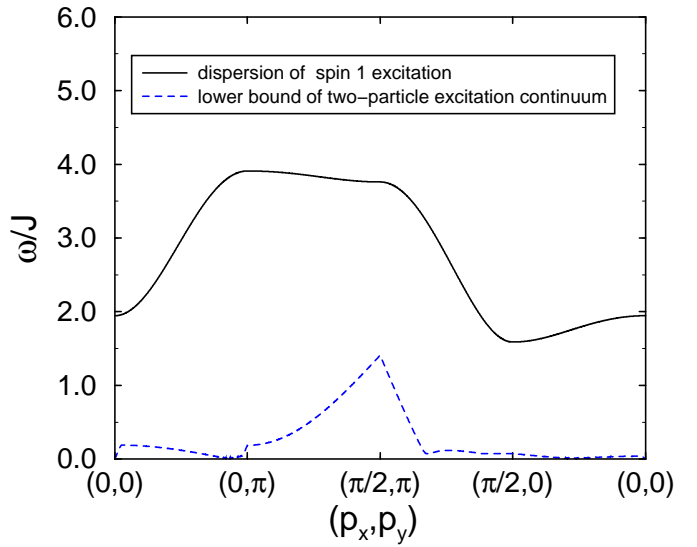}}
\caption{As in Fig~\protect\ref{tweak}, but in the strong pairing
region with nodal quasiparticles at $x=0.3$.} \label{tstrong}
\end{figure}

Finally, we consider the $S=1$ excitations, which are visible in
neutron scattering experiments, There are two categories of such
excitations. First, we have the $S=1$, $t_{\alpha}$ particles,
with a definite energy-momentum relation: these may be viewed as a
collective spin mode (or a triplet exciton) which goes soft at the
antiferromagnetic ordering transition. These are the only $S=1$
excitations in the undoped antiferromagnet, but, connect smoothly
to corresponding excitations in the doped antiferromagnet. The
second class of $S=1$ excitations are scattering states of the
fermionic $S=1/2$ excitations just discussed. These do not have a
definite energy momentum relation, but exist over a continuum in a
range of energies at any given momentum. We show a plot of the
spectra of these two $S=1$ excitations in Figs~\ref{tweak}
and~\ref{tstrong}.

At low doping, in Fig~\ref{tweak}, not that the $t_{\alpha}$
excitation has a well-formed minimum at $(0, \pi)$, and this
minimum is {\em below} the lower-bound of the two-particle
continuum. So the $t_{\alpha}$ excitations are absolutely stable
towards decay into 2 fermionic $S=1/2$ excitations over this
range of energy and momenta. This stable $S=1$ excitonic
excitation will lead to a resonance peak (in principle, infinitely
sharp) in the neutron scattering cross-section at the energy and
momentum of the $t_{\alpha}$ particle. Also, as the doping is
lowered, the minimum of the $t_{\alpha}$ dispersion gets lowered,
until it hits zero also in a region which has no overlap with the
fermionic two-particle continuum. This signals the transition to
the antiferromagnetic phase: the absence of low energy fermionic
excitations at near the minimum $t_{\alpha}$ energy implies that
the critical theory for the transition remains identical to that
for the corresponding transition in the insulator, which has been
discussed in much detail \cite{vojtaprl,book}. The condensation
of the $t_{\alpha}$ bosons occurs at momentum $(0,\pi)$, but note
that because of the doubling of the unit cell this momentum is
equivalent to $(\pi,\pi)$. Determination of the microscopic
configuration of the spins using the operator representation in
Section~\ref{sec:bo} shows that the mean moments indeed have the
staggered arrangement associated with the conventional,
two-sublattice N\'{e}el state.

The higher doping $S=1$ excitations are sketched in
Fig~\ref{tstrong}. Now the $t_{\alpha}$ excitations are well
within the two-fermionic-particle continuum, and so will quickly
decay and lose their identity. Note that the doping at which the
$t_{\alpha}$ excitations ceases to be stable at any momentum does
not (in general) co-incide with the point at which nodal
fermionic quasiparticles appear. As long as the nodal points are
not exactly at $(\pi/2, \pi/2)$, the $t_{\alpha}$ excitations can
be stable at very low energies near $(\pi, \pi)$. In the present
computation, the nodal points are quite close to $(\pi/2,\pi/2)$
and to the $t_{\alpha}$ excitation is quickly quenched. In a more
general model, with second neighbor hopping, the $t_{\alpha}$
excitations could be stable even in an isotropic $d$-wave
superconductor, provided we moved the nodal points sufficiently
far from $(\pi/2,\pi/2)$.

\section{Onset of gapless nodal fermionic excitations}
\label{nodal}

The phase diagram of Fig~\ref{phasediagram} shows a phase boundary
demarking superconductors with and without nodal quasiparticles.
This is a quantum phase transition in the sense that there is a
(weak) non-analyticity in the ground state energy as a function of
doping at this point. We will present a simplified treatment which
captures its essential universal features. Such a transition was
discussed earlier in Ref~\onlinecite{vojtaprl} and by Granath {\em
et al.} \cite{granath}; the latter authors also reached
conclusions on the universal properties which agree with our
discussion here. A related, but distinct, theory for the
annihilation of nodal particles in a $d$-wave superconductor was
also discussed by Duncan and S\'{a} de Melo \cite{carlos}. They
considered an isotropic superconductor as a function of electron
density, and found that the nodal points vanished when all four of
them collided at $k=0$. This transition is differs from that in
the anisotropic superconductors of interest here, where the nodal
points collide only in pairs, and the quantum critical points
belong to distinct universality classes.

We approach this transition from the side of the $d$-wave
superconductor with full square lattice symmetry. Here, the
fermionic excitations are described by the BCS Hamiltonain
\begin{equation}
H_{BCS} = \sum_{k} \left[ \epsilon (k) c_{k a}^{\dagger} c_{k a} +
\frac{\Delta (k)}{2} \left( \varepsilon_{ab} c^{\dagger}_{k a}
c^{\dagger}_{-k b} + \mbox{H. c.} \right) \right], \label{e12}
\end{equation}
where, in the simplest nearest-neighbor model $\epsilon_k = - 2t
(\cos k_x + \cos k_y) -\mu$, and the pairing energy $\Delta (k) =
\Delta_0 ( \cos k_x - \cos k_y )$. We now assume that there is an
onset of bond-centered charge order at wavevector $G = (\pi/a,
0)$ of amplitude $\psi_{\text{sp}}$---because this charge order
wave is in the $x$ direction, $\psi_{\text{sp}}$ is the real part
of the more general order parameter, $\Psi_{\text{sp}}$
considered in Section~\ref{intro} and Appendix~\ref{nematic}. This
order will lead to a modulation in the fermion hopping matrix
element and the pairing interaction at the wavevector $G$, and so
induce the following additional terms in the Hamiltonian for the
fermionic excitations
\begin{eqnarray}
H_{\text{sp}} =&& i \psi_{\text{sp}} \sum_{k} \biggl[ a (k) c_{k,
a}^{\dagger} c_{k+G, a} \nonumber \\
&&~~~~~~~~~~~~~+ \frac{b (k)}{2} \left( \varepsilon_{ab}
c^{\dagger}_{k a} c^{\dagger}_{-k+G, b} + \mbox{H. c.} \right)
\biggr], \label{e13}
\end{eqnarray}
where $a(k) = w_1 \sin k_x$ and $b(k) = w_2 \sin k_x$, with
$w_{1,2}$ some constants. These last factors of $\sin k_x$ are a
consequence of the bond-centered nature of the charge order
\cite{chetan}, but the results of this section are not crucially
dependent upon this fact; similar results will apply also to
site-centered charge orders.

It is possible to diagonalize the Hamitonian $H_{BCS} +
H_{\text{sp}}$ and determine the fermionic excitation spectrum of
the state with co-existing superconductivity and charge order.
The energy eigenvalues are
\begin{equation}
\left[ \Lambda_1 \pm \left( \Lambda_1^2 - \Lambda_2^2 -
\Lambda_3^2 \right)^{1/2} \right]^{1/2} \label{e14}
\end{equation}
where
\begin{eqnarray}
\Lambda_1 &\equiv& \left[\epsilon^2 (k) + \epsilon^2 (k+G) +
\Delta^2 (k) + \Delta^2 (k+G)\right]/2 \nonumber \\
&~&~~~~~~~~~~~+ \psi_{\text{sp}}^2
\left[a^2 (k) + b^2 (k)\right]  \nonumber \\
\Lambda_2 &\equiv & \Delta(k) \epsilon(k+G) + \Delta(k+G)
\epsilon(k) - 2 \psi_{\text{sp}}^2 a(k) b(k) \nonumber \\
\Lambda_3 &\equiv & \psi_{\text{sp}}^2 \left[a^2 (k) - b^2
(k)\right] \nonumber \\
&~&~~~~~~~~~~~+ \Delta(k) \Delta(k+G) - \epsilon(k) \epsilon(k+G)
\nonumber.
\end{eqnarray}

It is instructive to examine the evolution of the zeros of
(\ref{e14}) as a function of $\psi_{\text{sp}}$. For
$\psi_{\text{sp}} = 0$, there are four symmetric nodal points
determined by the solutions of $\epsilon(k)=0$ and $\Delta(k)=0$.
It is assumed (as is the case at non-zero doping and in the
absence of particle-hole symmetry), that the wavevector
separating any two of these nodal points is not equal to $G$. If
it was equal to $G$, then the nodal points would be gapped at an
infinitesimal value of $\psi_{\text{sp}}$, and a non-trivial
theory would describe the quantum critical fluctuations, as has
already been discussed in Ref~\onlinecite{vojtaprl}. Turning to
the more general situation, where the separation between the
nodal points is not equal to $G$, the quantum critical theory
describing the onset of a non-zero $\psi_{\text{sp}}$ is reviewed
in Appendix~\ref{nematic}---the leading critical singularities do
not involve the nodal fermions. As $\psi_{\text{sp}}$ increases,
examination of (\ref{e14}) shows that the four nodal points move
towards the new Brillouin zone boundary at $k_x = \pm \pi/2$.
Eventually, these points collide in pairs at
\begin{equation}
p_0 = (\pi/2, p_{0x}) \label{e14a}
\end{equation}
(and symmetry related points) and $\psi_{\text{sp}} =
\psi_{\text{sp}}^c$, and there are no nodal points for
$\psi_{\text{sp}} > \psi_{\text{sp}}^c$. We are interested here in
describing the quantum critical theory of this nodal point
collision. Clearly, the fluctuations of $\psi_{\text{sp}}$ will
not be critical at this point, as the transition occurs at a
non-zero value of $\psi_{\text{sp}}^c$.

\begin{figure}
\epsfxsize=3in \centerline{\epsffile{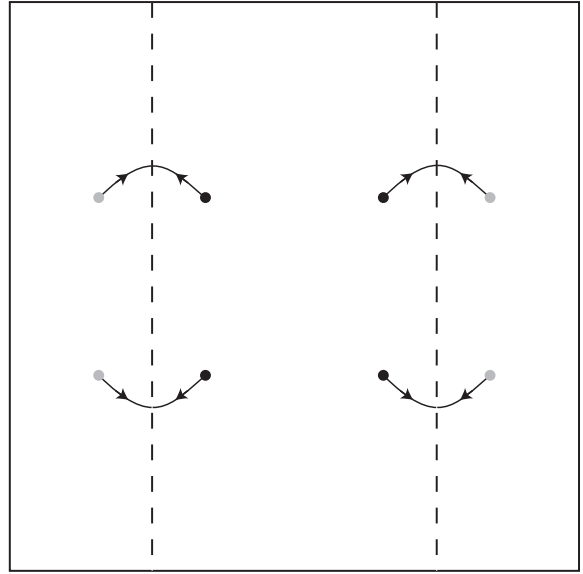}} \vspace{0.4in}
\caption{Evolution of the gapless nodal points in the fermionic
excitation spectrum as a function of $\psi_{\text{sp}}$. The
square contains the first Brillouin zone of the original square
lattice, extending between $k_{x,y} = \pm \pi$. The dark circles
are the positions of the nodal points in this $d$-wave
superconducvtor. Onset of a non-zero $\psi_{\text{sp}}$
introduces Bragg reflection planes at $k_x = \pm \pi/2$ and
images of the nodal points indicated by the grey circles. As
$\psi_{\text{sp}}$ increases, the nodal points move towards the
Bragg reflection planes and annihilate each other when they
collide \protect\cite{vojtaprl}. The critical theory of this
transition is discussed in Section~\protect\ref{nodal}.}
\label{fig:nodal}
\end{figure}

First, let us determine the values of $\psi_{\text{sp}}^c$ and
$p_0$. Using the facts that $\epsilon(p_0) = \epsilon(p_0 + G)$
and $\Delta(p_0) = \Delta(p_0 + G)$, the condition for the
presence of gapless nodal points in (\ref{e14}) becomes
\begin{equation}
\epsilon (p_0) = \psi_{\text{sp}}^c a(p_0)~~~~;~~~~\Delta (p_0) =
\psi_{\text{sp}}^c b(p_0). \label{e15}
\end{equation}
The solution of these two equations determines the two unknowns
$\psi_{\text{sp}}^c$ and $p_{0x}$. At this same point, the second
eigenvalue in (\ref{e14}) remains non-zero and finite. This
fermionic mode will play no role in the critical theory, and so
it pays to perform a canonical transformation at an early stage
to project it out. To leading order in the distance from the
critical point, this merely means that we have to take only the
linear combination of $c_{k a}$ and $c_{k+G, a}$ which appears in
gapless eigenvalue associated with (\ref{e15}). So, we introduce
a new fermionic degree of freedom, $f_{qa}^{+}$, where the small
momentum $q$ is measured as a deviation from $p_0$ (and a
corresponding fermionic mode $f_{qa}^{-}$ which resides at
momenta near $-p_0$); using the structure of the gapless
eigenvalue at $\psi_{\text{sp}}=\psi_{\text{sp}}^c$ and $k=p_0$,
we see that we should parameterize
\begin{eqnarray}
c_{ka} &=& f_{qa}^{+}/\sqrt{2} \nonumber \\
c_{k+G,a} &=& i \varepsilon_{ab}  f_{qb}^{+}/\sqrt{2}, \label{e16}
\end{eqnarray}
where $k=p_0+q$, and $q$ is small; a similar parameterization is
made near $-p_0$ with $f_a^{-}$. Finally, we insert (\ref{e16})
into $H_{BCS}+H_{\text{sp}}$, and expand in
$\psi_{\text{sp}}-\psi_{\text{sp}}^c$ and in gradients of
$f_{a}^{\pm}$. This leads to the following effective action for
the critical theory
\begin{eqnarray}
S_f &=& \int d^2 r d \tau \left[ f_a^{\pm \dagger} \left(
\frac{\partial}{\partial \tau} \pm i v_1 \frac{\partial}{\partial
y} - \frac{1}{2m_1} \frac{\partial^2}{\partial x^2} + \delta_1
\right)f_a^{\pm} \right. \nonumber \\
&+& \varepsilon_{ab} f_a^{-} \left( v_2 \frac{\partial}{\partial
y} - \frac{1}{2m_2} \frac{\partial^2}{\partial x^2} + \delta_2
\right) f_b^{+} + \mbox{H.c.} \label{e17}
\end{eqnarray}
where $\tau$ is imaginary time, $r=(x,y)$ is the spatial
co-ordinate, $\delta_{1,2} = w_{1,2}^{\prime} (\psi_{\text{sp}} -
\psi_{\text{sp}}^c)$, and $v_{1,2}$, $m_{1,2}$,
$w_{1,2}^{\prime}$ are constants dependent upon the detailed
momentum dependence of $\epsilon (k)$ and $\Delta (k)$. The
fermionic eigenenergy of (\ref{e17}) is easily determined; it is
\begin{equation}
\left[ \left( \delta_1 + v_1 q_y + \frac{q_x^2}{2 m_1} \right)^2 +
\left( \delta_2 + v_2 q_y + \frac{q_x^2}{2 m_2} \right)^2
\right]^{1/2}. \label{e18}
\end{equation}
The positions of the nodal points, if present, are easily
determined from (\ref{e18}). Assuming $v_{1,2}$, $m_{1,2}$,
$w_{1,2}^{\prime}$ are all positive, then for $w_1^{\prime}/v_1 >
w_2^{\prime}/v_2$ and $m_1 v_1 < m_2 v_2$, nodal points are
present for $\psi_{\text{sp}} < \psi_{\text{sp}}^c$, but not for
$\psi_{\text{sp}} > \psi_{\text{sp}}^c$ (similar results hold for
other signs and magnitudes of the various coupling constants).
The trajectory of the nodal points is sketched schematically in
Fig~\ref{fig:nodal}: they move on a parabolic trajectory with
$|q_y| \sim q_x^2$ before colliding along the Bragg reflection
planes at $k_x = \pm \pi/2$.

The availability of the action $S_f$ also allows one to determine
the consequences of the interactions near the quantum critical
point. Notice that at the critical point, $\delta_1=\delta_2=0$,
$S_f$ is invariant under the scale transformation $\tau
\rightarrow \tau/s$, $y \rightarrow y/s$, $x \rightarrow
x/\sqrt{s}$, and $f \rightarrow s^{3/4} f$. The simplest allowed
four fermion coupling is $\sim (f^{\dagger} f)^4$, and
power-counting shows that its co-efficient has a negative scaling
dimension of $-1/2$. So this interaction is an irrelevant
perturbation at the quantum critical point \cite{granath}. In a
similar manner, it is not difficult to show that all other
perturbations of $S_f$ are irrelevant, and so $S_f$ is the
complete critical theory of the transition involving the onset of
nodal fermionic excitations.

\section{Coupling of spin Peierls order to phonons} \label{phonon}

A central actor in all the considerations so far has been the
spin-Peierls order parameter, $\Psi_{\text{sp}}$. Experimental
observations of the fluctuations of this order parameter would
certainly be helpful in resolving the theoretical issues.
However, this is a spin-singlet, charge zero mode and is only
observable through its couplings to the ionic displacements. This
section will therefore present a simplified discussion of the
coupling between the dynamics of the spin Peierls order parameter
and the phonon modes of the CuO$_2$ plane. Indeed, it is probable
that the phonons are more than merely spectators of the spin
dynamics, and the spin-phonon coupling may play an important role
in selecting between different charge orderings and in
influencing the nature of the electronic ground state: the
important phonon and spin excitation energies are roughly of the
same order, and so a coupled dynamical model will be necessary
for a complete microscopic understanding.

\begin{figure}
\epsfxsize=3.5in \centerline{\epsffile{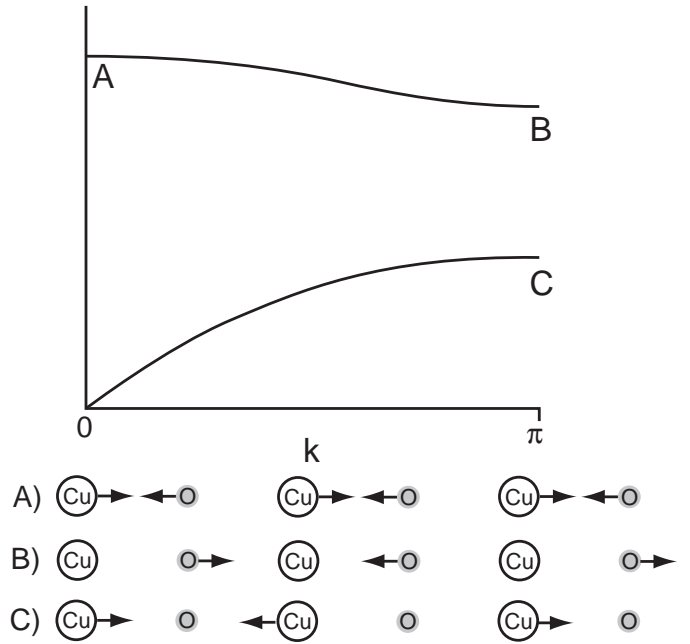}} \vspace{0.4in}
\caption{Phonon frequencies (vertical axis) of the one-dimensional
Cu-O chain. The frequency of the mode A is $[2 K (M_{\text{Cu}} +
M_{\text{O}})/(M_{\text{Cu}} M_{\text{O}})]^{1/2}$, that of B is
$[2K/M_{\text{O}}]^{1/2}$, and that of C is
$[2K/M_{\text{Cu}}]^{1/2}$. Only mode C is ``Peierls active'' {\em
i.e.} couples linearly to the order parameter $\Psi_{\text{sp}}$,
which measures the amplitude of a bond-centered charge order wave
at wavevector $(\pi/a,0)$.} \label{fig:phonon}
\end{figure}

The spin-Peierls order parameter couples most directly to the
``Peierls-active'' phonon mode sketched as mode C in
Fig~\ref{fig:phonon}: this is a staggered displacement of the Cu
ions at the wavevector $(\pi, 0)$.
Notice that the O ions are stationary in this phonon mode. As we
will describe below, the coupling between this Peierls-active
phonon and the spin-Peierls order parameter is likely to lead to
an additional low energy peak in the dynamic phonon structure
factor at the energy scale of the characteristic spin-Peierls
fluctuations. However, neutron scattering studies of the phonon
spectra have so far\cite{egami,braden,mook2,petrov} focused on
optical phonons involving motion primarily of the O ions (the mode
between A and B in Fig~\ref{fig:phonon}): the smaller O ion mass,
$M_{\text{O}} \approx M_{\text{Cu}}/4$, increases the displacement
of the O ions and makes these easier to observe. The O ion phonon
at wavevector $(\pi, 0)$ (mode B in Fig~\ref{fig:phonon}) is not
Peierls-active in the sense just noted. However, following the
suggestion of McQueeney {\em et al.} \cite{egami}, we will
describe below a non-linear coupling in the spin-phonon
Hamiltonian which leads to signatures of the spin-Peierls
fluctuations in the A-B phonon mode at the wavevector $(\pi/2,0)$.

It is also worth mentioning the recent work of Khaliullin and
Horsch~\cite{horsch}. In contrast to our focus on bond-centered
charge fluctuations, they considered Cu site-centered charge
density fluctuations at wavevectors near $(\pi, 0)$ (in the
notation of Nayak~\cite{chetan}, we are examining $p_x$ density
wave correlations at $(\pi, 0)$, while Ref~\onlinecite{horsch}
discusses $s$ density wave correlations at $(\pi, 0)$). These Cu
site-centered fluctuations do indeed couple linearly to the phonon
mode in the vicinity of the point B in Fig~\ref{fig:phonon}: this
led to a broadening and softening of the phonon near B. As we will
see below, our bond-centered spin-Peierls correlations instead
modify the A-B phonon mode near $(\pi/2, 0)$ in a manner which is
consistent with experimental observations.

We will work here with a simplified one-dimensional model of the
phonons in the CuO$_2$ plane. All the phonons we are interested in
are polarized both in wavevector and ionic displacement along the
$(1,0)$ axis, and so neglecting the second dimension is not too
serious. We will consider only a single chain of alternating Cu
and O ions, and consider displacements of the ions along the chain
direction (see Fig~\ref{fig:phonon}). A more complete treatment is
certainly possible, but we will not attempt it here: we hope that
with improved experimental resolution, a more precise and
microscopic theoretical study of the two-dimensional phonon modes
will be carried out, along the lines of the analyses of
Refs~\onlinecite{gros1,gros2} for CuGeO$_3$.

We begin by specifying our toy one-dimensional model of the
phonons in an alternating chain of Cu and O ions. We place the Cu
ions on the sites, $i$ (integer), of a chain with spacing $a$,
while the O ions are at the centers of the links of the chain. We
denote the displacement of these ions by $u_i$ and $v_{i}$
respectively, where the $i$'th O ion is taken to be to immediately
to the right of the $i$'th Cu ion. The harmonic action for the
phonon modes is (in imaginary time, $\tau$):
\begin{eqnarray}
S_{\text{ph}} &=& \int d \tau \sum_i \biggl[
\frac{M_{\text{Cu}}}{2} \left( \frac{ d u_i}{d \tau} \right)^2 +
\frac{M_{\text{O}}}{2} \left(
\frac{ d v_{i}}{d \tau} \right)^2 \nonumber \\
&+& \frac{K}{2} \left\{ (u_i - v_{i})^2 + (u_i - v_{i-1})^2
\right\} \biggr], \label{e19}
\end{eqnarray}
where $K$ is a ``spring constant'' which determines the phonon
frequencies. It is a simple matter to diagonalize $S_{\text{ph}}$
and obtain the phonon normal mode frequencies: they are
\begin{eqnarray}
\omega_{\pm}^2 &=& \frac{K}{M_{\text{Cu}} M_{\text{O}}} \Bigl[
M_{\text{Cu}} + M_{\text{O}} \pm \left(
(M_{\text{Cu}}-M_{\text{O}})^2 \right.  \nonumber \\
&~&~~~~~~~\left. + 4 M_{\text{Cu}} M_{\text{O}} \cos^2 (k/2)
\right)^{1/2} \Bigr], \label{e20}
\end{eqnarray}
and are sketched in Fig~\ref{fig:phonon}.

We also have to consider the dynamics of the spin Peierls order
$\Psi_{\text{sp}}$ introduced in Section~\ref{intro} and
considered in Section~\ref{nodal} and Appendix~\ref{nematic}. In
our present one-dimensional toy model, we need only consider
$\psi_{\text{sp}} = \mbox{Re}[\Psi_{\text{sp}}]$. For this order
parameter $\psi_{\text{sp}}$, the dynamics of an interacting
effective action like $S$ in (\ref{a1}) is assumed to be captured
by the following effective quadratic action \cite{vojtaprl}:
\begin{eqnarray}
S_{\text{sp}} &=& \frac{T}{2} \sum_{k, \omega_n} |\psi_{\text{sp}}
(k, \omega_n)|^2 \chi_{\text{sp}}^{-1} (k, i\omega_n) \label{e21}  \\
\chi_{\text{sp}}^{-1} (k, i\omega_n) &\equiv& \omega_n^2 + 2 a^2
c_1^2 (1-\cos(k))
 + \Delta_{\text{sp}}^2 +
\Gamma_{\text{sp}} |\omega_n|. \nonumber
\end{eqnarray}
Here $\omega_n$ is a Matsubara frequency, and $\Delta_{\text{sp}}$
and $\Gamma_{\text{sp}}$ are effective energy scales determining
the mean frequency and damping of the $\psi_{\text{sp}}$
fluctuations; the values of $\Delta_{\text{sp}}$ and
$\Gamma_{\text{sp}}$ are determined by the non-linear interactions
in (\ref{a1}). Also we have replaced the spatial gradient in
(\ref{a1}) by a nearest-neighbor lattice derivative in
(\ref{e21}). As one approaches the onset of spin-Peierls order
(approaching point C in Fig~\ref{phasediagram} from the right),
the value of $\Delta_{\text{sp}}$ will decrease to zero, while
$\Gamma_{\text{sp}}$ becomes of order $T$, representing the
damping of the order parameter mode in the quantum-critical
region. We will work here in the $\Delta_{\text{sp}} > 0$ regime,
staying to the right of C in Fig~\ref{phasediagram}. The velocity
$c_1$ should be order the spin-wave velocity, and this is about 10
times larger than the velocity of the acoustic phonon mode, C, in
Fig~\ref{fig:phonon}.

Finally, we have to couple $\psi_{\text{sp}}$ to the phonon modes.
This coupling \cite{gros1,gros2,girvin,zx} arises from the
dependence of the exchange constant $J$ between neighbor Cu spins
on the displacements of the Cu and O ions. We assume that $J \sim
t_{pd}^4$, where $t_{pd}$, the overlap between neighboring O and
Cu orbitals is a function of $(u_i - v_{i})$ and $(u_i-v_{i-1})$.
Expanding $J$ in derivatives of these variables, we obtain first
the simple linear coupling
\begin{equation}
S_{1c} = \int d \tau \sum_i \left[ \lambda (-1)^{i}
\psi_{\text{sp},i} \left(u_{i+1}-u_{i} \right) \right],
\label{e22}
\end{equation}
where $\lambda$ is the linear coupling constant, and
$\psi_{\text{sp},i}$ is the spin-Peierls order parameter in real
space; this naturally resides on the centers of the bonds, and we
locate $\psi_{\text{sp},i}$ on the same O site as $v_i$ . As we
noted earlier, $S_{1c}$ couples the spin-Peierls order most
strongly to the phonons in the vicinity of the point C in
Fig~\ref{fig:phonon}, and will lead to its broadening and
softening. We are interested here primarily in the modifications
of the optical phonon mode A-B, and so we will neglect $S_{1c}$ in
our computations below.

As indicated at the beginning of this section, the important
effect on the A-B optical phonon arises from a non-linear
spin-phonon coupling . Expanding to one higher order in the phonon
displacements, the coupling between $\psi_{\text{sp}}$ and the
phonon displacements can be written in the form
\begin{eqnarray}
S_{2c} &=& \int d \tau \sum_i \left[  (-1)^{i} \psi_{\text{sp},i}
\left\{\gamma_1 \left(u_{i+1}-u_{i}
\right)^2 \right. \right. \nonumber \\
&+& \left.\left. \gamma_2 \left(u_{i+1}+u_{i}-2v_{i} \right)^2
\right\} \right], \label{e23}
\end{eqnarray}
where $\gamma_{1,2}$ are the non-linear spin phonon coupling
constants.

\begin{figure}
\epsfxsize=3.5in \centerline{\epsffile{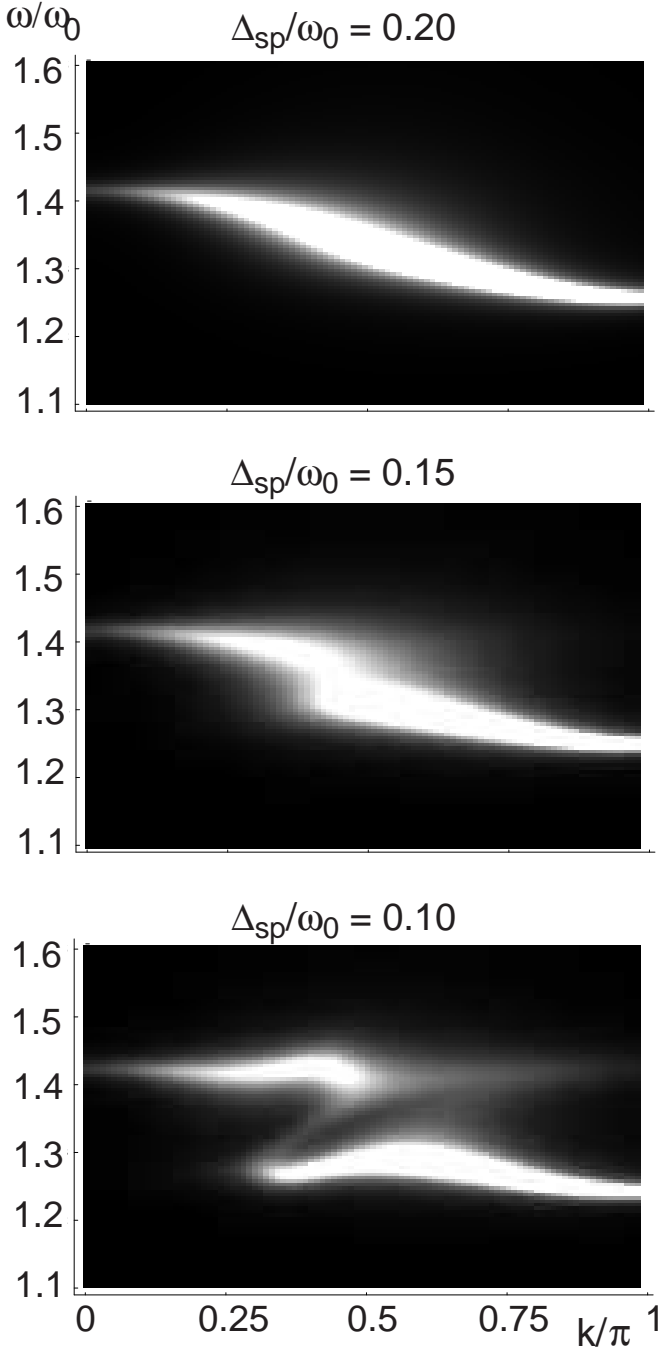}}
\vspace{0.4in} \caption{Plot of the phonon spectral density
$\mbox{Im} D(k, \omega)$ defined in (\protect\ref{e23a}). The
figure shows the influence of the spin-Peierls fluctuations
described by (\protect\ref{e21}) on the optical phonon mode A-B in
Fig~\protect\ref{fig:phonon}; the two degrees of freedom are
coupled by the non-linear terms in (\protect\ref{e23}). We used
the parameters $\omega_0 \equiv
\sqrt{K/(1/M_{\text{Cu}}+1/M_{\text{O}})}$, $\Gamma_{\text{sp}}=
0.5 \omega_0$, $T= 0.1 \omega_0$, $c= 10.0 \omega_0$, and
$\gamma_1=\gamma_2=0.5 K\omega_0$. The figures show the evolution
in the spectrum as a function of $\Delta_{\text{sp}}/\omega_0$.}
\label{fig:spectral1}
\end{figure}

\begin{figure}
\epsfxsize=3.5in \centerline{\epsffile{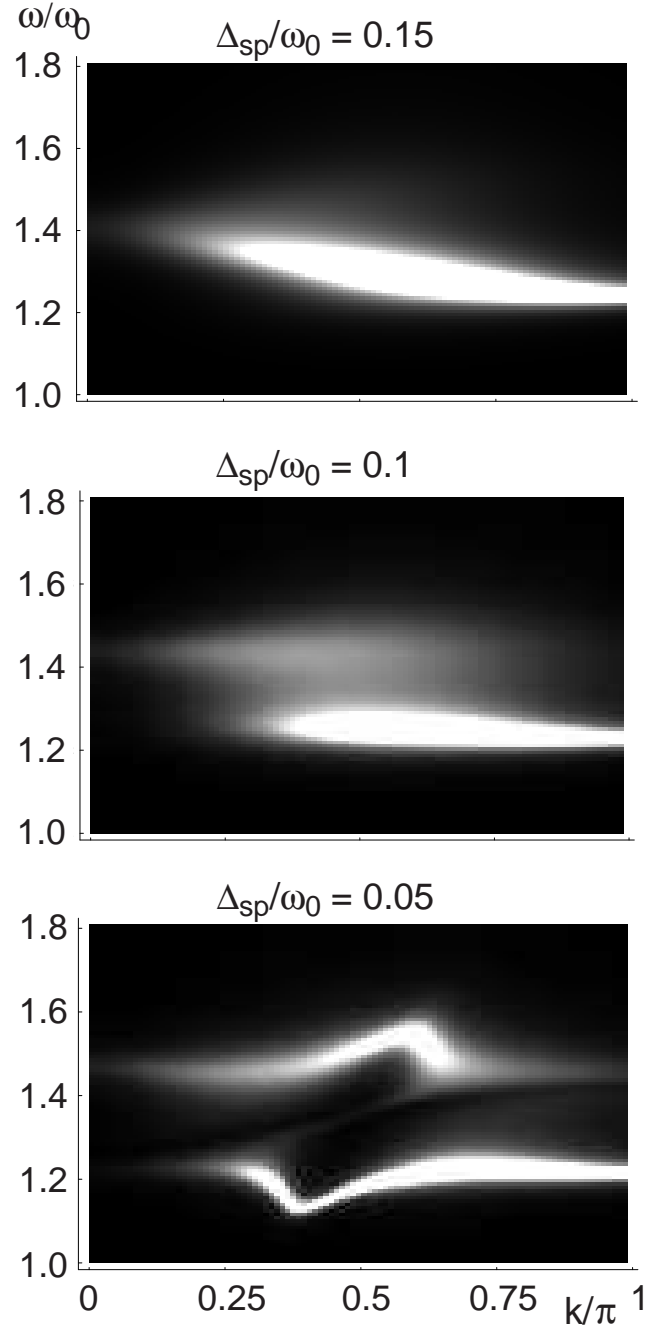}}
\vspace{0.4in} \caption{As in Fig~\protect\ref{fig:spectral1} but
for $\Gamma_{\text{sp}}= 0.2 \omega_0$, $T= 0.1 \omega_0$, $c= 5.0
\omega_0$, and $\gamma_1=\gamma_2=0.5 K\omega_0$.}
\label{fig:spectral2}
\end{figure}

We examined the properties of $S_{\text{ph}} + S_{\text{sp}} +
S_{2c}$ in a simple one-loop approximation: we neglect the
$\omega_-$ phonon in (\ref{e20}), computed the self-energy of the
$\omega_+$ optical phonon at order $\gamma_{1,2}^2$. Finally, to
compare to neutron scattering experiments, we computed
\begin{equation}
D(k, \omega) = \left\langle \left| b_{\text{Cu}} u (k, \omega) +
b_{\text{O}} v (k, \omega) e^{i k/2} \right|^2 \right\rangle,
\label{e23a}
\end{equation}
where $b_{\text{Cu}} = 7.718$ and $b_{\text{O}} = 5.803$ are the
neutron scattering lengths of Cu and O respectively\cite{neutron}.
The results are shown in Figs~\ref{fig:spectral1}
and~\ref{fig:spectral2} for a representative set of parameters.
As $\Delta_{\text{sp}}$ decreases, there is initially a broadening
and a sharpening of the phonon mode near $k=\pi/2$. For even
smaller $\Delta_{\text{sp}}$ we see the incipient Brillouin zone
boundary of the doubled unit cell developing near $k=\pi/2$: this
is reflected in the apparent discontinuity of the dispersion and
the appearance of ``shadow'' phonon bands reflected across the
incipient Brillouin zone boundary. The first of these features is
in excellent accord with all the available neutron scattering
experiments\cite{egami,braden,petrov}, and there has even been a
claim of the observation of shadow bands\cite{egami}, although
this has not been confirmed.

Another interesting feature of Figs~\ref{fig:spectral1}
and~\ref{fig:spectral2} is that the intensity of the phonon is
larger near the Brillouin zone boundary. This is a `form factor'
effect, and arises primarily from the interplay of the different
scattering lengths in (\ref{e23a}), and the distribution of the
phonon normal mode between the O and Cu sites. Curiously,
precisely such a dominance of intensity near $k=\pi$ is observed
in the experiments\cite{egami,petrov}.

\section{Conclusions}
\label{conc}

This paper has introduced a bond operator formalism to represent
the degrees of freedom of doped antiferromagnets. While the
approach bears some similarities to the popular ``slave'' boson
and fermion techniques, the implementation leads to theories with
a rather different structure: there are no long-range gauge forces
in the fluctuations about any reasonable saddle-point in the
latter formalism, and the quantum numbers of the true elementary
excitations are simply connected to those of the microscopic
operators. This is a powerful advantage of the bond operator
method, and allows much information to be gleaned from simple
mean-field Hatree-Fock-BCS computations. The main disadvantage of
the method is that it requires a pairing of the sites into bonds
at the outset. Such a pairing is naturally present in systems with
spontaneous or explicit bond-centered charge order (as in a
spin-Peierls states), and it is for these systems that the
approach is best suited.

The defining equations of the bond operator formalism were
presented in Section~\ref{sec:bo}, and are contained in Eqns
(\ref{e1}-\ref{e6}). Next, we applied this formalism to the
two-dimensional doped antiferromagnet sketched in Fig~\ref{fig5}b.
The main phase diagram of the model is sketched in
Fig~\ref{phasediagram}, and the important properties of the phases
are summarized in its caption. The remaining figures in
Fig~\ref{sec:bo} and their captions summarizing the excitation
spectra of the phases---the fermionic, $S=1/2$, spectra are in
Figs~\ref{hweak1}, \ref{hweak2}, \ref{hstrong1},
and~\ref{hstrong2}, while the $S=1$ excitations are in
Figs~\ref{tweak} and~\ref{tstrong}; the latter consist of a
bosonic, $S=1$ exciton, and the two-particle continua of the
fermionic $S=1/2$ excitations.

The connection of these result to recent neutron scattering
measurements of phonon spectra\cite{egami,braden,mook2,petrov} was
considered in Section~\ref{phonon}. This section did not use the
bond operator formalism. Instead, it considered the consequences
of incipient bond-centered charge order in an isotropic state, in
a simple perturbative calculation using a spin-phonon model. We
introduced an order parameter, $\Psi_{\text{sp}}$, characterizing
the ordering pattern associated with Fig~\ref{fig5}b, and wrote
down a phenomenological free energy\cite{vojtaprl,book} describing
its fluctuations in spacetime in the vicinity of the quantum
critical point labeled by the point C in Fig~\ref{phasediagram}.
On the symmetric side of C (the region with $\langle
\Psi_{\text{sp}} \rangle = 0$), these fluctuations were controlled
by an energy scale, $\Delta_{\text{sp}}$, which vanished as C was
approached, and we considered the evolution of the optical phonon
spectra as a function of decreasing $\Delta_{\text{sp}}$: these
results are contained in Figs~\ref{fig:spectral1}
and~\ref{fig:spectral2}.

In the unifying language introduced by Nayak~\cite{chetan}, the
bond-centered charge order parameter $\Psi_{\text{sp}}$ may be
considered as the amplitude of a $p_x$ density wave at wavevector
$(\pi, 0)$. The ordinary $s$ density wave at wavevector $(\pi, 0)$
is associated with Cu site-centered charge order, and these may
also be strong in the lightly doped antiferromagnet, especially in
the region with long-range magnetic order. Their influence on the
optical phonon spectra was considered in
Ref~\onlinecite{horsch}---the primary effect was a broadening of
the O optical phonon near $(\pi, 0)$. This should be contrasted
with the influence of the $\Psi_{\text{sp}}$ fluctuations
described above---the strongest effect was near $(\pi/2, 0)$ where
the phonon dispersion sharpened considerably, along with a
significant amount of broadening. The latter effects are clearly
seen in recent neutron scattering experiments \cite{egami,braden},
although some softening at $(\pi, 0)$ is also apparent
\cite{petrov}. We hope that higher precision and more detailed
neutron scattering experiments will be undertaken, and along with
more microscopic theoretical computations, these should help sort
out the relative roles of site- and bond- centered charge order as
a function of increasing doping.

In addition to fluctuating charge-order modes detected in phonon
scattering, it would also be useful to study systems in which the
charge order is required to be static; in such situations, atomic
resolution STM studies should yield much useful information on the
microstructure of the charge order. Our physical picture implies
that static charge order should be present in situations in which
both magnetic and superconducting order have been suppressed
(systems with one of these orders may only have fluctuating charge
order). A convenient way to achieve this is by application of a
strong magnetic field on underdoped samples \cite{greg}. A
phenomenological theory of the phase diagram in a magnetic field
has been provided recently in Ref~\onlinecite{dsz}: the ``normal''
state in this phase diagram is a very attractive candidate to
bond-centered charge order. It would be especially interesting to
conduct STM measurements on the strongly underdoped YBCO crystals
that have become available recently\cite{kam}, after
superconductivity has been suppressed by a static magnetic field.
An alternative is to look for charge order in STM studies in which
the superconductivity has only been locally suppressed, as is the
case in the cores of vortices in the superconducting order
\cite{snl,fischer}. However, the short-range nature of the
suppression means that charge order is not required to appear, and
may remain dynamic---this makes this approach less attractive.
Recent indications\cite{shuheng} of mesoscale self-segregation of
charge carriers in bulk samples also naturally raise the
possibility of bond charge order in the lower density regions
which (presumably) have suppressed superconductivity.

\subsection{Review of related work}

Before closing this paper, we present brief critical reviews of
some related work in the literature, and relate it our results.

Sushkov and collaborators have studied aspects of Mott insulators
both at zero and finite doping. At zero doping, they have examined
the square lattice $S=1/2$ Heisenbserg antiferromagnet with first
($J_1$) and second ($J_2$) neighbor exchange\cite{sushkov1}: with
increasing $J_2/J_1$ they find a transition from the familiar
N\'{e}el state to a paramagnetic state with bond-centered charge
order as in Fig~\ref{fig5}b, albeit with a small window of
parameters at which the two orders co-exist; such a co-existence
region is allowed in the framework of the earlier field-theoretic
analysis \cite{rodolfo}. At somewhat larger $J_2/J_1$ they find
further charge ordering which breaks an additional $Z_2$ lattice
symmetry: this pattern of charge ordering has also been seen in
recent DMRG studies \cite{leiden}. Sushkov \cite{sushkov2} has
also studied the evolution of the ground state as a function of
doping: by a series expansion method he finds that after
destruction of the Neel order, the ground state has the columnar
spin-Peierls order of Fig~\ref{fig5}b. Only at a significantly
larger coupling is the full square lattice symmetry restored. All
of these results are fully consistent with the approach outlined
in our paper.

Mazumdar, Clay, and Campbell \cite{campbell1,campbell2,campbell3}
have studied two-dimensional correlated electron models associated
with the organic charge-transfer solids
tetra\-methyl-tetra\-thia\-ful\-valene (TMTTF),
tetra\-methyl-tetra\-selena\-ful\-valene (TMTSF),
bis\-ethylene\-di\-thio-tetra\-thia\-ful\-valene (BEDT-TTF) and
bis\-ethylene\-di\-thio-tetra\-selena\-ful\-valene (BETS). They
have argued that these materials are characterized by
quarter-filled band, and produced evidence for ``bond order wave''
in their models at quarter filling. These states are closely
related to the period 4 bond-centered charge density wave state
studied in Ref~\onlinecite{vojtaprl}, and so also to the
insulating and superconducting states with the symmetry of
Fig~\ref{fig5}b studied here.

Kivelson, Fradkin and Emery \cite{kfe} have studied symmetry
breaking in two-dimensional electronic systems using a rather
general classification in terms of electronic liquid crystal
phases. Their formulation (and their earlier theory
\cite{proximity} of the ``spin gap proximity effect'') does not
distinguish between the nature of bond- and site- centered charge
order that we are paying particular attention to here. A
significant portion of the superconducting state in their phase
diagram overlaps with a region of nematic order. We believe that
the orthorhombic anisotropy of this nematic order is ultimately
driven by singlet bond correlations associated with the
spin-Peierls order in Fig~\ref{fig5}b; consequently this nematic
region should show bond-centered smectic correlations with a
period of 2 lattice spacings, as has been claimed in the
experiments of Ref~\onlinecite{egami}. We make this distinction
more precise by discussing how the spin-Peierls and nematic orders
are coupled. We can characterize the spin-Peierls order\cite{sp}
in Fig~\ref{fig5}b by a complex $Z_4$ order parameter
$\Psi_{\text{sp}}$ which takes the values $1$, $i$, $-1$, $-i$ on
the four states obtained by rotating Fig~\ref{fig5}b about any
lattice site successively by $90^{\circ}$. Similarly, we can
introduce a real Ising nematic order parameter $\Phi_{\text{n}}$
which is +1 (-1) for a nematic polarized along the $x$ ($y$)
direction. Then the symmetry properties of the two order
parameters show that\cite{sp}
\begin{equation}
\Phi_{\text{n}} \sim \Psi_{\text{sp}}^2 . \label{pn}
\end{equation}
A detailed Landau theory and fluctuation analysis of the interplay
between these two order parameters is presented in
Appendix~\ref{nematic}. It is our picture, based on the physical
arguments above, that $\Psi_{\text{sp}}$ is the primary order
parameter, and that $\Phi_{\text{n}}$ is tied to it via
(\ref{pn}).

An important ingredient in the work of Carlson, Orgad, Kivelson,
and Emery \cite{erica,proximity} is the crossover from the physics
of a one-dimensional electron gas moving in the vertical direction
in Fig~\ref{fig5}b at short scales, to a coupled two-dimensional
system at long scales. This should be contrasted from our
approach, in which we do not find any quasi-one-dimensional
regime. Although the lattice symmetry of our states is similar to
those considered by these authors, the physics is always
intrinsically two-dimensional, albeit with a spatial anisotropy.
Indeed, the arguments for confinement and bond-centered charge
order rely crucially on the two-dimensionality of the system.

Zaanen and collaborators\cite{jan}  have recently given a bold
physical picture of the microstructure of stripes and their
relationship to high temperature superconductivity. They take a
solitonic, low-dimensional perspective, in which special attention
is paid to excitations at various boundaries and dislocations in
the stripe order. In contrast, our perspective here is a higher
(two-) dimensional point of view, appropriate to systems not too
far below their upper-critical dimension. So {\em e.g.} we view
the spin-Peierls order parameter $\Psi_{\text{sp}}$ as a
``soft-spin'' field with large and continuous variations in its
local amplitude, rather than a field with 4 discrete possible
values in different regions of space separated by identifiable
domain walls. The advantage of our perspective is that it allows a
complete (in principle) and systematic treatment of the coupling
of fermionic degrees of freedom to the various order parameters in
a Landau-theory like scattering framework \cite{vojtaprl,qptd}.

White and Scalapino\cite{ws} have presented results for charge
order in DMRG studies of doped antiferromagnets. They find that
site and bond centered stripes are almost degenerate in their
energy. They do observe a period 4 bond-centered charge stripe
state \cite{vojtaprl}, but have not so far seen a state with the
period 2 symmetry of Fig~\ref{fig5}b.

Castro Neto\cite{neto} has recently presented an interesting
analysis of a striped superconductor, coupling transverse stripe
fluctuations between regions of high and low hole density.
Notably, he also finds pronounced tendency towards bond-centered
charge order with the symmetry of Fig~\ref{fig5}b.

Lannert, Fisher, and Senthil\cite{lannert} have considered Berry
phase effects in a $Z_2$ gauge theory of correlated electron
systems. Close to half-filling, they find that these Berry phases
induce bond-centered charge order in both insulating and
superconducting phases, in a manner closely analogous to that
found in Ref~\onlinecite{rs1} in paramagnetic Mott insulators.
They also discuss the critical properties of a transition between
superconducting and insulating states at half filling, with both
phases containing a background charge order with the symmetry of
Fig~\ref{fig5}b. They pay particular attention to the influence of
the gapless nodal fermions on the vortices in the superconducting
order. However, in the presence of the background charge order at
wavevector $(\pi, 0)$ , their nodal fermions (which are at $(\pm
\pi/2, \pm \pi/2)$) should immediately acquire a gap, and their
theory then reduces to a dualized version of familiar bosonic
theory of the superconductor-insulator transition.

\acknowledgements We thank G.~Aeppli, M.~Braden, D.~Campbell,
E.~Carlson, T.~Egami, S.~Girvin, E.~Fradkin, S.~Kivelson, A.
Lanzara, A.~Polkovnikov, T.~Senthil, Z.-X. Shen, M.~Vojta and
J.~Zaanen for useful discussions. We are especially grateful to
S.~Kivelson for a critical reading of the manuscript. This
research was supported by US NSF Grant DMR 0098226.


\appendix

\section{Interplay of spin Peierls and nematic orders}
\label{nematic}

In Section~\ref{intro} we discussed the relationship between
states with spin-Peierls order, with the symmetry of
Fig~\ref{fig5}b, and the nematic electronic states discussed by
Kivelson, Fradkin and Emery~\cite{kfe}. We noted that the
spin-Peierls state was characterized by a complex order
parameter\cite{sp} $\Psi_{\text{sp}}$ (which took the values $1$,
$i$, $-1$, $-i$ on the four states obtained by rotating
Fig~\ref{fig5}b about any lattice site), the nematic was
characterized by a real Ising order parameter $\Phi_{\text{n}}$,
and the symmetry properties of these two order parameters implied
the relationship (\ref{pn}) between them. More generally, we can
write down the following simple effective action for them, keeping
all low order terms consistent with the underlying symmetries:
\begin{eqnarray}
S_{\text{sp-n}} &=&\int_0^{1/T} d\tau \int d^2 x \Biggl[
|\partial_{\tau} \Psi_{\text{sp}} |^2 +  c_1^2 |\nabla_x
\Psi_{\text{sp}}|^2 + r_1 | \Psi_{\text{sp}}|^2 \nonumber \\ &+&
\frac{1}{2} (\partial_{\tau} \Phi_{\text{n}})^2+ \frac{1}{2} c_2^2
(\nabla_x \Phi_{\text{n}})^2
+ \frac{r_2}{2} \Phi_{\text{n}}^2 \nonumber \\
&-& g \Phi_{\text{n}} \left( \Psi_{\text{sp}}^2 +
\Psi_{\text{sp}}^{\ast 2} \right) + \frac{u_1}{2} |
\Psi_{\text{sp}} |^4 - \frac{v}{4} \left(\Psi_{\text{sp}}^4 +
\Psi_{\text{sp}}^{\ast 4} \right) \nonumber \\ &~&~~~~~~~~~~~~-
\frac{w}{2} \Phi_n^2 |\Psi_{\text{sp}}|^2 +  \frac{u_2}{4}
\Phi_{\text{n}}^4 \Biggr], \label{a1}
\end{eqnarray}
where $\tau$ is imaginary time, and $u_1>v>0$, $u_2 > 0$, and
$u_2 (u_1 - v)
> w^2$ required for stability. We expect that the appearance of
nematic order will enhance the probability of spin Peierls order,
and so $w > 0$. We have, for now, neglected the couplings of
these order to the fermionic excitations, and will consider the
consequences of these later in this section.

\begin{figure}
\epsfxsize=2.5in \centerline{\epsffile{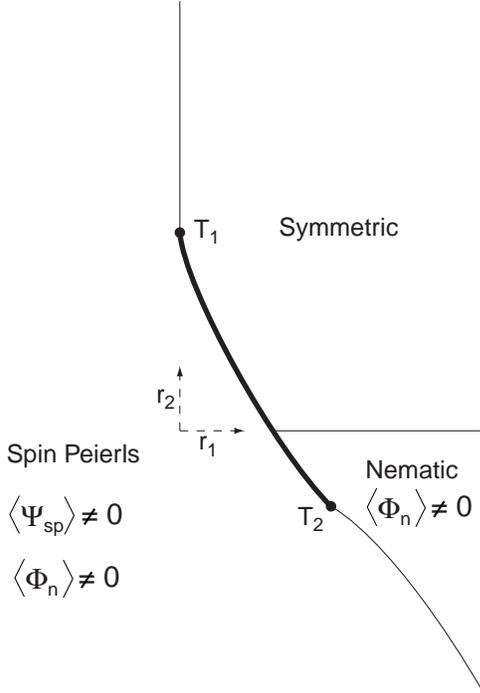}}
\caption{Mean-field phase diagram of the action $S_{\text{sp-n}}$
in (\protect\ref{a1}). The symmetric phase has no broken lattice
symmetries. The thin (thick) lines represent second (first) order
transitions. The positions of the second order transition lines is
indicated in (\protect\ref{a2}) and above. The tricritical point
$T_1$ is at $r_1=0$, $r_2 = 4 g^2 /(u_1 -v )$, while the
tricritical point $T_2$ is at the $r_2<0$ which is a solution of
$2 u_2 g^2 + 4 g w \sqrt{-r_2/u_2}+ r_2 ( u_2 (u_1 -v) - w^2 /2 )
=0 $, and $r_1$ given by (\protect\ref{a2}) for this value of
$r_2$. At $T=0$, the coupling to the nodal fermions in a $d$-wave
superconductor drives the transition from the symmetric to the
nematic phase first order \protect\cite{qptd}. } \label{figphase}
\end{figure}

First, let us analyze $S_{\text{sp-n}}$ in mean-field theory. The
results of such an analysis are shown in Fig~\ref{figphase}.
There are three phases: (i) the symmetric phase, where no lattice
symmetry is broken, (ii) the spin Peierls phase, where both the
spin Peierls and nematic order parameters are non-zero, and (iii),
the nematic phase, where only the nematic order is non-zero. There
can be a second-order phase transition between any two of these
three phases, as indicated in Fig~\ref{figphase}. The position of
these second-order boundaries can be determined by a conventional
Landau theory analysis in powers of the associated order
parameter. In this manner, we find that the second-order line
between the symmetric phase and the spin-Peierls phase is at
$r_1=0$, while that between the symmetric phase and the nematic
phase is at $r_2=0$. Finally, the second-order transition from the
nematic to the spin-Peierls phase is at
\begin{equation}
r_1 = 2 g \sqrt{-r_2/u_2} - w r_2 /(2 u_2). \label{a2}
\end{equation}
Notice that these three second-order lines all appear to meet at
the origin $r_1=0$, $r_2=0$. However, as indicated in
Fig~\ref{figphase}, this is pre-empted by a line of {\em first
order} transitions close to the origin. The reason for this may
be seen by the following simple argument. Imagine integrating out
the $\Phi_{\text{n}}$ fluctuations to derive an effective action
for the $\Psi_{\text{sp}}$: this always induces an effective
quartic term $\sim |\Psi_{\text{sp}}|^4$ with a coefficient $\sim
- g^2/|r_2|$. So for small enough $|r_2|$, the net coefficient of
$|\Psi_{\text{sp}}|^4$ always becomes negative, and this drives
the transitions involving onset of non-zero $\langle
\Psi_{\text{sp}} \rangle$ first order. A consequence of this line
of first-order transitions is that there are two tricritical
points $T_1$, $T_2$, and their positions are indicated in the
caption to Fig~\ref{figphase}.

Let us now consider the nature of the fluctuations in the
vicinity of the second-order phase transitions. We examine first
the transition from the symmetric to the nematic phase, which has
an Ising order parameter. For $T>0$, these transitions will remain
in the universality class of the two-dimensional Ising model.
However, at $T=0$, we have to also consider the influence of the
fermionic excitations. A general analysis of such effects has been
presented in Ref~\onlinecite{qptd}, and we recall the results
relevant to our discussion here. The $T=0$ ground state is either
an insulator or a superconductor, and only the gapless nodal
fermionic excitations can influence the nature of the critical
properties of a zero temperature quantum phase transition. In an
anisotropic, $d$-wave superconductor, the four nodal points are
described by a Dirac-like Hamiltonian for 2, four-component,
fermionic Nambu spinors $\Psi_1$, $\Psi_2$ (we will follow the
notation of Ref~\onlinecite{qptd}. The nematic order parameter
has zero net momentum, and so has a simple linear coupling to
these fermionic excitations \cite{qptd}:
\begin{equation}
\Phi_{\text{n}} \left( \Psi_1^{\dagger} \tau^x \Psi_1 + \Psi_2
\tau^x \Psi_2 \right), \label{a3}
\end{equation}
where $\tau^{x,y,z}$ are Pauli matrices in the Nambu space
(again, in the notation of Ref~\onlinecite{qptd}). A
renormalization group analysis of the consequences of (\ref{a3})
has been carried out, and it is found that the couplings to the
fermions drives the $T=0$ transition between the symmetric and
nematic phases first order.

Now let us consider the remaining two second-order transitions in
Fig~\ref{figphase}, for which $\Psi_{\text{sp}}$ is the order
parameter. In contrast to $\Phi_{\text{n}}$, the spin-Peierls
order has momentum $(\pi, 0)$ or $(0, \pi)$, and so has no simple
linear coupling to the nodal points, barring the exceptional case
in which the nodal points are exactly at $(\pi/2,\pi/2)$. In
general, the couplings between $\Psi_{\text{sp}}$ and the
fermions are of the form
\begin{equation}
\Psi_{\text{sp}}^2 \left( \Psi_1^{\dagger} \tau^x \Psi_1 + \Psi_2
\tau^x \Psi_2 \right), \label{a4}
\end{equation}
as can be expected from (\ref{pn}) and (\ref{a3}), and also of
the form
\begin{equation}
|\Psi_{\text{sp}}|^2 \left( \Psi_1^{\dagger} \tau^z \Psi_1 +
\Psi_2 \tau^z \Psi_2 \right), \label{a5}
\end{equation}
Unlike (\ref{a3}), the couplings in (\ref{a4},\ref{a5}) are
irrelevant, and so the fermions do not modify the leading
critical properties of any transition in which $\Psi_{\text{sp}}$
is the order parameter.

For the transition between the nematic and spin-Peierls phases,
the order parameter is either the real or imaginary part of
$\Psi_{\text{sp}}$: the presence of nematic order in both phases
therefore makes the residual order parameter Ising-like. So this
transition is in the universality class of three-dimensional
(two-dimensional) Ising model at $T=0$ ($T>0$).

Finally, we consider the fluctuations near the second-order
transition between the symmetric and the spin-Peierls phase. As
we have just noted, the fermionic excitations can be neglected
even at $T=0$. Further, as $\Psi_{\text{sp}}$ is the primary
order parameter, we can integrate out the $\Phi_{\text{n}}$
fluctuations at the cost of modifying the couplings in the
effective action for the $\Psi_{\text{sp}}$ alone. The resulting
action has the same form as $S$, after dropping all the terms
involving $\Phi_{\text{n}}$; so near the critical point, we need
only follow the flow of the nonlinear couplings $u_1$ and $v$.
For the $T=0$ transition, the standard Wilson-Fisher analysis of
a phase transition in three spacetime dimension leads to the
renormalization group flow equations
\begin{eqnarray}
\frac{du_1}{d\ell} &=& u_1 - 5 u_1^2 - 9 v^2 \nonumber \\
\frac{dv}{d\ell} &=& v - 6 u_1 v,
\end{eqnarray}
where we have absorbed a phase-space factor by a rescaling of the
couplings in a standard manner. The only stable fixed point of
these equations is $u_1 = 1/5$, $v=0$. At the point $v=0$, the
effective action for $\Psi_{\text{sp}}$ has the additional
symmetry under the global $U(1)$ change in the phase of
$\Psi_{\text{sp}}$. So the $T=0$ critical point is in the
universality class of three dimensional XY model. Related
considerations can be applied to the $T>0$ transition between the
symmetric and spin-Peierls phases: there, the anisotropy
associated with $v$ leads to phase transitions with continuously
varying exponents, associated with physics of the Ashkin-Teller
model \cite{jkkn}.

We conclude this section by summarizing the main physical
implications of Fig~\ref{figphase}. Starting from the symmetric
phase, one may break a square lattice symmetry either by lowering
temperature, or by reducing doping. One route to this is a direct
second-order transition from the symmetric to the spin-Peierls
phase; the latter phase also has nematic order. The implications
of the physical arguments we have presented in
Section~\ref{intro} is that this is the preferred route.
Alternatively, we can first undergo a transition to the nematic
phase, and possibly have a further symmetry breaking into the
spin-Peierls phase.

\section{Computations for the two-leg ladder}
\label{2leg}

We begin a bond operator theory of doped two-leg ladder by writing
down the Hamiltonian for the pure t-J model. The effect of Coulomb
repulsion will be considered in Appendix~\ref{coulomb}.

\begin{eqnarray}
H &=& H_t + H_J
\nonumber \\
&=& -t \sum_{i} \left\{ c^{\dagger}_{1ia} c_{2ia} +
c^{\dagger}_{2ia} c_{1ia} \right\}
\nonumber \\
&-& t \sum_{\langle i,j \rangle} \left\{ c^{\dagger}_{1ja} c_{1ia}
 + c^{\dagger}_{2ja} c_{2ia} + (h.c.) \right\}
\nonumber \\
&+& J \sum_i {\bf S}_{1i}\cdot{\bf S}_{2i} \nonumber \\
&+& \lambda J \sum_{\langle i,j \rangle} \left\{ {\bf
S}_{1i}\cdot{\bf S}_{1j}+{\bf S}_{2i}\cdot{\bf S}_{2j} \right\}
\end{eqnarray}
where $i$ indicates the position of the $i$-th dimer along the
ladder direction and $\langle i,j \rangle$ stands for the
summation over the nearest neighbor, i.e. $j=i+1$.

First, let us take a look at the hopping Hamiltonian $H_t$.
Hopping between the sites inside a given dimer is given by:

\begin{equation}
c^{\dagger}_{1ia} c_{2ia} + c^{\dagger}_{2ia} c_{1ia} =
h^{\dagger}_{1ia} h_{2ia} + h^{\dagger}_{2ia} h_{1ia}
\end{equation}

Hopping between the sites of adjacent dimers is obtained as
follows.

\begin{eqnarray}
c^{\dagger}_{1ia} c_{1ja} &=& \left\{
h^{\dagger}_{1ia}d_i+\frac{1}{\sqrt{2}}
\varepsilon_{ab}s^{\dagger}_i h_{2ib}
-\frac{1}{\sqrt{2}}\varepsilon_{ac}\sigma^{\alpha}_{cb}t^{\dagger}_{i
\alpha}
h_{2ib} \right\} \nonumber \\
&\times& \left\{ d^{\dagger}_j h_{1ja}+\frac{1}{\sqrt{2}}
\varepsilon_{ac}h^{\dagger}_{2jc} s_j
-\frac{1}{\sqrt{2}}\varepsilon_{ae}\bar{\sigma}^{\beta}_{ed}h^{\dagger}_{2jd}
t_{j \beta} \right\} \nonumber \\
&\Rightarrow& h^{\dagger}_{1ia}h_{1ja} d_i d^{\dagger}_j
+\frac{1}{2}\varepsilon_{ab}\varepsilon_{ac} s^{\dagger}_i s_j
h_{2ib} h^{\dagger}_{2jc} \nonumber \\
&+&\frac{1}{\sqrt{2}} d_i s_j \varepsilon_{ac} h^{\dagger}_{1ia}
h^{\dagger}_{2jc} +\frac{1}{\sqrt{2}} d^{\dagger}_j s^{\dagger}_i
\varepsilon_{ab}h_{2ib}
h_{1ja} \nonumber \\
&+& \frac{1}{2}
\varepsilon_{ac}\varepsilon_{ae}\sigma^{\alpha}_{cb}
\bar{\sigma}^{\beta}_{ed}t^{\dagger}_{i \alpha}t_{j
\beta}h_{2ib}h^{\dagger}_{2jd}
\end{eqnarray}
where any term containing a sigle $t$ boson operator is ignored
assuming that there is no magnetic order. Similarly,

\begin{eqnarray}
c^{\dagger}_{2ia} c_{2ja} &=& \left\{
h^{\dagger}_{2ia}d_i+\frac{1}{\sqrt{2}}
\varepsilon_{ab}s^{\dagger}_i h_{1ib}
+\frac{1}{\sqrt{2}}\varepsilon_{ac}\sigma^{\alpha}_{cb}t^{\dagger}_{i
\alpha}
h_{1ib} \right\} \nonumber \\
&\times& \left\{ d^{\dagger}_j h_{2ja}+\frac{1}{\sqrt{2}}
\varepsilon_{ac}h^{\dagger}_{1jc} s_j
+\frac{1}{\sqrt{2}}\varepsilon_{ae}\bar{\sigma}^{\beta}_{ed}h^{\dagger}_{1jd}
t_{j \beta} \right\} \nonumber \\
&\Rightarrow& h^{\dagger}_{2ia}h_{2ja} d_i d^{\dagger}_j
+\frac{1}{2}\varepsilon_{ab}\varepsilon_{ac} s^{\dagger}_i s_j
h_{1ib} h^{\dagger}_{1jc} \nonumber \\
&+&\frac{1}{\sqrt{2}} d_i s_j \varepsilon_{ac} h^{\dagger}_{2ia}
h^{\dagger}_{1jc} +\frac{1}{\sqrt{2}} d^{\dagger}_j s^{\dagger}_i
\varepsilon_{ab}h_{1ib}
h_{2ja} \nonumber \\
&+& \frac{1}{2}
\varepsilon_{ac}\varepsilon_{ae}\sigma^{\alpha}_{cb}
\bar{\sigma}^{\beta}_{ed}t^{\dagger}_{i \alpha}t_{j
\beta}h_{1ib}h^{\dagger}_{1jd}
\end{eqnarray}
By combining the above two equations, we get

\begin{eqnarray}
&&c^{\dagger}_{1ia} c_{1ja} + c^{\dagger}_{2ia} c_{2ja}+(h.c.)
\nonumber \\
&=& d_i d^{\dagger}_j \{ h^{\dagger}_{1ia}h_{1ja}+
h^{\dagger}_{2ia}h_{2ja} \}
\nonumber \\
&+&\frac{1}{2}s^{\dagger}_i s_j \varepsilon_{ab}\varepsilon_{ac}
\{ h_{1ib} h^{\dagger}_{1jc}+h_{2ib} h^{\dagger}_{2jc} \}
\nonumber \\
&+&\frac{1}{\sqrt{2}} d_i s_j \varepsilon_{ac} \{
 h^{\dagger}_{2ia}h^{\dagger}_{1jc}+ h^{\dagger}_{1ia}
h^{\dagger}_{2jc} \}
\nonumber \\
&+&\frac{1}{\sqrt{2}}d^{\dagger}_j s^{\dagger}_i \varepsilon_{ab}
\{ h_{1ib}h_{2ja}+h_{2ib}h_{1ja} \}
\nonumber \\
&+&\frac{1}{2}
\varepsilon_{ac}\varepsilon_{ae}\sigma^{\alpha}_{cb}
\bar{\sigma}^{\beta}_{ed}t^{\dagger}_{i \alpha}t_{j \beta} \{
h_{1ib}h^{\dagger}_{1jd}+h_{2ib}h^{\dagger}_{2jd} \} +(h.c.)
\end{eqnarray}

Assuming $s$ and $d$ bosons are condensed, i.e. $s_i = \bar{s}$
and $d_i = \bar{d}$, one is able to express $H_t$ as follows.

\begin{eqnarray}
H_t &=& -t \sum_i \{
h^{\dagger}_{1ia}h_{2ia}+h^{\dagger}_{2ia}h_{1ia} \}
\nonumber \\
&-& t (\bar{d}^2 - \frac{1}{2} \bar{s}^2) \sum_{\langle i,j
\rangle} \{ h^{\dagger}_{1ia}h_{1ja}+h^{\dagger}_{1ja}h_{1ia}
\nonumber \\
&&+h^{\dagger}_{2ia}h_{2ja}+h^{\dagger}_{2ja}h_{2ia} \}
\nonumber \\
&-&  \sqrt{2} t \bar{d}\bar{s} \sum_{\langle i,j \rangle}
\varepsilon_{ab}\{ h^{\dagger}_{1ia}h^{\dagger}_{2jb}
\nonumber \\
&&+h^{\dagger}_{2ia}h^{\dagger}_{1jb}
+h_{2jb}h_{1ia}+h_{1jb}h_{2ia} \}
\nonumber \\
&+&H_{t^2 h^2}
\end{eqnarray}
where
\begin{eqnarray}
H_{t^2 h^2} &=& - \frac{t}{2} \varepsilon_{ac}\varepsilon_{ae}
\sigma^{\alpha}_{cb} \sigma^{\beta}_{ed} \sum_{\langle i,j
\rangle} \{ t^{\dagger}_{i \alpha}t_{j \beta}
(h_{1ib}h^{\dagger}_{1jd}+h_{2ib}h^{\dagger}_{2jd})
\nonumber \\
&+&(h.c.) \}
\end{eqnarray}

Now let us turn our attention to the Hamiltonian $H_J$.
Contribution from the coupling inside the dimer has been shown
previously. That is,

\begin{equation}
{\bf S}_{1i}\cdot{\bf S}_{2i} = -\frac{3}{4}s^{\dagger}_i s_i
+\frac{1}{4} t^{\dagger}_{i\alpha} t_{i\alpha}
\end{equation}

Coupling term between the spins of different dimers is given by:

\begin{eqnarray}
S_{1i\alpha} S_{1j\alpha} &=& \tilde{S}_{1i\alpha}
\tilde{S}_{1j\alpha} +\frac{1}{4}
\sigma^{\alpha}_{ab}\sigma^{\alpha}_{cd}
h^{\dagger}_{1ia}h_{1ib}h^{\dagger}_{1jc}h_{1jd} \nonumber \\
&+&\frac{1}{2}h^{\dagger}_{1ia}\sigma^{\alpha}_{ab} h_{1ib}
\tilde{S}_{1j\alpha}+\frac{1}{2}h^{\dagger}_{1ja}\sigma^{\alpha}_{ab}
h_{1jb} \tilde{S}_{1i\alpha} .
\end{eqnarray}
Remember that $\tilde{S}_{1i\alpha} = \frac{1}{2}(s^{\dagger}_i
t_{i\alpha}+t^{\dagger}_{i\alpha}s_i -i
\epsilon_{\alpha\beta\gamma} t^{\dagger}_{i\beta}t_{i\gamma})$ and
$\tilde{S}_{2i\alpha} = -\frac{1}{2}(s^{\dagger}_i
t_{i\alpha}+t^{\dagger}_{i\alpha}s_i +i
\epsilon_{\alpha\beta\gamma} t^{\dagger}_{i\beta}t_{i\gamma})$.
The cross term can be ignored assuming that there is no
contribution from the term containing a single $t$ operator or
$t^{\dagger}_{\alpha}t_{\beta}$ with $\alpha \neq \beta$. In other
words, it is assumed that there is no magnetic ordering.
Therefore,

\begin{eqnarray}
&&S_{1i\alpha} S_{1j\alpha} + S_{2i\alpha} S_{2j\alpha}
\nonumber \\
&=& \tilde{S}_{1i\alpha} \tilde{S}_{1j\alpha}
+\tilde{S}_{2i\alpha} \tilde{S}_{2j\alpha} +\frac{1}{\lambda
J}H_{h^4}
\end{eqnarray}
where
\begin{eqnarray}
H_{h^4} &=& \frac{\lambda J}{4} \sum_{\langle i,j \rangle}
\sigma^{\alpha}_{ab}\sigma^{\alpha}_{cd} \{
h^{\dagger}_{1ia}h_{1ib}h^{\dagger}_{1jc}h_{1jd}
\nonumber \\
&+&h^{\dagger}_{2ia}h_{2ib}h^{\dagger}_{2jc}h_{2jd} \}
\end{eqnarray}

Coupling terms involving $\tilde{S}$ are evaluated below.

\begin{eqnarray}
4(\tilde{S}_{1i\alpha}\tilde{S}_{1j\alpha}
&+&\tilde{S}_{2i\alpha}\tilde{S}_{2j\alpha})
\nonumber \\
&=& ( s^{\dagger}_i t_{i\alpha}+t^{\dagger}_{i\alpha}s_i -i
\epsilon_{\alpha\beta\gamma} t^{\dagger}_{i\beta}t_{i\gamma})
\nonumber \\
&&\times(s^{\dagger}_j t_{j\alpha}+t^{\dagger}_{j\alpha}s_j
-i \epsilon_{\alpha\mu\nu} t^{\dagger}_{j\mu}t_{j\nu}) \nonumber \\
&+& ( s^{\dagger}_i t_{i\alpha}+t^{\dagger}_{i\alpha}s_i +i
\epsilon_{\alpha\beta\gamma} t^{\dagger}_{i\beta}t_{i\gamma})
\nonumber \\
&&\times(s^{\dagger}_j t_{j\alpha}+t^{\dagger}_{j\alpha}s_j
+i \epsilon_{\alpha\mu\nu} t^{\dagger}_{j\mu}t_{j\nu}) \nonumber \\
&=& 2 \{ t^{\dagger}_{i\alpha}t^{\dagger}_{j\alpha}s_i s_j +
t_{i\alpha}t_{j\alpha}s^{\dagger}_i s^{\dagger}_j
\nonumber \\
&&+t^{\dagger}_{i\alpha}t_{j\alpha}s^{\dagger}_j s_i
+ t^{\dagger}_{j\alpha}t_{i\alpha}s^{\dagger}_i s_j \} \nonumber \\
&-& 2\epsilon_{\alpha\beta\gamma}\epsilon_{\alpha\mu\nu}
t^{\dagger}_{i\beta}t_{i\gamma}t^{\dagger}_{j\mu}t_{j\nu}
\end{eqnarray}

Therefore $H_J$ is summarized as follows.

\begin{eqnarray}
H_J &=& J \sum_i \{ -\frac{3}{4}s^{\dagger}_i s_i +\frac{1}{4}
t^{\dagger}_{i\alpha}t_{i\alpha} \}
\nonumber \\
&+& \frac{\lambda J}{2} \sum_{\langle i,j \rangle} \{
t^{\dagger}_{i\alpha}t^{\dagger}_{j\alpha}s_i s_j +
t_{i\alpha}t_{j\alpha}s^{\dagger}_i s^{\dagger}_j
\nonumber \\
&&+t^{\dagger}_{i\alpha}t_{j\alpha}s^{\dagger}_j s_i +
t^{\dagger}_{j\alpha}t_{i\alpha}s^{\dagger}_i s_j \}
\nonumber \\
&& \underbrace{-\frac{\lambda J}{2} \sum_{\langle i,j \rangle}
\epsilon_{\alpha\beta\gamma}\epsilon_{\alpha\mu\nu}
t^{\dagger}_{i\beta}t_{i\gamma}t^{\dagger}_{j\mu}t_{j\nu}}_{\equiv
H_{t^4}} + H_{h^4}
\nonumber \\
&\Rightarrow& -\frac{3}{4}J N \bar{s}^2 +\frac{J}{4}\sum_i
t^{\dagger}_{i\alpha}t_{i\alpha}
\nonumber \\
&+& \frac{\lambda J}{2} \bar{s}^2 \sum_{\langle i,j \rangle} \{
t^{\dagger}_{i\alpha}t^{\dagger}_{j\alpha} +
t_{i\alpha}t_{j\alpha} + t^{\dagger}_{i\alpha}t_{j\alpha} +
t^{\dagger}_{j\alpha}t_{i\alpha} \}
\nonumber \\
&+& H_{t^4} + H_{h^4}
\end{eqnarray}
where $N$ is the number of dimers and $s$ boson is condensed so
that $s_i=\bar{s}$. The total Hamiltonian (without constraints) is
given by the sum of $H_t$ and $H_J$.

\begin{eqnarray}
H &=& H_t + H_J \nonumber \\
&=&  -t \sum_i \{
h^{\dagger}_{1ia}h_{2ia}+h^{\dagger}_{2ia}h_{1ia} \}
\nonumber \\
&-& t (\bar{d}^2 - \frac{1}{2} \bar{s}^2) \sum_{\langle i,j
\rangle} \{ h^{\dagger}_{1ia}h_{1ja}+h^{\dagger}_{1ja}h_{1ia}
\nonumber \\
&&+h^{\dagger}_{2ia}h_{2ja}+h^{\dagger}_{2ja}h_{2ia} \}
\nonumber \\
&-&  \sqrt{2} t \bar{d}\bar{s} \sum_{\langle i,j \rangle}
\varepsilon_{ab}\{ h^{\dagger}_{1ia}h^{\dagger}_{2jb}
+h^{\dagger}_{2ia}h^{\dagger}_{1jb}
\nonumber \\
&&+h_{2jb}h_{1ia}+h_{1jb}h_{2ia} \} \nonumber \\
&-&\frac{3}{4}J N \bar{s}^2 +\frac{J}{4}\sum_i
t^{\dagger}_{i\alpha}t_{i\alpha}
\nonumber \\
&+& \frac{\lambda J}{2} \bar{s}^2 \sum_{\langle i,j \rangle} \{
t^{\dagger}_{i\alpha}t^{\dagger}_{j\alpha} +
t_{i\alpha}t_{j\alpha} + t^{\dagger}_{i\alpha}t_{j\alpha}
+ t^{\dagger}_{j\alpha}t_{i\alpha} \} \nonumber \\
&+& H_{t^2 h^2} + H_{t^4} + H_{h^4}
\end{eqnarray}

As one can see in the following sections, a convenient
simplification is obtained by expressing the Hamiltonian in terms
of the bonding and anti-bonding fermionic operators which are
defined as follows:

\begin{eqnarray}
h_{i+a} \equiv \frac{1}{\sqrt{2}}(h_{1ia}+h_{2ia}) \nonumber \\
h_{i-a} \equiv \frac{1}{\sqrt{2}}(h_{1ia}-h_{2ia})
\end{eqnarray}

\subsection{Mean-Field Hamiltonian without Quartic Terms}

In this section we will ignore the terms from $H_{t^4}$, $H_{h^4}$
and $H_{t^2 h^2}$. Then, the Hamiltonian is given in momentum
space representation as follows.

\begin{eqnarray}
&H& = -N\frac{3}{4} J \bar{s}^2 + \sum_{k_y} \left(
\frac{J}{4}+\lambda J \bar{s}^2 \cos{k_y} \right)
t^{\dagger}_{\alpha}(k_y) t_{\alpha}(k_y)
\nonumber \\
&+&\sum_{k_y} \frac{\lambda J \bar{s}^2}{2} \cos{k_y} \left\{
t^{\dagger}_{\alpha} (k_y) t^{\dagger}_{\alpha}(-k_y)+
t_{\alpha}(k_y) t_{\alpha}(-k_y)
\right\}  \nonumber \\
&+& \sum_{k_y} \{t(\bar{s}^2 -2\bar{d}^2)\cos{k_y}
-t\}h^{\dagger}_{+a}(k_y)
h_{+a}(k_y) \nonumber \\
&+& \sum_{k_y} \{t(\bar{s}^2 -2\bar{d}^2)\cos{k_y}
+t\}h^{\dagger}_{-a}(k_y)
h_{-a}(k_y) \nonumber \\
&-&2\sqrt{2} t \bar{d}\bar{s}\sum_{k_y} \cos{k_y} \{
h^{\dagger}_{+\uparrow}(k_y)h^{\dagger}_{+\downarrow}(-k_y)
\nonumber \\
&&-h^{\dagger}_{-\uparrow}(k_y)h^{\dagger}_{-\downarrow}(-k_y)
+(h.c.) \}
\end{eqnarray}

As mentioned previously, it is necessary to impose certain
constraints. In this study, constraints are introduced in Lagrange
multiplier method. In other words, ``constraint Hamiltonian''
($H_c$) is added to the original Hamiltonian. And then the unknown
parameters are determined by the saddle point condition of the
ground state energy.

\begin{eqnarray}
&H&_c = -\mu \sum_i ( s^{\dagger}_i s_i +
t^{\dagger}_{i\alpha}t_{i\alpha} +h^{\dagger}_{1ia}h_{1ia}
+h^{\dagger}_{2ia}h_{2ia}
\nonumber \\
&&+d^{\dagger}_i d_i - 1 ) \nonumber \\
&-& \xi\sum_i \big( h^{\dagger}_{1ia}h_{1ia}
+h^{\dagger}_{2ia}h_{2ia} +2d^{\dagger}_i d_i - 2x \big) \nonumber \\
&\Rightarrow& -\mu \sum_{k_y} \left( t^{\dagger}_{\alpha}(k_y)
t_{\alpha}(k_y)
+\bar{s}^2-\bar{d}^2 +2x -1 \right) \nonumber \\
&-& \xi \sum_{k_y} \big( h^{\dagger}_{+a}(k_y)
h_{+a}(k_y)+h^{\dagger}_{-a}(k_y)h_{-a}(k_y)
\nonumber \\
&&+2\bar{d}^2 -2x \big)
\end{eqnarray}
where $x$ is the hole concentration.

Therefore the total Hamiltonian with constraints is given by:
\begin{eqnarray}
&H& = N \epsilon_0 \nonumber \\
&+&\sum_{k_y} \{ A_{k_y} t^{\dagger}_{\alpha}(k_y) t_{\alpha}(k_y)
\nonumber \\
&&+B_{k_y} \left(
t^{\dagger}_{\alpha}(k_y)t^{\dagger}_{\alpha}(-k_y)
+t_{\alpha}(k_y)t_{\alpha}(-k_y) \right) \} \nonumber \\
&+& \sum_{k_y} \{
\epsilon_{+}(k_y)h^{\dagger}_{+a}(k_y)h_{+a}(k_y)
\nonumber \\
&-&D(k_y) (
h^{\dagger}_{+\uparrow}(k_y)h^{\dagger}_{+\downarrow}(-k_y)
+h_{+\downarrow}(-k_y)h_{+\uparrow}(k_y) ) \} \nonumber \\
&+& \sum_{k_y} \{
\epsilon_{-}(k_y)h^{\dagger}_{-a}(k_y)h_{-a}(k_y)
\nonumber \\
&+&D(k_y) (
h^{\dagger}_{-\uparrow}(k_y)h^{\dagger}_{-\downarrow}(-k_y)
+h_{-\downarrow}(-k_y)h_{-\uparrow}(k_y) ) \}
\end{eqnarray}
where
\begin{eqnarray}
\epsilon_0 &=& -\frac{3}{4}J \bar{s}^2 -\mu (\bar{s}^2 -\bar{d}^2
+2x-1)
\nonumber \\
&-&\xi (2\bar{d}^2 -2x)\\
A_{k_y} &=& \frac{J}{4}-\mu+\lambda J\bar{s}^2 \cos{k_y} \\
B_{k_y} &=& \frac{\lambda J}{2} \bar{s}^2 \cos{k_y} \\
\epsilon_{+}(k_y) &=& t(\bar{s}^2-2\bar{d}^2)\cos{k_y}-t-\xi \\
\epsilon_{-}(k_y) &=& t(\bar{s}^2-2\bar{d}^2)\cos{k_y}+t-\xi \\
D(k_y) &=& 2\sqrt{2} t \bar{d}\bar{s} \cos{k_y}
\end{eqnarray}

The above Hamiltonian can be diagonalized by using Bogoliubov
transformation. That is,

\begin{eqnarray}
\gamma_{\alpha}(k_y) &=& u_t(k_y) t_{\alpha}(k_y) +v_t (k_y) t^{\dagger}_{\alpha}(-k_y)\\
\beta_{\pm a}(k_y) &=& u_{\pm}(k_y)h_{\pm a}(k_y)+v_{\pm}(k_y)
\varepsilon_{ab}h^{\dagger}_{\pm b}(-k_y)
\end{eqnarray}
where
\begin{eqnarray}
u_{t}^2 (k_y) &=& \frac{1}{2} \left( \frac{A_{k_y}}{\omega_{k_y}}+1 \right) \\
v_{t}^2 (k_y) &=& \frac{1}{2} \left( \frac{A_{k_y}}{\omega_{k_y}}-1 \right) \\
u_t (k_y) v_t (k_y) &=& \frac{B_{k_y}}{\omega_{k_y}}\\
\omega_{k_y} &=& \sqrt{A_{k_y}^2 -4B_{k_y}^2}
\end{eqnarray}
and
\begin{eqnarray}
u_{\pm}^2 (k_y) &=& \frac{1}{2} \left(1+
\frac{\epsilon_{\pm}(k_y)}{\Omega_{\pm}(k_y)}\right) \\
v_{\pm}^2 (k_y) &=& \frac{1}{2} \left(1-
\frac{\epsilon_{\pm}(k_y)}{\Omega_{\pm}(k_y)} \right) \\
u_{\pm}(k_y)v_{\pm}(k_y) &=& \mp \frac{D(k_y)}{\Omega_{\pm}(k_y)} \\
\Omega_{\pm}(k_y) &=& \sqrt{\epsilon_{\pm}^{2}(k_y) +D^{2}(k_y)}
\end{eqnarray}

In terms of Bogoliubov variables, Hamiltonian is written as
follows.
\begin{eqnarray}
&H& = N \epsilon_0 +\sum_{k_y} \omega_{k_y}
\gamma^{\dagger}_{\alpha}(k_y) \gamma_{\alpha}(k_y)
+\frac{3}{2}\sum_{k_y} \left( \omega_{k_y} - A_{k_y} \right)
\nonumber \\
&+& \sum_{k_y} \Omega_{+}(k_y) \beta^{\dagger}_{+a}(k_y)
\beta_{+a}(k_y)
\nonumber \\
&+&\sum_{k_y} \left\{
\epsilon_{+}(k_y)-\sqrt{\epsilon^2_{+}(k_y)+D^2 (k_y)} \right\}
\nonumber \\
&+& \sum_{k_y} \Omega_{-}(k_y) \beta^{\dagger}_{-a}(k_y)
\beta_{-a}(k_y)
\nonumber \\
&+&\sum_{k_y} \left\{
\epsilon_{-}(k_y)-\sqrt{\epsilon^2_{-}(k_y)+D^2 (k_y)} \right\}
\end{eqnarray}

\subsubsection{Ground State Energy and Saddle-Point Equations}

Now the ground state energy per particle is given by:
\begin{eqnarray}
\epsilon_{gr}
&=& \frac{\langle H \rangle_{gr}}{N} \nonumber \\
&=& \epsilon_0 +\frac{3}{2} \int^{\pi}_0 \frac{dk_y}{\pi}
(\omega_{k_y} - A_{k_y})
\nonumber \\
&&+\int^{\pi}_0 \frac{dk_y}{\pi} \left\{ \epsilon_{+}(k_y) -
\Omega_{+}(k_y) \right\}
\nonumber \\
&&+\int^{\pi}_0 \frac{dk_y}{\pi} \left\{ \epsilon_{-}(k_y) -
\Omega_{-}(k_y) \right\}
\nonumber \\
&=& \epsilon^{'}_0 +\frac{3}{2} \int^{\pi}_0 \frac{dk_y}{\pi}
\omega_{k_y}
\nonumber \\
&& -\int^{\pi}_0 \frac{dk_y}{\pi} \Omega_{+}(k_y) -\int^{\pi}_0
\frac{dk_y}{\pi} \Omega_{-}(k_y)
\end{eqnarray}
where
\begin{equation}
\epsilon^{'}_0 = -\frac{3}{4}J \bar{s}^2 -\mu (\bar{s}^2
-\bar{d}^2 +2x -\frac{5}{2}) -\xi (2 +2\bar{d}^2 -2x) .
\end{equation}
Note that constants obtained from integrating $A_{k_y}$ and
$\epsilon_{\pm}(k_y)$ are ignored.

The term contributed from the triplet bosons ($t_\alpha$) can be
intergrated explicitly by using the elliptic function ${\bf E}$.
\begin{eqnarray}
\epsilon_{gr} &=&  \epsilon^{'}_0 +\frac{3}{\pi}J \left(
\frac{1}{4}-\frac{\mu}{J} \right) \sqrt{1+\eta} {\bf E}
\left(\sqrt{\frac{2\eta}{1+\eta}} \right)
\nonumber \\
&-&\int^{\pi}_0 \frac{dk_y}{\pi} \Omega_{+}(k_y) -\int^{\pi}_0
\frac{dk_y}{\pi} \Omega_{-}(k_y)
\end{eqnarray}
where
\begin{equation}
\eta = \frac{2\lambda \bar{s}^2}{\frac{1}{4} - \frac{\mu}{J}}.
\end{equation}

Parameters in the Hamiltonian are determined by the saddle point
condition in the mean-field approximation. In other words,

\begin{eqnarray}
\frac{\partial \epsilon_{gr}}{\partial \bar{s}} = \frac{\partial
\epsilon_{gr}}{\partial \mu} = \frac{\partial
\epsilon_{gr}}{\partial \bar{d}} = \frac{\partial
\epsilon_{gr}}{\partial \xi} = 0 .
\end{eqnarray}

Explicit saddle-point equations are written as follows.
\begin{eqnarray}
\frac{\partial (\epsilon_{gr}/J)}{\partial \bar{s}^2} &=&
-\frac{3}{4} -\frac{\mu}{J} +\frac{3 \lambda}{\pi\eta} \Bigg\{
\sqrt{1+\eta} {\bf E} \left(\sqrt{\frac{2\eta}{1+\eta}} \right)
\nonumber \\
&-&\frac{1}{ \sqrt{1+\eta}}  {\bf K}
\left(\sqrt{\frac{2\eta}{1+\eta}} \right) \Bigg\}
\nonumber \\
&-&\frac{t/J}{\bar{s}}\int^{\pi}_0 \frac{dk_y}{\pi} \cos{k_y}
\Bigg\{
\frac{\bar{s}\epsilon_{+}(k_y)+\sqrt{2}\bar{d}D(k_y)}{\Omega_{+}(k_y)}
\nonumber \\
&+&\frac{\bar{s}\epsilon_{-}(k_y)+\sqrt{2}\bar{d}D(k_y)}{\Omega_{-}(k_y)}
\Bigg\} = 0
\\
\frac{\partial (\epsilon_{gr}/J)}{\partial (\mu/J)} &=& \bar{d}^2
-\bar{s}^2 +\frac{5}{2} -2x
\nonumber \\
&-&\frac{3}{2\pi} \Bigg\{ \sqrt{1+\eta} {\bf E}
\left(\sqrt{\frac{2\eta}{1+\eta}} \right)
\nonumber \\
&+&\frac{1}{ \sqrt{1+\eta}}  {\bf K}
\left(\sqrt{\frac{2\eta}{1+\eta}} \right) \Bigg\} = 0
\\
\frac{\partial (\epsilon_{gr}/J)}{\partial \bar{d}^2} &=&
\frac{\mu}{J} -2\frac{t}{J}\frac{\xi}{t} \nonumber \\
&+&\frac{t/J}{\bar{d}}\int^{\pi}_0 \frac{dk_y}{\pi} \cos{k_y}
\Bigg\{ \frac{2\bar{d}\epsilon_{+}(k_y)-\sqrt{2}\bar{s}
D(k_y)}{\Omega_{+}(k_y)}
\nonumber \\
&+&\frac{2\bar{d}\epsilon_{-}(k_y)-\sqrt{2}\bar{s}
D(k_y)}{\Omega_{-}(k_y)} \Bigg\} = 0
\\
\frac{\partial (\epsilon_{gr}/J)}{\partial (\xi/J)} &=& -2
-2\bar{d}^2+2x
\nonumber \\
&+&\int^{\pi}_0 \frac{dk_y}{\pi} \left\{
\frac{\epsilon_{+}(k_y)}{\Omega_{+}(k_y)}
+\frac{\epsilon_{-}(k_y)}{\Omega_{-}(k_y)} \right\} = 0
\end{eqnarray}

\subsubsection{Limiting Case with $\bar{d}=0$}

The above saddle-point equations are highly non-linear as a
function of parameters $\bar{s}$, $\mu$, $\bar{d}$, and $\xi$.
Since analytic solution is not accessible in this case, it is
natural to make use of a numerical method. In general, solving a
set of non-linear equations numerically is quite sensitive to the
initial guesses of parameters. Therefore, it is very desirable to
have some limiting situation where an explicit, analytic solution
is available. Such a situation is realized when we set $\bar{d} =
0$, i.e. turning off the pairing terms. Then, the ground state
energy is given as follows.

\begin{eqnarray}
\epsilon_{gr} &=&  \epsilon_0 -\frac{3}{2} J \left(
\frac{1}{4}-\frac{\mu}{J} \right) \left\{ 1- \frac{2}{\pi}
\sqrt{1+\eta} {\bf E} \left(\sqrt{\frac{2\eta}{1+\eta}} \right)
\right\}
\nonumber \\
&+&\int^{\pi}_0 \frac{dk_y}{\pi} \left\{ \epsilon_{+}(k_y) -
\Omega_{+}(k_y) \right\}
\end{eqnarray}
where the contribution from the anti-bonding fermion ($h_{-}$) is
ignored because, in the case of $t/J \geq 1$,
 the anti-bonding fermion will have a reasonably larger energy
than the bonding one, separated by roughly $2t$. And therefore its
contribution to the ground state energy is very small.

The fermion contribution can be simplified further.
\begin{eqnarray}
{\int}^{\pi}_0 \frac{dk_y}{\pi}
\Big\{\epsilon_{+}(k_y)&-&\Omega_{+}(k_y)\Big\}
\nonumber \\
&=& \frac{2}{\pi} \int^{\pi}_{k_F} dk_y \left[ \epsilon_{+}(k_y)
\right]_{\bar{d}=0}
\nonumber \\
&=& \frac{2}{\pi} \int^{\pi}_{k_F} dk_y ( t\bar{s}^2 \cos{k_y} -t
-\xi )
\nonumber \\
&=& -\frac{2}{\pi} \left\{ (t+\xi)(\pi-k_F) +t\bar{s}^2 \sin{k_F}
\right\}
\nonumber \\
&=& \frac{2}{\pi} t \bar{s}^2 \left\{ \pi x \cos{(\pi x)} -
\sin{(\pi x)} \right\}
\end{eqnarray}
where the chemical potential $\xi$ and the Fermi wavevector $k_F$
are determined by:
\begin{eqnarray}
\sum_{k_y} h^{\dagger}_{+ a}(k_y) h_{+ a}(k_y) = x
&\Leftrightarrow&
\frac{2}{\pi}(\pi-k_F ) = 2x \\
\epsilon_{+}(k_F) = 0 &\Leftrightarrow& \xi = t\bar{s}^2 \cos{k_F}
-t .
\end{eqnarray}

Now the saddle-point equations are reduced to the following
equations.
\begin{eqnarray}
\frac{\partial (\epsilon_{gr}/J)}{\partial \bar{s}^2} &=& -\left(
\frac{3}{4} +\frac{\mu}{J} \right) + \frac{3 \lambda}{\pi\eta}
\Bigg\{ \sqrt{1+\eta} {\bf E} \left(\sqrt{\frac{2\eta}{1+\eta}}
\right)
\nonumber \\
&-&\frac{1}{ \sqrt{1+\eta}}  {\bf K}
\left(\sqrt{\frac{2\eta}{1+\eta}} \right) \Bigg\}
\nonumber \\
&+& \frac{2t}{\pi} \left( \pi x \cos{(\pi x)} -\sin{(\pi x)}
\right) =0
\label{saddleeq1}\\
\frac{\partial (\epsilon_{gr}/J)}{\partial (\mu/J)} &=&
\frac{5}{2} -2x -\bar{s}^2 -\frac{3}{2\pi} \Bigg\{ \sqrt{1+\eta}
{\bf E} \left(\sqrt{\frac{2\eta}{1+\eta}} \right)
\nonumber \\
&+&\frac{1}{ \sqrt{1+\eta}}  {\bf K}
\left(\sqrt{\frac{2\eta}{1+\eta}} \right) \Bigg\} = 0
\label{saddleeq2}
\end{eqnarray}
Note that the other saddle-point equations are explicitly solved.
In other words,
\begin{eqnarray}
\bar{d} &=& 0 \\
\xi &=& -t\bar{s}^2 \cos{(\pi x)} -t
\end{eqnarray}

Convenient simplification is realized from the fact that
Eq.(\ref{saddleeq1}) and (\ref{saddleeq2}) can be combined to
reduce to a equation containing only one parameter $\eta$. That
is,

\begin{eqnarray}
\frac{\eta}{2\lambda} \Bigg\{ &1& -\frac{2}{\pi}\frac{t}{J}\left(
\pi x \cos{(\pi x)} -\sin{(\pi x)} \right) \Bigg\}
\nonumber \\
&=& \frac{5}{2} -2x -\frac{3}{\pi}\frac{1}{\sqrt{1+\eta}} {\bf
K}\left( \sqrt{\frac{2\eta}{1+\eta}} \right)
\end{eqnarray}
where, once again, $\eta = \frac{2\lambda\bar{s}^2}{1/4-\mu/J}$.

\subsection{Mean-Field Hamiltonian with Quartic Terms}

In this section, effects of quartic terms such as $H_{t^4}$,
$H_{h^4}$, and $H_{t^2 h^2}$ are included approximately by using
quadratic decoupling. Let us start from $H_{t^4}$.
\begin{eqnarray}
H_{t^4} &=& -\frac{\lambda J}{2} \sum_{\langle i,j \rangle}
\epsilon_{\alpha\beta\gamma}\epsilon_{\alpha\mu\nu}
t^{\dagger}_{i\beta}t_{i\gamma}t^{\dagger}_{j\mu}t_{j\nu}
\nonumber \\
&=& -\frac{\lambda J}{2} \sum_{\langle i,j \rangle} \{
t^{\dagger}_{i\alpha}t_{i\beta}t^{\dagger}_{j\alpha}t_{j\beta} -
t^{\dagger}_{i\alpha}t_{i\beta}t^{\dagger}_{j\beta}t_{j\alpha} \}
\end{eqnarray}

Quadratic decoupling is carried out as follows.

\begin{eqnarray}
t^{\dagger}_{i\alpha}t^{\dagger}_{j\alpha}t_{i\beta}t_{j\beta}
&\simeq& \langle t^{\dagger}_{i\alpha}t^{\dagger}_{j\alpha}
\rangle t_{i\beta}t_{j\beta}
+t^{\dagger}_{i\alpha}t^{\dagger}_{j\alpha} \langle
t_{i\beta}t_{j\beta} \rangle
\nonumber \\
&-& \langle t^{\dagger}_{i\alpha}t^{\dagger}_{j\alpha}\rangle
\langle t_{i\beta}t_{j\beta}\rangle
\\
t^{\dagger}_{i\alpha}t^{\dagger}_{j\beta}t_{i\beta}t_{j\alpha}
&\simeq& \langle t^{\dagger}_{i\alpha}t_{j\alpha} \rangle
t^{\dagger}_{j\beta}t_{i\beta} +t^{\dagger}_{i\alpha}t_{j\alpha}
\langle t^{\dagger}_{j\beta}t_{i\beta} \rangle
\nonumber \\
&-& \langle t^{\dagger}_{i\alpha}t_{j\alpha}\rangle \langle
t^{\dagger}_{j\beta}t_{i\beta}\rangle
\end{eqnarray}

Therefore, $H_{t^4}$ is given by:
\begin{eqnarray}
H_{t^4} &=& \frac{\lambda J}{2} \sum_{\langle i,j \rangle} \big\{
P_y \left(t^{\dagger}_{i\alpha}t_{j\alpha}
+t^{\dagger}_{j\alpha}t_{i\alpha} \right)
\nonumber \\
&-&Q_y \left( t^{\dagger}_{i\alpha}t^{\dagger}_{j\alpha}
+t_{j\alpha}t_{i\alpha} \right) \big\}
\nonumber \\
&-& N \frac{\lambda J}{2}( P_y^2 -Q_y^2 )
\nonumber \\
&=& \frac{\lambda J}{2} \sum_{k_y} \cos{k_y} \big\{ 2P_y
t^{\dagger}_{\alpha}(k_y)t_{\alpha}(k_y)
\nonumber \\
&-&Q_y \left( t^{\dagger}_{\alpha}(k_y)t^{\dagger}_{\alpha}(-k_y)
+t_{\alpha}(-k_y)t_{\alpha}(k_y) \right) \big\}
\nonumber \\
&-& N \frac{\lambda J}{2}( P_y^2 -Q_y^2 )
\end{eqnarray}
where
\begin{eqnarray}
P_y &\equiv& \langle t^{\dagger}_{i\alpha}t_{j\alpha} \rangle =
\sum_{k_y} \cos{k_y} \langle
t^{\dagger}_{\alpha}(k_y)t_{\alpha}(k_y) \rangle
\\
Q_y &\equiv& \langle t^{\dagger}_{i\alpha}t^{\dagger}_{j\alpha}
\rangle = \sum_{k_y} \cos{k_y}  \langle
t^{\dagger}_{\alpha}(k_y)t^{\dagger}_{\alpha}(-k_y) \rangle .
\end{eqnarray}
Note that the translational symmetry is assumed. Now let us turn
to the Hamiltonian $H_{t^2 h^2}$.

\begin{eqnarray}
H_{t^2 h^2} &=& -\frac{t}{2} \varepsilon_{ac}\varepsilon_{ae}
\sigma^{\alpha}_{cb}\sigma^{\beta}_{ed} \sum_{\langle i,j \rangle}
\{ t^{\dagger}_{i \alpha}t_{j \beta}
(h_{1ib}h^{\dagger}_{1jd}+h_{2ib}h^{\dagger}_{2jd})
\nonumber \\
&+&(h.c.) \}
\end{eqnarray}
Remembering that, in the present approximation, non-zero quadratic
contraction is obtained only when $\alpha = \beta$ for $\langle
t^{\dagger}_{\alpha}t_{\beta} \rangle$, and $a = b$ for $ \langle
h^{\dagger}_a h_b \rangle$, the Hamiltonian $H_{t^2 h^2}$ is
quadratically decoupled as follows.

\begin{eqnarray}
&+&\varepsilon_{ac}\varepsilon_{ae}
\sigma^{\alpha}_{cb}\bar{\sigma}^{\beta}_{ed} \sum_{\langle i,j
\rangle} t^{\dagger}_{i \alpha}t_{j \beta} \{
h_{1ib}h^{\dagger}_{1jd} +(h.c.) \}
\nonumber \\
&\simeq& \varepsilon_{ac}\varepsilon_{ae}
\sigma^{\alpha}_{cb}\bar{\sigma}^{\alpha}_{eb} \sum_{\langle i,j
\rangle} t^{\dagger}_{i \alpha}t_{j \alpha} \{
h_{1ib}h^{\dagger}_{1jb} +(h.c.) \}
\nonumber \\
&\simeq& -\langle t^{\dagger}_{i\alpha}t_{j\alpha} \rangle
h^{\dagger}_{1ja}h_{1ia} -t^{\dagger}_{i\alpha}t_{j\alpha} \langle
h^{\dagger}_{1ja}h_{1ia} \rangle
\nonumber \\
&&+ \langle t^{\dagger}_{i\alpha}t_{j\alpha} \rangle \langle
h^{\dagger}_{1ja}h_{1ia} \rangle +(h.c.)
\nonumber \\
&=& -P_y \left( h^{\dagger}_{1ja}h_{1ia} +
h^{\dagger}_{1ia}h_{1ja} \right) -\Pi_y \left(
t^{\dagger}_{i\alpha}t_{j\alpha} +t^{\dagger}_{i\alpha}t_{j\alpha}
\right)
\nonumber \\
&&+2 P_y\Pi_y
\end{eqnarray}
where
\begin{equation}
\Pi_y \equiv \langle h^{\dagger}_{1ja}h_{1ia} \rangle = \sum_{k_y}
\cos{k_y} \langle h^{\dagger}_{1a}(k_y)h_{1a}(k_y) \rangle .
\end{equation}

Once again it is convenient to use the bonding and anti-bonding
fermion representation. By using the reflection symmetry, $\Pi_y$
is given by:

\begin{eqnarray}
\Pi_y &=& \sum_{k_y} \cos{k_y} \langle
h^{\dagger}_{1a}(k_y)h_{1a}(k_y) \rangle
\nonumber \\
&=& \sum_{k_y} \cos{k_y} \langle h^{\dagger}_{2a}(k_y)h_{2a}(k_y)
\rangle
\nonumber \\
&=& \frac{1}{2} \sum_{k_y} \cos{k_y} \big\{ \langle
h^{\dagger}_{+a}(k_y)h_{+a}(k_y) \rangle
\nonumber \\
&&+ \langle h^{\dagger}_{-a}(k_y)h_{-a}(k_y) \rangle \big\}
\end{eqnarray}

Finally, the Hamiltonian $H_{t^2 h^2}$ is given by:
\begin{eqnarray}
H_{t^2 h^2} &=& t P_y \sum_{k_y} \cos{k_y}  \left(
h^{\dagger}_{+a}(k) h_{+a}(k) + h^{\dagger}_{-a}(k) h_{-a}(k)
\right)
\nonumber \\
&+&2t\Pi_y \sum_{k_y} \cos{k_y} t^{\dagger}_{\alpha}(k_y)
t_{\alpha}(k_y)
\nonumber \\
&-&4 t N P_y \Pi_y
\end{eqnarray}

Now, quadratic decoupling of $H_{h^4}$ is performed. First,
$H_{h^4}$ is given by:
\begin{eqnarray}
H_{h^4} &=& \frac{\lambda J}{4} \sum_{\langle i,j, \rangle}
\sigma^{\alpha}_{ab}\sigma^{\alpha}_{cd} \{
h^{\dagger}_{1ia}h_{1ib}h^{\dagger}_{1jc}h_{1jd}
\nonumber \\
&&+h^{\dagger}_{2ia}h_{2ib}h^{\dagger}_{2jc}h_{2jd} \}
\end{eqnarray}
Suppressing the site index within a dimer, i.e. $1$ or $2$, let us
first consider the following term inside summation.

\begin{eqnarray}
&&\sigma^{\alpha}_{ab}\sigma^{\alpha}_{cd}
h^{\dagger}_{ia}h_{ib}h^{\dagger}_{jc}h_{jd}
\nonumber \\
&=& 2 \{ h^{\dagger}_{i\uparrow}h_{i\downarrow}
h^{\dagger}_{j\downarrow}h_{j\uparrow}
+h^{\dagger}_{i\downarrow}h_{i\uparrow}
h^{\dagger}_{j\uparrow}h_{j\downarrow} \}
\nonumber \\
&&+\{ h^{\dagger}_{i\uparrow}h_{i\uparrow}
-h^{\dagger}_{i\downarrow}h_{i\downarrow} \} \{
h^{\dagger}_{j\uparrow}h_{j\uparrow}
-h^{\dagger}_{j\downarrow}h_{j\downarrow} \}
\nonumber \\
&=& -\frac{3}{2} \Pi_y ( h^{\dagger}_{ja}h_{ia}
+h^{\dagger}_{ia}h_{ja} ) +\frac{3}{2} \Pi_y^2
\nonumber \\
&-&3 \Delta_y ( h^{\dagger}_{i\uparrow}h^{\dagger}_{j\downarrow}
+h_{j\downarrow}h_{i\uparrow} ) -3 \Delta_y (
h^{\dagger}_{j\uparrow}h^{\dagger}_{i\downarrow}
+h_{i\downarrow}h_{j\uparrow} )
\nonumber \\
&&+6 \Delta_y^2
\end{eqnarray}
where, once again, $\Pi_y$ is given by:

\begin{eqnarray}
\Pi_y &=& \frac{1}{2} \sum_{k_y} \cos{k_y} \bigg\{ \langle
h^{\dagger}_{+a}(k_y)h_{+a}(k_y) \rangle
\nonumber \\
&&+ \langle h^{\dagger}_{-a}(k_y)h_{-a}(k_y) \rangle \bigg\},
\end{eqnarray}
and also $\Delta_y$ is defined as follows:

\begin{eqnarray}
\Delta_y &=& \frac{1}{2} \sum_{k_y} \cos{k_y} \bigg\{ \langle
h^{\dagger}_{+\uparrow}(k_y)h^{\dagger}_{+\downarrow}(-k_y)
\rangle
\nonumber \\
&&+\langle
h^{\dagger}_{-\uparrow}(k_y)h^{\dagger}_{-\downarrow}(-k_y)
\rangle \bigg\}.
\end{eqnarray}

Therefore $H_{h^4}$ can be written as follows.
\begin{eqnarray}
H_{h^4} &=& -\frac{3}{4}\lambda J \sum_{k_y} \cos{k_y}\bigg\{
\Pi_y h^{\dagger}_{+a}(k_y)h_{+a}(k_y)
\nonumber \\
&&+2\Delta_y \left(
h^{\dagger}_{+\uparrow}(k_y)h^{\dagger}_{+\downarrow}(-k_y)
+h_{+\downarrow}(-k_y)h_{+\uparrow}(k_y) \right) \bigg\}
\nonumber \\
&-& \frac{3}{4}\lambda J \sum_{k_y} \cos{k_y}\bigg\{ \Pi_y
h^{\dagger}_{-a}(k_y)h_{-a}(k_y)
\nonumber \\
&&+2\Delta_y \left(
h^{\dagger}_{-\uparrow}(k_y)h^{\dagger}_{-\downarrow}(-k_y)
+h_{-\downarrow}(-k_y)h_{-\uparrow}(k_y) \right) \bigg\}
\nonumber \\
&+&\frac{3}{4}N \lambda J \left( \Pi_y^2 +4\Delta_y^2 \right)
\end{eqnarray}

Including all the quartic terms, the full, mean-field Hamiltonian
is finally given by:
\begin{eqnarray}
&H& =  N \epsilon_0 +\sum_{k_y} \bigg\{ A^{'}_{k_y}
t^{\dagger}_{\alpha}(k_y) t_{\alpha}(k_y)
\nonumber \\
&&+B^{'}_{k_y} \left(
t^{\dagger}_{\alpha}(k_y)t^{\dagger}_{\alpha}(-k_y)
+t_{\alpha}(k_y)t_{\alpha}(-k_y) \right) \bigg\}
\nonumber \\
&+& \sum_{k_y} \bigg\{
\epsilon^{'}_{+}(k_y)h^{\dagger}_{+a}(k_y)h_{+a}(k_y)
\nonumber \\
&&-D^{'}_{+}(k_y)
\left(h^{\dagger}_{+\uparrow}(k_y)h^{\dagger}_{+\downarrow}(-k_y)
+h_{+\downarrow}(-k_y)h_{+\uparrow}(k_y) \right) \bigg\}
\nonumber \\
&+& \sum_{k_y} \bigg\{
\epsilon^{'}_{-}(k_y)h^{\dagger}_{-a}(k_y)h_{-a}(k_y)
\nonumber \\
&&+D^{'}_{-} (k_y)
\left(h^{\dagger}_{-\uparrow}(k_y)h^{\dagger}_{-\downarrow}(-k_y)
+h_{-\downarrow}(-k_y)h_{-\uparrow}(k_y) \right) \bigg\}
\nonumber \\
\end{eqnarray}
where
\begin{eqnarray}
\epsilon_0 &=& -\frac{3}{4}J \bar{s}^2 -\mu (\bar{s}^2 -\bar{d}^2
+2x -1)
\nonumber \\
&&-\xi (2\bar{d}^2 -2x)
\nonumber \\
A^{'}_{k_y} &=& A_{k_y} +\lambda J P_y \cos{k_y} +2t\Pi_y
\cos{k_y}
\nonumber \\
&=& \frac{J}{4} -\mu +\lambda J (\bar{s}^2+ P_y)\cos{k_y}
+2t\Pi_y\cos{k}
\nonumber \\
B^{'}_{k_y} &=& B_{k_y} -\frac{\lambda J}{2}Q_y \cos{k_y} \nonumber \\
&=& \frac{\lambda J}{2} (\bar{s}^2-Q_y) \cos{k_y}
\nonumber \\
\epsilon^{'}_{+}(k_y) &=& \epsilon_{+}(k_y) -\frac{3}{4}\lambda J
\Pi_y \cos{k_y}
+tP_y\cos{k_y} \nonumber \\
&=& t(\bar{s}^2-2\bar{d}^2+P_y)\cos{k_y}-\frac{3}{4}\lambda J
\Pi_y \cos{k_y} -t-\xi
\nonumber \\
\epsilon^{'}_{-}(k_y) &=& \epsilon_{-}(k_y) -\frac{3}{4}\lambda J
\Pi_y \cos{k_y}
+tP_y\cos{k_y} \nonumber \\
&=& t(\bar{s}^2-2\bar{d}^2+P_y)\cos{k_y}-\frac{3}{4}\lambda J
\Pi_y \cos{k_y} +t-\xi
\nonumber \\
D^{'}_{+}(k_y) &=& D(k_y) +\frac{3}{2}\lambda J \Delta_y \cos{k_y} \nonumber \\
&=& 2\sqrt{2}t\bar{d}\bar{s}\cos{k_y}  +\frac{3}{2}\lambda J
\Delta_y \cos{k_y}
\nonumber \\
D^{'}_{-}(k_y) &=& D(k_y) -\frac{3}{2}\lambda J \Delta_y \cos{k_y} \nonumber \\
&=& 2\sqrt{2}t\bar{d}\bar{s}\cos{k_y}  -\frac{3}{2}\lambda J
\Delta_y \cos{k_y}
\nonumber \\
\end{eqnarray}
Note that all the constants which are not explicitly dependent on
$\bar{s}$, $\mu$, $\bar{d}$, and $\xi$ are ignored. As previously,
the above Hamiltonian can be diagonalized by using Bogoliubov
transformation. That is,

\begin{eqnarray}
H &=& N\epsilon_0 +\sum_{k_y} \omega^{'}_{k_y}
\gamma^{\dagger}_{\alpha}(k_y) \gamma_{\alpha}(k_y)
+\frac{3}{2}\sum_{k_y} (\omega^{'}_{k_y} - A^{'}_{k_y})
\nonumber \\
&+&\sum_{k_y}
\Omega^{'}_{+}(k_y)\beta^{\dagger}_{+a}(k_y)\beta_{+a}(k_y)
\nonumber \\
&&+\sum_{k_y} \left( \epsilon^{'}_{+}(k_y) - \Omega^{'}_{+}(k_y)
\right)
\nonumber \\
&+&\sum_{k_y}
\Omega^{'}_{-}(k_y)\beta^{\dagger}_{-a}(k_y)\beta_{-a}(k_y)
\nonumber \\
&&+\sum_{k_y} \left( \epsilon^{'}_{-}(k_y) - \Omega^{'}_{-}(k_y)
\right)
\nonumber \\
\end{eqnarray}

Saddle-point equations of the full Hamiltonian with quartic terms
are similar to those without quartic terms except the fact that
even the $t$ boson part is not expressed explicitly.

\begin{eqnarray}
\frac{\partial (\epsilon_{gr}/J)}{\partial \bar{s}^2} &=&
-\frac{3}{4} -\frac{\mu}{J} +\frac{3}{2}\lambda\int^{\pi}_{0}
\frac{dk_y}{\pi} \cos{k_y} \frac{A^{'}_{k_y}
-2B^{'}_{k_y}}{\omega^{'}_{k_y}}
\nonumber \\
&-&\frac{t/J}{\bar{s}}\int^{\pi}_0 \frac{dk_y}{\pi} \cos{k_y}
\bigg\{
\frac{\bar{s}\epsilon^{'}_{+}(k_y)+\sqrt{2}\bar{d}D^{'}_{+}(k_y)}{\Omega^{'}_{+}(k_y)}
\nonumber \\
&&+\frac{\bar{s}\epsilon^{'}_{-}(k_y)+\sqrt{2}\bar{d}D^{'}_{-}(k_y)}{\Omega^{'}_{-}(k_y)}
\bigg\} = 0
\\
\frac{\partial (\epsilon_{gr}/J)}{\partial (\mu/J)} &=& \bar{d}^2
-\bar{s}^2 +\frac{5}{2} -2x -\frac{3}{2} \int^{\pi}_0
\frac{dk_y}{\pi} \frac{A^{'}_{k_y}}{\omega^{'}_{k_y}} = 0
\\
\frac{\partial (\epsilon_{gr}/J)}{\partial \bar{d}^2} &=&
\frac{\mu}{J} -2\frac{t}{J}\frac{\xi}{t} \nonumber \\
&+&\frac{t/J}{\bar{d}}\int^{\pi}_0 \frac{dk_y}{\pi} \cos{k_y}
\bigg\{ \frac{2\bar{d}\epsilon^{'}_{+}(k_y)-\sqrt{2}\bar{s}
D^{'}_{+}(k_y)}{\Omega^{'}_{+}(k_y)}
\nonumber \\
&&+\frac{2\bar{d}\epsilon^{'}_{-}(k_y)-\sqrt{2}\bar{s}
D^{'}_{-}(k_y)}{\Omega^{'}_{-}(k_y)} \bigg\} = 0
\\
\frac{\partial (\epsilon_{gr}/J)}{\partial (\xi/J)} &=& -2
-2\bar{d}^2 +2x
\nonumber \\
&&+\int^{\pi}_0 \frac{dk_y}{\pi} \left\{
\frac{\epsilon^{'}_{+}(k_y)}{\Omega^{'}_{+}(k_y)}
+\frac{\epsilon^{'}_{-}(k_y)}{\Omega^{'}_{-}(k_y)} \right\} = 0
\end{eqnarray}

However, the above saddle-point equations are not meaningful
without the knowledge of $P_y$, $Q_y$, $\Pi_y$, and $\Delta_y$
used in $A^{'}_{k_y}$, $B^{'}_{k_y}$, $\epsilon^{'}_{\pm}(k_y)$,
and $D^{'}_{\pm}(k_y)$. These are the formulas for them.

\begin{eqnarray}
P_y &=& \frac{3}{2}\int^{\pi}_0 \frac{dk_y}{\pi} \cos{k_y}
\frac{A^{'}_{k_y}}{\omega^{'}_{k_y}} \\
Q_y &=& -3 \int^{\pi}_0 \frac{dk_y}{\pi} \cos{k_y}
\frac{B^{'}_{k_y}}{\omega^{'}_{k_y}} \\
\Pi_y &=& -\frac{1}{2} \int^{\pi}_0 \frac{dk_y}{\pi} \cos{k_y}
\left\{ \frac{\epsilon^{'}_{+}(k_y)}{\Omega^{'}_{+}(k_y)}
+\frac{\epsilon^{'}_{-}(k_y)}{\Omega^{'}_{-}(k_y)}
\right\} \\
\Delta_y &=& \frac{1}{2} \int^{\pi}_0 \frac{dk_y}{\pi} \cos{k_y}
\left\{ \frac{D^{'}_{+}(k_y)}{\Omega^{'}_{+}(k_y)}
-\frac{D^{'}_{-}(k_y)}{\Omega^{'}_{-}(k_y)} \right\}
\end{eqnarray}

Upon solving the saddle-point equations, it is convenient to view
$P_y$, $Q_y$, $\Pi_y$, and $\Delta_y$ as unknown parameters
similar to $\bar{s}$, $\bar{d}$, $\mu$, and $\xi$. Then, the
original saddle-point conditions are equivalent to solving four
saddle-point equations and four equations for $P_y$, $Q_y$,
$\Pi_y$, and $\Delta_y$ simultaneously.

\section{Computations for the square lattice}
\label{square}

This section is devoted to the two-dimensional array of coupled
t-J ladders. Hamiltonian for the coupling between two-leg t-J
ladders can be written as follows.

\begin{eqnarray}
H_{\perp} &=& -t^{'} \sum_i \sum_{\langle n,m \rangle} \{
c^{\dagger}_{2ina}c_{1ima} +c^{\dagger}_{1ima}c_{2ina} \}
\nonumber \\
&+&\lambda^{'} J \sum_i \sum_{\langle n,m \rangle}
S^{\alpha}_{1im}S^{\alpha}_{2in}
\end{eqnarray}
where the index $i$ indicates the $i$-th dimer within a given
ladder and the indices $n$ ($m \equiv n+1$) indicate the $n$-th
($m$-th) ladder. Derivation of the mean-field Hamiltonian for the
dynamics perpenticular to the ladder direction, is rather similar
to that of the parallel direction. But caution should be used in
keeping track of signs of various terms in Hamiltonian. Let us
start from the hopping term between ladders.

\begin{eqnarray}
&&c^{\dagger}_{2ina} c_{1ima}
\nonumber \\
&=& \left\{ h^{\dagger}_{2ina}d_{in} +\frac{1}{\sqrt{2}}
\varepsilon_{ab}s^{\dagger}_{in} h_{1inb}
+\frac{1}{\sqrt{2}}\varepsilon_{ac}\sigma^{\alpha}_{cb}
t^{\dagger}_{in \alpha}
h_{1inb} \right\} \nonumber \\
&\times& \left\{ d^{\dagger}_{im} h_{1ima}+\frac{1}{\sqrt{2}}
\varepsilon_{ac}h^{\dagger}_{2imc} s_{im}
-\frac{1}{\sqrt{2}}\varepsilon_{ae}\bar{\sigma}^{\beta}_{ed}
h^{\dagger}_{2imd}
t_{im \beta} \right\} \nonumber \\
&\Rightarrow& \bar{d}^2 h^{\dagger}_{2ina}h_{1ima}
-\frac{1}{2}\varepsilon_{ab}\varepsilon_{ac} \bar{s}^2
h^{\dagger}_{2imc}h_{1inb} \nonumber \\
&+&\frac{1}{\sqrt{2}} \bar{d}\bar{s} \varepsilon_{ac}
h^{\dagger}_{2ina} h^{\dagger}_{2imc} +\frac{1}{\sqrt{2}}
\bar{d}\bar{s} \varepsilon_{ab}h_{1inb}
h_{1ima} \nonumber \\
&-& \frac{1}{2}
\varepsilon_{ac}\varepsilon_{ae}\sigma^{\alpha}_{cb}
\bar{\sigma}^{\beta}_{ed}t^{\dagger}_{in \alpha}t_{im \beta}
h_{1inb}h^{\dagger}_{2imd}
\end{eqnarray}
Also, assuming that there is no magnetic ordering, the product of
spin operators is given by:

\begin{eqnarray}
&&S_{1im\alpha} S_{2in\alpha}
\nonumber \\
&=& \tilde{S}_{1im\alpha} \tilde{S}_{2in\alpha} +\frac{1}{4}
\sigma^{\alpha}_{ab}\sigma^{\alpha}_{cd}
h^{\dagger}_{1ima}h_{1imb}h^{\dagger}_{2inc}h_{2ind}
\nonumber \\
&=& -\frac{1}{4} ( s^{\dagger}_{im}
t_{im\alpha}+t^{\dagger}_{im\alpha}s_{im} -i
\epsilon_{\alpha\beta\gamma} t^{\dagger}_{im\beta}t_{im\gamma})
\nonumber \\
&&\times(s^{\dagger}_{in}
t_{in\alpha}+t^{\dagger}_{in\alpha}s_{in} +i
\epsilon_{\alpha\mu\nu} t^{\dagger}_{in\mu}t_{in\nu})
\nonumber \\
&+&\frac{1}{4} \sigma^{\alpha}_{ab}\sigma^{\alpha}_{cd}
h^{\dagger}_{1ima}h_{1imb}h^{\dagger}_{2inc}h_{2ind}
\nonumber \\
&\Rightarrow& -\frac{1}{4} \bar{s}^2 (
t^{\dagger}_{im\alpha}t_{in\alpha}+t^{\dagger}_{in\alpha}t_{im\alpha}
+t^{\dagger}_{im\alpha}t^{\dagger}_{in\alpha}
+t_{im\alpha}t_{in\alpha} )
\nonumber \\
&&+\frac{1}{4} (
t^{\dagger}_{im\alpha}t_{in\alpha}t^{\dagger}_{in\beta}t_{im\beta}
-
t^{\dagger}_{im\alpha}t^{\dagger}_{in\alpha}t_{im\beta}t_{in\beta}
)
\nonumber \\
&&+\frac{1}{4} \sigma^{\alpha}_{ab}\sigma^{\alpha}_{cd}
h^{\dagger}_{1ima}h_{1imb}h^{\dagger}_{2inc}h_{2ind}
\end{eqnarray}

Performing the quadratic decoupling and ignoring irrelevent
constants, the Hamiltonian $H_{\perp}$ is writtten as follows.
(Note that $k_x$ is the momentum associated with the perpendicular
direction to ladders, while $k_y$ is with the parallel direction.
Also, in laboratory unit, momentum in the parallel direction is
given by $p_y = k_y /a$, and that in the perpendicular direction
is given by $p_x= k_x /2a$ where $a$ is the lattice spacing and
$-\pi < k_x, \; k_y  \leq \pi$.)

\begin{eqnarray}
H_{\perp} = H_{\perp,h^2} +H_{\perp,t^2 h^2}+ H_{\perp,t^2}
+H_{\perp,t^4} +H_{\perp,h^4}
\end{eqnarray}

\begin{eqnarray}
H_{\perp,h^2} &=& \frac{t^{'}}{2} (\bar{s}^2-2\bar{d}^2) \sum_{\bf
k} \cos{k_x} \big\{ h^{\dagger}_{+a}({\bf k})h_{+a}({\bf k})
\nonumber \\
&&-h^{\dagger}_{-a}({\bf k})h_{-a}({\bf k}) \big\}
\nonumber \\
&-&\sqrt{2}t^{'}\bar{d}\bar{s} \sum_{\bf k} \cos{k_x} \big\{
h^{\dagger}_{+\uparrow}({\bf k})h^{\dagger}_{+\downarrow}(-{\bf
k})
\nonumber \\
&&+h_{+\downarrow}(-{\bf k})h_{+\uparrow}({\bf k}) \big\}
\nonumber \\
&-&\sqrt{2}t^{'}\bar{d}\bar{s} \sum_{\bf k} \cos{k_x} \big\{
h^{\dagger}_{-\uparrow}({\bf k})h^{\dagger}_{-\downarrow}(-{\bf
k})
\nonumber \\
&&+h_{-\downarrow}(-{\bf k})h_{-\uparrow}({\bf k}) \big\}
\\
H_{\perp,t^2 h^2} &=& -\frac{t^{'}}{2} P_x \sum_{\bf k}\cos{k_x}
\big\{
 h^{\dagger}_{+a}({\bf k})h_{+a}({\bf k})
\nonumber \\
&&- h^{\dagger}_{-a}({\bf k})h_{-a}({\bf k}) \big\}
\nonumber \\
&-&t^{'} \Pi_x \sum_{\bf k}\cos{k_x} t^{\dagger}_{\alpha}({\bf
k})t_{\alpha}({\bf k})
\\
H_{\perp,t^2} &=& -\frac{\lambda^{'} J}{4}\bar{s}^2 \sum_{\bf k}
\cos{k_x} \big\{ 2t^{\dagger}_{\alpha}({\bf k})t_{\alpha}({\bf k})
\nonumber \\
&&+t^{\dagger}_{\alpha}({\bf k})t^{\dagger}_{\alpha}(-{\bf k})
+t_{\alpha}({\bf k})t_{\alpha}(-{\bf k}) \big\}
\\
H_{\perp,t^4} &=& \frac{\lambda^{'} J}{4} \sum_{\bf k} \cos{k_x}
\big\{ 2 P_x t^{\dagger}_{\alpha}({\bf k})t_{\alpha}({\bf k})
\nonumber \\
&&-Q_x \left(t^{\dagger}_{\alpha}({\bf
k})t^{\dagger}_{\alpha}(-{\bf k}) +t_{\alpha}({\bf
k})t_{\alpha}(-{\bf k})\right)  \big\}
\\
H_{\perp,h^4} &=& -\frac{3}{8}\lambda^{'} J \Pi_x \sum_{\bf k}
\cos{k_x} \big\{ h^{\dagger}_{+a}({\bf k})h_{+a}({\bf k})
\nonumber \\
&&-h^{\dagger}_{-a}({\bf k})h_{-a}({\bf k}) \big\}
\nonumber \\
&-&\frac{3}{4}\lambda^{'} J \Delta_x \sum_{\bf k}\cos{k_x} \big\{
h^{\dagger}_{+\uparrow}({\bf k})h^{\dagger}_{+\downarrow}(-{\bf
k})
\nonumber \\
&&+h_{+\downarrow}(-{\bf k})h_{+\uparrow}({\bf k}) \big\}
\nonumber \\
&+& \frac{3}{4}\lambda^{'} J \Delta_x \sum_{\bf k}\cos{k_x} \big\{
h^{\dagger}_{-\uparrow}({\bf k})h^{\dagger}_{-\downarrow}(-{\bf
k})
\nonumber \\
&&+h_{-\downarrow}(-{\bf k})h_{-\uparrow}({\bf k}) \big\}
\end{eqnarray}
where
\begin{eqnarray}
P_x &=& \sum_{\bf k} \cos{k_x} \langle t^{\dagger}_{\alpha}({\bf
k}) t_{\alpha}({\bf k}) \rangle
\nonumber \\
Q_x &=& \sum_{\bf k} \cos{k_x}  \langle t^{\dagger}_{\alpha}({\bf
k}) t^{\dagger}_{\alpha}(-{\bf k}) \rangle .
\nonumber \\
\Pi_x &=& \frac{1}{2} \sum_{\bf k} \cos{k_x} \left\{ \langle
h^{\dagger}_{+a}({\bf k})h_{+a}({\bf k}) \rangle - \langle
h^{\dagger}_{-a}({\bf k})h_{-a}({\bf k}) \rangle \right\}
\nonumber \\
\Delta_x &=& \frac{1}{2} \sum_{\bf k} \cos{k_x} \bigg\{ \langle
h^{\dagger}_{+\uparrow}({\bf k})h^{\dagger}_{+\downarrow}(-{\bf
k}) \rangle
\nonumber \\
&&- \langle h^{\dagger}_{-\uparrow}({\bf
k})h^{\dagger}_{-\downarrow}(-{\bf k}) \rangle \bigg\}
\nonumber \\
\end{eqnarray}
where the bonding and anti-bonding fermions are assumed to be well
separated in energy so that the mixing terms between them are
ignored.

Combining $H_{\perp}$ with the Hamiltonian describing the dynamics
inside ladder, the mean-field Hamiltonian for the two-dimensional
array of coupled t-J ladders is obtained.

\begin{eqnarray}
&H& =  N^2 \epsilon_0 /2 +\sum_{\bf k} \big\{ \tilde{A}_{\bf k}
t^{\dagger}_{\alpha}({\bf k}) t_{\alpha}({\bf k})
\nonumber \\
&&+\tilde{B}_{{\bf k}} \left( t^{\dagger}_{\alpha}({\bf
k})t^{\dagger}_{\alpha}(-{\bf k}) +t_{\alpha}({\bf
k})t_{\alpha}(-{\bf k}) \right) \big\}
\nonumber \\
&+& \sum_{{\bf k}} \big\{ \tilde{\epsilon}_{+}({\bf
k})h^{\dagger}_{+a}({\bf k}) h_{+a}({\bf k})
\nonumber \\
&&-\tilde{D}_{+}({\bf k}) (h^{\dagger}_{+\uparrow}({\bf k})
h^{\dagger}_{+\downarrow}(-{\bf k}) +h_{+\downarrow}(-{\bf
k})h_{+\uparrow}({\bf k}) ) \big\}
\nonumber \\
&+& \sum_{{\bf k}} \big\{ \tilde{\epsilon}_{-}({\bf k})
h^{\dagger}_{-a}({\bf k})h_{-a}({\bf k})
\nonumber \\
&&+\tilde{D}_{-} ({\bf k}) (h^{\dagger}_{-\uparrow}({\bf k})
h^{\dagger}_{-\downarrow}(-{\bf k}) +h_{-\downarrow}(-{\bf
k})h_{-\uparrow}({\bf k}) ) \big\}
\end{eqnarray}
where
\begin{eqnarray}
\epsilon_0 &=& -\frac{3}{4}J \bar{s}^2 -\mu (\bar{s}^2 -\bar{d}^2
+2x -1)
\nonumber \\
&&-\xi (2\bar{d}^2 -2x)
\\
\tilde{A}_{\bf k} &=& \frac{J}{4} -\mu +\lambda J \bar{s}^2 \left(
\cos{k_y}-\frac{\lambda^{'}}{2\lambda}\cos{k_x} \right)
\nonumber \\
&+& \lambda J \left( P_y\cos{k_y}
+\frac{\lambda^{'}}{2\lambda}P_x\cos{k_x} \right)
\nonumber \\
&+&2t \left( \Pi_y\cos{k_y} -\frac{t^{'}}{2t} \Pi_x\cos{k_x}
\right)
\\
\tilde{B}_{\bf k} &=& \frac{\lambda J}{2} \bar{s}^2 \left(
\cos{k_y} -\frac{\lambda^{'}}{2\lambda}\cos{k_x} \right)
\nonumber \\
&-&\frac{\lambda J}{2} \left( Q_y\cos{k_y}
+\frac{\lambda^{'}}{2\lambda} Q_x\cos{k_x} \right)
\\
\tilde{\epsilon}_{+}({\bf k}) &=& t(\bar{s}^2-2\bar{d}^2) \left(
\cos{k_y} +\frac{t^{'}}{2t}\cos{k_x} \right) -t-\xi
\nonumber \\
&-&\frac{3}{4}\lambda J \left(
\Pi_y\cos{k_y}+\frac{\lambda^{'}}{2\lambda} \Pi_x\cos{k_x} \right)
\nonumber \\
&+& t \left( P_y\cos{k_y} - \frac{t^{'}}{2t}P_x \cos{k_x} \right)
\\
\tilde{\epsilon}_{-}({\bf k}) &=& t(\bar{s}^2-2\bar{d}^2)\left(
\cos{k_y} -\frac{t^{'}}{2t}\cos{k_x} \right) +t-\xi
\nonumber \\
&-&\frac{3}{4}\lambda J \left(
\Pi_y\cos{k_y}-\frac{\lambda^{'}}{2\lambda} \Pi_x\cos{k_x} \right)
\nonumber \\
&+& t \left( P_y\cos{k_y} + \frac{t^{'}}{2t}P_x\cos{k_x} \right)
\\
\tilde{D}_{+}({\bf k}) &=& 2\sqrt{2}t\bar{d}\bar{s}\left(
\cos{k_y} +\frac{t^{'}}{2t}\cos{k_x} \right)
\nonumber \\
&+&\frac{3}{2}\lambda J \left(
\Delta_y\cos{k_y}+\frac{\lambda^{'}}{2\lambda} \Delta_x\cos{k_x}
\right)
\\
\tilde{D}_{-}({\bf k}) &=& 2\sqrt{2}t\bar{d}\bar{s}\left(
\cos{k_y} -\frac{t^{'}}{2t}\cos{k_x} \right)
\nonumber \\
&-&\frac{3}{2}\lambda J \left(
\Delta_y\cos{k_y}-\frac{\lambda^{'}}{2\lambda} \Delta_x\cos{k_x}
\right)
\end{eqnarray}

Diagonalization via Bogoliubov transformation and the
correspondent saddle-point equations are similar to those of
single ladder case. That is to say,

\begin{eqnarray}
\epsilon_{gr} &=& \epsilon^{'} +\frac{3}{2}\int\int \frac{d^2 {\bf
k}}{\pi^2} \tilde{\omega}({\bf k})
\nonumber \\
&-&\int\int\frac{d^2 {\bf k}}{\pi^2} \tilde{\Omega}_{+}({\bf k})
-\int\int\frac{d^2 {\bf k}}{\pi^2} \tilde{\Omega}_{-}({\bf k})
\end{eqnarray}
where

\begin{eqnarray}
\epsilon^{'}_0 &=& -\frac{3}{4}J \bar{s}^2 -\mu (\bar{s}^2
-\bar{d}^2 +2x -\frac{5}{2})
\nonumber \\
&-&\xi (2 +2\bar{d}^2 -2x) \\
\tilde{\omega}({\bf k}) &=& \sqrt{\tilde{A}^2_{\bf k}-4\tilde{B}^2_{\bf k}} \\
\tilde{\Omega}_{\pm}({\bf k}) &=&
\sqrt{\tilde{\epsilon}^2_{\pm}({\bf k}) +\tilde{D}^2_{\pm}({\bf
k})} .
\end{eqnarray}
where the upper and lower limit of integral, $\pi$ and $0$, are
not explicitly written. Saddle-point equations are given as
follows.

\begin{eqnarray}
\frac{\partial (\epsilon_{gr}/J)}{\partial \bar{s}^2} &=&
-\frac{3}{4} -\frac{\mu}{J}
\nonumber \\
&+&\frac{3}{2}\lambda\int\int \frac{d^2 {\bf k}}{\pi^2} \left(
\cos{k_y} -\frac{\lambda^{'}}{2\lambda}\cos{k_x} \right)
\nonumber \\
&&\times \frac{\tilde{A}_{\bf k} -2\tilde{B}_{\bf
k}}{\tilde{\omega}_{\bf k}}
\nonumber \\
&-&\frac{t/J}{\bar{s}}\int\int\frac{d^2 {\bf k}}{\pi^2}
\left(\cos{k_y}+\frac{t^{'}}{2t}\cos{k_x}\right)
\nonumber \\
&&\times \frac{\bar{s}\tilde{\epsilon}_{+}({\bf k})
+\sqrt{2}\bar{d}\tilde{D}_{+}({\bf k})}{\tilde{\Omega}_{+}({\bf
k})}
\nonumber \\
&-&\frac{t/J}{\bar{s}}\int\int \frac{d^2 {\bf k}}{\pi^2}
\left(\cos{k_y}-\frac{t^{'}}{2t}\cos{k_x}\right)
\nonumber \\
&&\times \frac{\bar{s}\tilde{\epsilon}_{-}({\bf k})
+\sqrt{2}\bar{d}\tilde{D}_{-}({\bf k})}{\tilde{\Omega}_{-}({\bf
k})} = 0
\\
\frac{\partial (\epsilon_{gr}/J)}{\partial (\mu/J)} &=& \bar{d}^2
-\bar{s}^2 +\frac{5}{2} -2x
\nonumber \\
&-&\frac{3}{2} \int\int \frac{d^2 {\bf k}}{\pi^2}
\frac{\tilde{A}_{\bf k}}{\tilde{\omega}_{\bf k}} = 0
\\
\frac{\partial (\epsilon_{gr}/J)}{\partial \bar{d}^2} &=&
\frac{\mu}{J} -2\frac{t}{J}\frac{\xi}{t}
\nonumber \\
&+&\frac{t/J}{\bar{d}}\int\int\frac{d^2 {\bf k}}{\pi^2}
\left(\cos{k_y}+\frac{t^{'}}{2t}\cos{k_x}\right)
\nonumber \\
&& \times \frac{2\bar{d}\tilde{\epsilon}_{+}({\bf k})
-\sqrt{2}\bar{s}\tilde{D}_{+}({\bf k})}{\tilde{\Omega}_{+}({\bf
k})}
\nonumber \\
&+&\frac{t/J}{\bar{d}}\int\int \frac{d^2 {\bf k}}{\pi^2}
\left(\cos{k_y}-\frac{t^{'}}{2t}\cos{k_x}\right)
\nonumber \\
&&\times\frac{2\bar{d}\tilde{\epsilon}_{-}({\bf k})
-\sqrt{2}\bar{s}\tilde{D}_{-}({\bf k})}{\tilde{\Omega}_{-}({\bf
k})} = 0
\\
\frac{\partial (\epsilon_{gr}/J)}{\partial (\xi/J)} &=& -2
-2\bar{d}^2 +2x
\nonumber \\
&+&\int\int \frac{d^2 {\bf k}}{\pi^2} \left\{
\frac{\tilde{\epsilon}_{+}({\bf k})}{\tilde{\Omega}_{+}({\bf k})}
+\frac{\tilde{\epsilon}_{-}({\bf k})}{\tilde{\Omega}_{-}({\bf k})}
\right\} = 0
\end{eqnarray}
Also, the parameters due to quartic terms are defined as follows.
\begin{eqnarray}
P_y &=& \frac{3}{2}\int\int\frac{d^2 {\bf k}}{\pi^2} \cos{k_y}
\frac{\tilde{A}_{\bf k}}{\tilde{\omega}_{\bf k}} \\
Q_y &=& -3 \int\int \frac{d^2 {\bf k}}{\pi^2} \cos{k_y}
\frac{\tilde{B}_{\bf k}}{\tilde{\omega}_{\bf k}} \\
\Pi_y &=& -\frac{1}{2} \int\int \frac{d^2 {\bf k}}{\pi^2}
\cos{k_y} \left\{ \frac{\tilde{\epsilon}_{+}({\bf
k})}{\tilde{\Omega}_{+}({\bf k})} +\frac{\tilde{\epsilon}_{-}({\bf
k})}{\tilde{\Omega}_{-}({\bf k})}
\right\} \\
\Delta_y &=& \frac{1}{2} \int\int \frac{d^2 {\bf k}}{\pi^2}
\cos{k_y} \left\{ \frac{\tilde{D}_{+}({\bf
k})}{\tilde{\Omega}_{+}({\bf k})} -\frac{\tilde{D}_{-}({\bf
k})}{\tilde{\Omega}_{-}({\bf k})} \right\}
\\
P_x &=& \frac{3}{2}\int\int\frac{d^2 {\bf k}}{\pi^2} \cos{k_x}
\frac{\tilde{A}_{\bf k}}{\tilde{\omega}_{\bf k}} \\
Q_x &=& -3 \int\int \frac{d^2 {\bf k}}{\pi^2} \cos{k_x}
\frac{\tilde{B}_{\bf k}}{\tilde{\omega}_{\bf k}} \\
\Pi_x &=& -\frac{1}{2} \int\int \frac{d^2 {\bf k}}{\pi^2}
\cos{k_x} \left\{ \frac{\tilde{\epsilon}_{+}({\bf
k})}{\tilde{\Omega}_{+}({\bf k})} -\frac{\tilde{\epsilon}_{-}({\bf
k})}{\tilde{\Omega}_{-}({\bf k})}
\right\} \\
\Delta_x &=& \frac{1}{2} \int\int \frac{d^2 {\bf k}}{\pi^2}
\cos{k_x} \left\{ \frac{\tilde{D}_{+}({\bf
k})}{\tilde{\Omega}_{+}({\bf k})} +\frac{\tilde{D}_{-}({\bf
k})}{\tilde{\Omega}_{-}({\bf k})} \right\}
\end{eqnarray}

As previously, it is convenient to have some limiting case when
the explicit solution is available. We will take the case of
$x=0$, i.e. no-doping case.

\begin{eqnarray}
\frac{\partial (\epsilon_{gr}/J)}{\partial \bar{s}^2} &=&
-\frac{3}{4} -\frac{\mu}{J} +\frac{3}{2}\lambda \int\int\frac{d^2
{\bf k}}{\pi^2} \frac{f_{\bf k}}{\sqrt{1+\eta f_{\bf k}}} =0
\\
\frac{\partial (\epsilon_{gr}/J)}{\partial (\mu/J)} &=&
\frac{5}{2} -\bar{s}^2 -\frac{3}{2}\int\int\frac{d^2 {\bf
k}}{\pi^2} \frac{1+\frac{\eta}{2} f_{\bf k}}{\sqrt{1+\eta f_{\bf
k}}} =0
\end{eqnarray}
where $\eta= \frac{2\lambda\bar{s}^2}{1/4-\mu/J}$ and $f_{\bf k} =
\cos{k_y} - \frac{\lambda^{'}}{2\lambda}\cos{k_x}$. As before, it
is possible to combine the above two equations to get a single
equation only for $\eta$. That is to say,

\begin{equation}
\frac{\eta}{2\lambda} = \frac{5}{2} -\frac{3}{2} \int\int
\frac{d^2 {\bf k}}{\pi^2} \frac{1}{\sqrt{1+\eta f_{\bf k}}}
\end{equation}

\subsection{Superconducting Order Parameters}
\label{sec:scop}

This subsection relates the parameters in the above computations
to the physical order parameters describing the nature of the
superconducting state. In the case of two-dimensional array of
coupled ladders, there are three order parameters to be defined:
\begin{eqnarray}
\chi_0 &\equiv& \varepsilon_{ab} \langle
c^{\dagger}_{1ina}c^{\dagger}_{2inb} \rangle
\\
\chi_{\parallel} &\equiv& \varepsilon_{ab}\langle
c^{\dagger}_{1,i+1,na}c^{\dagger}_{1inb} \rangle
\\
\chi_{\perp} &\equiv& \varepsilon_{ab} \langle
c^{\dagger}_{1i,n+1,a}c^{\dagger}_{2inb} \rangle
\end{eqnarray}

The first one, $\chi_0$, is the pairing order parameter within
dimer. Using constraints, it is written as follows.
\begin{eqnarray}
\chi_0 &=& \varepsilon_{ab} \langle \{ h^{\dagger}_{1ina}d_{in}
+\frac{1}{\sqrt{2}} \varepsilon_{ac}s^{\dagger}_{in} h_{2inc}
-\frac{1}{\sqrt{2}}\varepsilon_{ac}\sigma^{\alpha}_{cd}
t^{\dagger}_{in \alpha}
h_{2ind} \} \nonumber \\
&\times& \{h^{\dagger}_{2inb}d_{in} +\frac{1}{\sqrt{2}}
\varepsilon_{bc} s^{\dagger}_{in} h_{1inc}
+\frac{1}{\sqrt{2}}\varepsilon_{bc}\bar{\sigma}^{\beta}_{cd}
t^{\dagger}_{in\beta} h_{1ind} \} \rangle
\nonumber \\
&=& \varepsilon_{ab} \cdot \frac{1}{\sqrt{2}}
\langle\varepsilon_{ac} s^{\dagger}_{in} d_{in} h_{2inc}
h^{\dagger}_{2inb} \rangle  = \sqrt{2} \bar{s}\bar{d}
\end{eqnarray}

The pairing order parameter in the parallel direction,
$\chi_{\parallel}$, is evaluated as follows.
\begin{eqnarray}
\chi_{\parallel} &=& \varepsilon_{ab} \{ \bar{d}^2 \langle
h^{\dagger}_{1,i+1,na}h^{\dagger}_{1inb} \rangle +\frac{1}{2}
\varepsilon_{ac}\varepsilon_{bd}\bar{s}^2 \langle
h_{2,i+1,na}h_{2ind} \rangle
\nonumber \\
&+&\frac{1}{\sqrt{2}}\varepsilon_{bc}\bar{d}\bar{s} \langle
h^{\dagger}_{1,i+1,na}h_{2inc}\rangle
+\frac{1}{\sqrt{2}}\varepsilon_{ac}\bar{d}\bar{s} \langle
h_{2,i+1,nc}h^{\dagger}_{1inb}\rangle
\nonumber \\
&+&\frac{1}{2}\varepsilon_{ac}\varepsilon_{be}\sigma^{\alpha}_{cd}
\sigma^{\beta}_{ef} \langle t^{\dagger}_{i+1,n\alpha}
t^{\dagger}_{in\beta} h_{2,i+1,nd}h_{2inf} \rangle \}
\nonumber \\
&=& (\bar{d}^2 -\frac{1}{2}\bar{s}^2 +Q)\Gamma
-\sqrt{2}\bar{d}\bar{s}\tilde{\Pi}
\end{eqnarray}
where $Q$ and $\Gamma$ are defined previously, and $\tilde{\Pi}$
is given by:

\begin{equation}
\tilde{\Pi} = -\frac{1}{2} \int\int \frac{dkdq}{\pi^2} \cos{k}
\left\{ \frac{\tilde{\epsilon}_{+}(k,q)}{\tilde{\Omega}_{+}(k,q)}
-\frac{\tilde{\epsilon}_{-}(k,q)}{\tilde{\Omega}_{-}(k,q)}
\right\}.
\end{equation}

Similarly, pairing order parameter in the perpendicular direction,
$\chi_{\perp}$, is computed to give the following formula.
\begin{eqnarray}
\chi_{\perp} &=& \varepsilon_{ab} \{ \bar{d}^2 \langle
h^{\dagger}_{2ina}h^{\dagger}_{1i,n+1,b} \rangle +\frac{1}{2}
\varepsilon_{ac}\varepsilon_{bd}\bar{s}^2 \langle
h_{1inc}h_{2i,n+1,d} \rangle
\nonumber \\
&+&\frac{1}{\sqrt{2}}\varepsilon_{bc}\bar{d}\bar{s} \langle
h^{\dagger}_{2ina}h_{2i,n+1,c}\rangle
+\frac{1}{\sqrt{2}}\varepsilon_{ac}\bar{d}\bar{s} \langle
h_{1inc}h^{\dagger}_{1i,n+1,b}\rangle
\nonumber \\
&-&\frac{1}{2}\varepsilon_{ac}\varepsilon_{be}\sigma^{\alpha}_{cd}
\sigma^{\beta}_{ef} \langle t^{\dagger}_{in\alpha}
t^{\dagger}_{i,n+1,\beta} h_{1ind}h_{2i,n+1,f} \rangle \}
\nonumber \\
&=& (\bar{d}^2 -\frac{1}{2}\bar{s}^2 -Q_{\perp})\Gamma_{\perp}
-\sqrt{2}\bar{d}\bar{s}\tilde{\Pi}_{\perp}
\end{eqnarray}
where $Q_{\perp}$ and $\Gamma_{\perp}$ are defined previously, and
$\tilde{\Pi}_{\perp}$ is given by:

\begin{equation}
\tilde{\Pi}_{\perp} = -\frac{1}{2} \int\int \frac{dkdq}{\pi^2}
\cos{q} \left\{
\frac{\tilde{\epsilon}_{+}(k,q)}{\tilde{\Omega}_{+}(k,q)}
+\frac{\tilde{\epsilon}_{-}(k,q)}{\tilde{\Omega}_{-}(k,q)}
\right\}.
\end{equation}

\section{Effect of Coulomb Interaction}
\label{coulomb}

This section is devoted to the effect of Coulomb repulsion.
Hamiltonian due to the Coulomb interaction between the nearest
neighbors is written in terms of the bond operators as follows.

\begin{eqnarray}
H_{Coulomb} &=& V \sum_i\sum_n n_{1in}n_{2in}
\nonumber \\
&+&V \sum_{\langle i,j \rangle} \sum_{n} \{ n_{1in} n_{1jn} +
n_{2in} n_{2jn} \}
\nonumber \\
&+& V\sum_{i} \sum_{\langle n,m \rangle} n_{2in} n_{1im}
\end{eqnarray}
where

\begin{eqnarray}
n_{1in} &\equiv& c^{\dagger}_{1ina} c_{1ina}
\nonumber \\
&=& h^{\dagger}_{1ina}h_{1ina} +s^{\dagger}_{in}s_{in}
+t^{\dagger}_{in\alpha}t_{in\alpha}
\nonumber \\
&=& 1-h^{\dagger}_{2ina}h_{2ina} -d^{\dagger}_{in}d_{in}
\end{eqnarray}
and
\begin{eqnarray}
n_{2in} &\equiv& c^{\dagger}_{2ina} c_{2ina}
\nonumber \\
&=& h^{\dagger}_{2ina}h_{2ina} +s^{\dagger}_{in}s_{in}
+t^{\dagger}_{in\alpha}t_{in\alpha}
\nonumber \\
&=& 1-h^{\dagger}_{1ina}h_{1ina} -d^{\dagger}_{in}d_{in}.
\end{eqnarray}

The first term is evaluated as follows.
\begin{eqnarray}
&&\sum_{i}\sum_{n} n_{1in}n_{2in}
\nonumber \\
&=&\sum_{i}\sum_{n} (1-h^{\dagger}_{2ina}h_{2ina}
-d^{\dagger}_{in}d_{in})
\nonumber \\
&&\times(1-h^{\dagger}_{1inb}h_{1inb} -d^{\dagger}_{in}d_{in})
\nonumber \\
&=& \sum_{i}\sum_{n} \{1- d^{\dagger}_{in}d_{in}
-(h^{\dagger}_{1ina}h_{1ina}+ h^{\dagger}_{2ina}h_{2ina} ) \}
\nonumber \\
&=& \sum_{i}\sum_{n} \{1 -2x +d^{\dagger}_{in}d_{in} \}
\nonumber \\
&\Rightarrow& \frac{N^2}{2} (1 -2x +\bar{d}^2 )
\end{eqnarray}
where $N^2 /2$ is the total number of dimers. Note that the
constraint conditions are used.

The second term is due to the Coulomb interaction between the
dimers along the ladder direction, which is computed by:
\begin{eqnarray}
&&\sum_{\langle i,j \rangle} \sum_{n}
\{ n_{1in} n_{1jn} + n_{2in} n_{2jn} \} \nonumber \\
&\Rightarrow& \sum_{\langle i,j \rangle} \sum_{n} \big\{
(1-\bar{d}^2-h^{\dagger}_{2ina}h_{2ina})
(1-\bar{d}^2-h^{\dagger}_{2jnb}h_{2jnb})
\nonumber \\
&+& (1-\bar{d}^2-h^{\dagger}_{1ina}h_{1ina})
(1-\bar{d}^2-h^{\dagger}_{1jnb}h_{1jnb}) \big\}
\nonumber \\
&=& N^2 (1-\bar{d}^2)(1-2x+\bar{d}^2)
\nonumber \\
&+& \sum_{\langle i,j \rangle} \sum_{n} \{
h^{\dagger}_{1ina}h_{1ina}h^{\dagger}_{1jnb}h_{1jnb} +
h^{\dagger}_{2ina}h_{2ina}h^{\dagger}_{2jnb}h_{2jnb} \}.
\nonumber \\
\end{eqnarray}

Similarly, the third term is evaluated as follows.
\begin{eqnarray}
&&\sum_{i} \sum_{\langle n,m \rangle} n_{2in} n_{1im}
\nonumber \\
&=& \frac{N^2}{2} (1-\bar{d}^2)(1-2x+\bar{d}^2)
\nonumber \\
&+& \sum_{i} \sum_{\langle n,m \rangle}
h^{\dagger}_{1ina}h_{1ina}h^{\dagger}_{2imb}h_{2imb}
\end{eqnarray}

Therefore,
\begin{eqnarray}
&H&_{Coulomb} = \frac{N^2}{2} V (4-3\bar{d}^2)(1-2x+\bar{d}^2)
\nonumber \\
&+&V\sum_{\langle i,j \rangle} \sum_{n} \{
h^{\dagger}_{1ina}h_{1ina}h^{\dagger}_{1jnb}h_{1jnb} +
h^{\dagger}_{2ina}h_{2ina}h^{\dagger}_{2jnb}h_{2jnb} \}
\nonumber \\
&+& V\sum_{i} \sum_{\langle n,m \rangle}
h^{\dagger}_{1ina}h_{1ina}h^{\dagger}_{2imb}h_{2imb}
\nonumber \\
&=&  \frac{N^2}{2} V (4-3\bar{d}^2)(1-2x+\bar{d}^2) + H_{Coul,\;
h^4}
\end{eqnarray}

The full treatment of $H_{Coul,\; h^4}$ is beyond the scope of our
present paper. However, we will follow the philosophy of BCS
theory in that electrons interacting with the Coulomb repulsion
under the neutralizing background are described in terms of the
Landau-Fermi liquid. Assuming that the hopping parameter, $t$, and
the spin-coupling parameter, $J$, are already renoramilzed
quatities due to the Coulomb repulsion, one has only to consider
the Hartree contribution of $H_{Coul,\; h^4}$, which is given by:
\begin{eqnarray}
H^{Hartree}_{Coul,\; h^4} = \frac{N^2}{2} 3 V (x-\bar{d}^2)^2
\end{eqnarray}
which is obtained by using the constraint conditions. Also, note
that the mixing between  the bonding and anti-bonding fermions is
ignored.

Therefore, in our approximation, the Hamiltonian (per site) due to
the Coulomb repulsion is given as follows:

\begin{eqnarray}
\frac{H_{Coulomb}}{N^2 /2} &\simeq& V
(4-3\bar{d}^2)(1-2x+\bar{d}^2) + H^{Hartree}_{Coul,\; h^4}
\nonumber \\
&=& V(4-8x+3x^2) +V\bar{d}^2 \label{hcoul}
\end{eqnarray}

Since Eq.(\ref{hcoul}) depends only on $\bar{d}^2$, the
modification of saddle-point equations is straightforward. That
is,

\begin{eqnarray}
\frac{\partial (\epsilon_{gr}/J)}{\partial \bar{d}^2} = 0
\rightarrow \frac{\partial (\epsilon_{gr}/J)}{\partial \bar{d}^2}
+ V = 0
\end{eqnarray}
All the other saddle-point equations are the same as before.

Now it is reasonable to assume the limit of large Coulomb
repulsion which not only captures the essential physics that we
are interested in, but also will simplify explicit computations.
So we will take the limit $V \rightarrow \infty$; consequently
 $\bar{d}=0$. While this limit  is not physical
for large doping, our primary interest lies in the case of small
doping.



\begin{thebibliography}{}

\bibitem{science} See {\em e.g.} S. Sachdev, Science, {\bf 288}, 475 (2000)
and references therein.

\bibitem{bobroff} J.~Bobroff, H.~Alloul, W.~A.~MacFarlane, P.~Mendels,
N.~Blanchard, G.~Collin, and J.-F.~Marucco, Phys. Rev. Lett. {\bf
86}, 4116 (2001).

\bibitem{alloulold} H.~Alloul {\em et al.}, Phys.
Rev. Lett. {\bf 67}, 3140 (1991); A.~V.~Mahajan {\em et al.}, \prl
{\bf 72}, 3100 (1994); J.~Bobroff {\em et al.}, \prl {\bf 83},
4381 (1999); P.~Mendels {\em et al.}, Europhys. Lett. {\bf 46},
678 (1999); A.~V.~Mahajan {\em et al.} Eur. Phys. J. B {\bf 13},
457 (2000); M.-H.~Julien {\em et al.}, Phys. Rev. Lett. {\bf 84},
3422 (2000).

\bibitem{rs1} N.~Read and S.~Sachdev, Phys. Rev. Lett. {\bf 62}, 1694
(1989); Phys. Rev. B {\bf 42}, 4568 (1990).

\bibitem{sr} S.~Sachdev and N.~Read, Int. J. Mod. Phys. B {\bf 5}, 219
(1991).

\bibitem{rodolfo} S. Sachdev and R. Jalabert,
Mod. Phys. Lett. B {\bf 4}, 1043 (1990).

\bibitem{vojtaprl} M.~Vojta and S.~Sachdev, Phys. Rev. Lett. {\bf 83}, 3916
(1999); M.~Vojta, Y.~Zhang, and S.~Sachdev, Phys. Rev. B {\bf 62},
6721 (2000).

\bibitem{sbv} S.~Sachdev, C.~Buragohain, and M.~Vojta, Science
{\bf 286}, 2479 (1999); M. Vojta, C.~Buragohain, and S.~Sachdev,
Phys. Rev. B {\bf 61}, 15152 (2000).

\bibitem{lt22} S.~Sachdev and M.~Vojta, Physica B {\bf 280}, 333 (2000).

\bibitem{icmp} S.~Sachdev and M.~Vojta, Proceedings of the XIII
International Congress on Mathematical Physics, July 2000, London,
cond-mat/0009202.

\bibitem{fradkiv} E.~Fradkin and S.~Kivelson, Mod. Phys. Lett. B {\bf 4}, 225 (1990).

\bibitem{troyer} H.~Tsunetsugu, M.~Troyer, and T.~M.~Rice, Phys.
Rev. B {\bf 49}, 16078 (1994); M.~Troyer, H.~Tsunetsugu, and
T.~M.~Rice, Phys. Rev. B {\bf 53}, 251 (1996).

\bibitem{poil} D.~Poilblanc, O.~Chiappa, J.~Riera, S.~R.~White, and
D.~J.~Scalapino, cond-mat/0005403.

\bibitem{rossat1} J.~Rossat-Mignod, L.~P.~Regnault, C.~Vettier,
P.~Bourges, P.~Burlet, J.~Bossy, J.~Y.~Henry, and G.~Lapertot,
Physica C {\bf 185-189}, 86 (1991).

\bibitem{mook1} H.~A.~Mook, M.~Yethiraj, G.~Aeppli, T.~E.~Mason, and
T.~Armstrong, Phys. Rev. Lett. {\bf 70}, 3490 (1993).

\bibitem{tony3} H.~F.~Fong, B.~Keimer, D.~Reznik, D.~L.~Milius,
and I.~A.~Aksay, Phys. Rev. B {\bf 54}, 6708 (1996); H.~F.~Fong,
B.~Keimer, D.~L.~Milius, and I.~A.~Aksay, Phys. Rev. Lett. {\bf
78}, 713 (1997).

\bibitem{bourges} P.~Bourges in {\it The Gap Symmetry and
Fluctuations in High Temperature Superconductors} ed. J.~Bok,
G.~Deutscher, D.~Pavuna, and S.~A.~Wolf (Plenum, New York, 1998);
cond-mat/9901333.

\bibitem{he} H.~He, Y.~Sidis, P.~Bourges, G.~D.~Gu, A.~Ivanov,
N.~Koshizuka, B.~Liang, C.~T.~Lin, L.~P.~Regnault, E.~Schoenherr,
and B.~Keimer, Phys. Rev. Lett. {\bf 86}, 1610 (2001).

\bibitem{period4} J.~M.~Tranquada, J.~D.~Axe, N.~Ichikawa,
A.~R.~Moodenbaugh, Y.~Nakamura, and S.~Uchida, Phys. Rev. Lett.
{\bf 78}, 338 (1997)

\bibitem{ek} V.~J.~Emery, S.~A.~Kivelson, J.~M.~Tranquada, {\it Proc. Natl.
Acad. Sci} {\bf 96}, 8814 (1999) and references therein.

\bibitem{egami} R.~J.~McQueeney, Y.~Petrov, T.~Egami, M.~Yethiraj,
G.~Shirane, and Y.~Endoh,  Phys. Rev. Lett. {\bf 82}, 628 (1999);
T.~Egami, cond-mat/0102449; R.~J.~McQueeney, J.~L.~Sarrao,
P.~G.~Pagliuso, P.~W.~Stephens, and R.~Osborn, cond-mat/0104118;
R.~J.~McQueeney, T.~Egami, J.-H.~Chung, Y.~Petrov, M.~Yethiraj,
M.~Arai, Y.~Inamura, Y.~Endoh, C.~Frost, and F.~Dogan,
cond-mat/0105593.

\bibitem{kondo}
D.~Withoff and E.~Fradkin, Phys. Rev. Lett. {\bf 64}, 1835 (1990);
L.~S.~Borkowski and P.~J.~Hirschfeld, Phys. Rev. B {\bf 46}, 9274
(1992); K.~Chen and C.~Jayaprakash, J. Phys.: Condens. Matter {\bf
7}, L491 (1995); K.~Ingersent, Phys. Rev. B {\bf 54}, 11936
(1996); C.~R.~Cassanello and E.~Fradkin, Phys. Rev. B {\bf 53},
15079 (1996) and {\bf 56}, 11246 (1997); R.~Bulla, Th.~Pruschke,
and A.~C.~Hewson, J. Phys.: Condens. Matter {\bf 9}, 10463 (1997);
K.~Ingersent and Q.~Si, cond-mat/9810226; C.~Gonzalez-Buxton and
K.~Ingersent, \prb {\bf 57}, 14254 (1998).

\bibitem{tolya} A.~Polkovnikov, S.~Sachdev, and M.~Vojta,
Phys. Rev. Lett. {\bf 86}, 296 (2001).

\bibitem{zhu} J-X.~Zhu and C.~S.~Ting, cond-mat/0008156.

\bibitem{csy} A.~V.~Chubukov, S.~Sachdev, and J.~Ye,
Phys.  Rev. B {\bf 49}, 11919 (1994).

\bibitem{nagaosa} N.~Nagaosa and P.~A.~Lee, Phys. Rev. B {\bf 61}, 9166
(2000).

\bibitem{sb} S.~Sachdev and R.~N.~Bhatt, \prb {\bf 41}, 9323 (1990).

\bibitem{andrey} A.~V.~Chubukov and Th.~Jolicoeur,
Phys. Rev. B {\bf 44}, 12050 (1991)

\bibitem{rgs} T.~M.~Rice, S.~Gopalan, M.~Sigrist, Europhys. Lett.
{\bf 23}, 445 (1993).

\bibitem{grs} S.~Gopalan, T.~M.~Rice, and M.~Sigrist, \prb {\bf
49}, 8901 (1994).

\bibitem{ps} J~ Piekarewicz and J.~R.~Shepard, \prb {\bf 60}, 9456
(1999).

\bibitem{valeri} V.~N.~Kotov, O.~Sushkov, Zheng Weihong, and
J.~Oitmaa, \prl {\bf 80}, 5790 (1998); V.~N.~Kotov, J.~Oitmaa,
O.~Sushkov, and Zheng Weihong, Phil. Mag. B {\bf 80}, 1483 (2000).

\bibitem{mgi} Y.~Matsushita, M.~P.~Gelfand, and C.~Ishii,
J. Phys. Soc. Jpn., {\bf 68} 247 (1999).

\bibitem{tmu} K.~Totsuka, S.~Miyahara, K.~Ueda, \prl {\bf 86},
520 (2001).

\bibitem{cb} D.~Carpentier and L.~Balents, cond-mat/0102218.

\bibitem{so5} R.~Eder, A.~Dorneich, M.~G.~Zacher, W.~Hanke, and S.-C.~Zhang,
Phys. Rev. B {\bf 59}, 561 (1999); A. Furusaki, S.-C.~Zhang, Phys.
Rev. B {\bf 60}, 1175 (1999).

\bibitem{eugene} E.~Demler and S.~Das Sarma,
Phys. Rev. Lett. {\bf 82}, 3895 (1999).

\bibitem{sommer}  T.~Sommer, M.~Vojta, and K.~W.~Becker,
cond-mat/0104356.

\bibitem{lee} Y.~L.~Lee, Y.~W.~Lee, C.-Y.~Mou, and Z.~Y.~Weng,
Phys. Rev. B {\bf 60}, 13418 (1999).

\bibitem{eder} R.~Eder, Phys. Rev. B {\bf 57}, 12832 (1998).

\bibitem{sushkov2} O.~P.~Sushkov, cond-mat/0002421.

\bibitem{vb} M.~Vojta and K.~W.~Becker, \prb {\bf 60}, 15201
(1999).

\bibitem{brenig} C.~Jurecka and W.~Brenig, cond-mat/0103511.

\bibitem{book} S.~Sachdev, {\it Quantum Phase Transitions},
Cambridge University Press, Cambridge (1999).

\bibitem{zacher} M.~G.~Zacher, R.~Eder, E.~Arrigoni, and W.~Hanke,
cond-mat/0103030.

\bibitem{granath} M.~Granath, V.~Oganesyan, S.~A.~Kivelson, E.~Fradkin,
and V.~J.~Emery, cond-mat/0010350.

\bibitem{carlos} R.~D.~Duncan and C.~A.~R.~S\'{a} de Melo, Phys.
Rev. B {\bf 62}, 9675 (2000).

\bibitem{chetan} C.~Nayak, Phys. Rev. B {\bf 62}, 4880 (2000).

\bibitem{braden} L.~Pintschovius and M.~Braden, Phys. Rev. B {\bf 60}, R15039
(1999).

\bibitem{mook2} H.~A.~Mook and F.~Dogan, Nature {\bf 401}, 145 (1999);
H.~A.~Mook, P.~Dai, F.~Dogan, and R.~D.~Hunt, Nature {\bf 404},
729 (2000).

\bibitem{petrov} Y.~Petrov, T.~Egami, R.~J.~McQueeney, M.~Yethiraj,
H.~A.~Mook, and F.~Dogan, cond-mat/0003414.

\bibitem{horsch} G.~Khaliullin and P.~Horsch, Physica C {\bf
282-287}, 1751 (1997); P.~Horsch, G.~Khaliullin, and V.~Oudovenko,
Physica C {\bf 341-348}, 117 (2000).

\bibitem{gros1} C.~Gros and R.~Werner, Phys. Rev. B {\bf 58},
R14677 (1998).

\bibitem{gros2} R.~Werner, C.~Gros, and M.~Braden, Phys. Rev. B {\bf 59},
14356 (1999).

\bibitem{girvin} M. J. Massey, R. Merlin, and S. M. Girvin, Phys.
Rev. Lett. {\bf 69}, 2299 (1992).

\bibitem{zx}  Z.-X.~Shen, A.~Lanzara, and N.~Nagaosa,
cond-mat/0102224.

\bibitem{neutron} G.~L.~Squires, {\em Introduction to the Theory
of Thermal Neutron Scattering}, Dover Publications, New York
(1978); see also http://rrdjazz.nist.gov/resources/n-lengths/.

\bibitem{greg} S.~Ono, Yoichi Ando, T.~Murayama, F.~F.~Balakirev,
J.~B.~Betts, and G.~S.~Boebinger, Phys. Rev. Lett. {\bf 85}, 638
(2000).

\bibitem{dsz} E.~Demler, S.~Sachdev, and Y.~Zhang,
Phys. Rev. Lett. {\bf 87}, 067202 (2001).

\bibitem{kam} J.~C.~Wynn, D.~A.~Bonn, B.~W.~Gardner, Yu-Ju Lin,
Ruixing Liang, W.~N.~Hardy, J.~R.~Kirtley, and K.~A.~Moler,
preprint.

\bibitem{snl} S. Sachdev, Phys. Rev. B {\bf 45}, 389 (1992);
N.~Nagaosa and P.~A.~Lee, Phys. Rev. B {\bf 45}, 966 (1992); Jung
Hoon Han and Dung-Hai Lee, Phys. Rev. Lett. {\bf 85}, 1100 (2000);
M.~Franz and Z.~Tesanovic, Phys. Rev. B {\bf 63}, 064516 (2001).

\bibitem{fischer} Ch.~Renner, B.~Revaz, K.~Kadowaki,
I.~Maggio-Aprile, and O.~Fischer, Phys. Rev. Lett. {\bf 80}, 3606
(1998); S.~H.~Pan, E.~W.~Hudson, A.~K.~Gupta, K.-W.~Ng, H.~Eisaki,
S.~Uchida, and J.~C.~Davis, Phys. Rev. Lett. {\bf 85}, 1536
(2000).

\bibitem{shuheng} S.~H.~Pan, J.~P.~O'Neal, R~.L.~Badzey, C.~Chamon,
H.~Ding, J.~R.~Engelbrecht, Z.~Wang, H.~Eisaki, S.~Uchida,
A.~K.~Gupta, K.-W.~Ng, E.~W.~Hudson, K.~M.~Lang, and J.~C.~Davis,
preprint.

\bibitem{sushkov1}  O.~P.~Sushkov, J.~Oitmaa, and Zheng Weihong,
Phys. Rev. B {\bf 63}, 104420 (2001).

\bibitem{leiden} M.~S.~L.~du Croo de Jongh, J.~M.~J. van Leeuwen, W.~van
Saarloos, Phys. Rev. B {\bf 62}, 14844 (2000).


\bibitem{campbell1} S.~Mazumdar, S.~Ramasesha, R.~T.~Clay, and
D.~K.~Campbell, Phys. Rev. Lett. {\bf 82}, 1522 (1999).

\bibitem{campbell2} S.~Mazumdar, R.~T.~Clay, and D.~K.~Campbell,
cond-mat/9910164.

\bibitem{campbell3} S.~Mazumdar, R.~T.~Clay, and D.~K.~Campbell,
Phys. Rev. B 62, 13400 (2000).

\bibitem{kfe} S.~A.~Kivelson, E.~Fradkin, and V.~J.~Emery, Nature
{\bf 393}, 550 (1998).

\bibitem{proximity} V.~J.~Emery, S.~A.~Kivelson, and O.~Zachar,
Phys. Rev. B {\bf 56}, 6120 (1997).

\bibitem{sp} S. Sachdev, Phys. Rev. B {\bf 40}, 5204 (1989).

\bibitem{erica} E.~W.~Carlson, D.~Orgad, S.~A.~Kivelson, and V.~J.~Emery,
Phys. Rev. B {\bf 62}, 3422 (2000).

\bibitem{jan} J.~Zaanen, O.~Y.~Osman, H.~V.~Kruis, Z.~Nussinov, and
J.~Tworzydlo, cond-mat/0102103.

\bibitem{qptd} M.~Vojta, Y.~Zhang, and S.~Sachdev, Phys. Rev. Lett.
{\bf 85}, 4940 (2000); Int. J. Mod. Phys. B {\bf 14}, 3719 (2000).

\bibitem{ws} S.~R.~White and D.~J~Scalapino, Phys. Rev. Lett.
{\bf 80}, 1272 (1998); {\bf 81}, 3227 (1998); \prb {\bf 60}, R753
(1999).

\bibitem{neto} A.~H.~Castro Neto, cond-mat/0102281.

\bibitem{lannert} C.~Lannert, M.~P.~A.~Fisher, and T.~Senthil,
Phys. Rev. B {\bf 63}, 134510 (2001).

\bibitem{jkkn} J.~V.~Jose, L.~P.~Kadanoff, S.~Kirkpatrick, and D.~R.~Nelson,
Phys. Rev. B {\bf 16}, 1217 (1977).

\end{thebibliography}
\end{document}